\documentclass[11pt]{article}
\usepackage[margin=1in]{geometry} 
\usepackage[T1]{fontenc}
\usepackage[utf8]{inputenc} 
\usepackage{lmodern}
\usepackage{microtype}

\usepackage{authblk}
\usepackage{graphicx}
\usepackage{amsmath}
\usepackage{amssymb}
\usepackage{amsthm}
\usepackage{color}
\usepackage{url}
\usepackage{hyperref}
\usepackage{enumitem}
\usepackage{braket}
\usepackage{enumitem}
\usepackage{booktabs}
\usepackage{stmaryrd}
\usepackage{tikz}
\usetikzlibrary{quantikz2}
\usepackage{setspace}
\usepackage[table,xcdraw]{xcolor}
\usepackage{bm}

\usepackage{caption}
\usepackage[labelformat=simple]{subcaption}

\usepackage{circuitikz}
\usetikzlibrary{shapes.misc}
\usetikzlibrary{shapes.geometric}
\usetikzlibrary{decorations.pathmorphing}

\usepackage{algpseudocode}
\usepackage[linesnumbered,ruled,vlined]{algorithm2e}

\newcommand{\be}{\begin{equation}}
\newcommand{\ee}{\end{equation}}
\newcommand{\ba}{\begin{array}}
\newcommand{\ea}{\end{array}}
\newcommand{\bea}{\begin{eqnarray}}
\newcommand{\eea}{\end{eqnarray}}

\newtheorem{dfn}{Definition}

\newcommand{\app}[1]{\hyperref[app:#1]{Appendix~\ref*{app:#1}}}

\newcommand{\eq}[1]{Eq.~(\ref{eq:#1})}
\renewcommand{\sec}[1]{\hyperref[sec:#1]{Section~\ref*{sec:#1}}}
\newcommand{\ssec}[1]{\hyperref[ssec:#1]{Subsection~\ref*{ssec:#1}}}
\newcommand{\fig}[1]{\hyperref[fig:#1]{Figure~\ref*{fig:#1}}}
\newcommand{\tab}[1]{\hyperref[tab:#1]{Table~\ref*{tab:#1}}}
\newcommand{\lem}[1]{\hyperref[lem:#1]{Lemma~\ref*{lem:#1}}}
\newcommand{\propos}[1]{\hyperref[propos:#1]{Proposition~\ref*{propos:#1}}}
\newcommand{\thm}[1]{\hyperref[thm:#1]{Theorem~\ref*{thm:#1}}}
\newcommand{\alg}[1]{\hyperref[alg:#1]{Algorithm~\ref*{alg:#1}}}

\SetCommentSty{mycommfont}

\SetKwInput{KwInput}{Input}                
\SetKwInput{KwOutput}{Output}              

\title{FPGA-tailored algorithms for real-time decoding of quantum LDPC codes}

\author[1]{Satvik Maurya}
\author[2]{Thilo Maurer}
\author[2]{Markus B{\"u}hler}
\author[2]{Drew Vandeth}
\author[2]{Michael E. Beverland}
\affil[1]{University of Wisconsin-Madison}
\affil[2]{IBM Quantum}

\begin{document}
\maketitle

\begin{abstract}
Real-time decoding is crucial for fault-tolerant quantum computing but likely requires specialized hardware such as field-programmable gate arrays (FPGAs), whose parallelism can alter relative algorithmic performance.
We analyze FPGA-tailored versions of three decoder classes for quantum low-density parity-check (qLDPC) codes: message passing, ordered statistics, and clustering.
For message passing, we analyze the recently introduced Relay decoder and its FPGA implementation; for ordered statistics decoding (OSD), we introduce a filtered variant that concentrates computation on high-likelihood fault locations; and for clustering, we design an FPGA-adapted generalized union-find decoder.
We design a systolic algorithm for Gaussian elimination on rank-deficient systems that runs in linear parallel time, enabling fast validity checks and local corrections in clustering and eliminating costly full-rank inversion in filtered-OSD.
Despite these improvements, both remain far slower and less accurate than Relay, suggesting message passing is the most viable route to real-time qLDPC decoding.
\end{abstract}

\newpage
\tableofcontents

\newpage

\section{Introduction and summary of main results}
\label{sec:intro}

Fault-tolerant quantum computation (FTQC) enables large-scale quantum algorithms by using quantum error correction (QEC) to protect logical information from the inevitable noise in physical qubits and operations.
A central challenge in realizing FTQC is the implementation of a \emph{real-time decoder}, a classical algorithm that corrects errors fast enough to keep pace with successive QEC cycles.
For superconducting qubits, with microsecond QEC cycle times~\cite{Battistel_2023}, this will likely require specialized computing hardware such as field-programmable gate arrays (FPGAs) rather than central processing units (CPUs), which in turn influences the choice of decoder because the platforms differ fundamentally. 
Unlike CPUs, which are the target platforms for most decoders, FPGAs comprise millions of small, interconnected processing elements pre-configured to execute fixed local functions in parallel to collectively realize a specified computation.

The appropriate choice of decoder can also depend on the underlying QEC code.
Recent theoretical advances in quantum low-density parity-check (qLDPC) codes, combined with improving qubit connectivity, have made these codes increasingly attractive alternatives to the surface code, offering comparable protection with far fewer physical qubits.
However, most existing FPGA decoders~\cite{Battistel_2023,vittal2023astrea,wu2025micro,ziad2024local,liyanage2023scalable} exploit specific features of the surface code that do not apply to qLDPC codes.
This motivates a comparative study of decoder classes suitable for real-time FPGA implementation of general qLDPC codes.

\begin{figure}[h]
    \centering
    \includegraphics[width=0.99\linewidth]{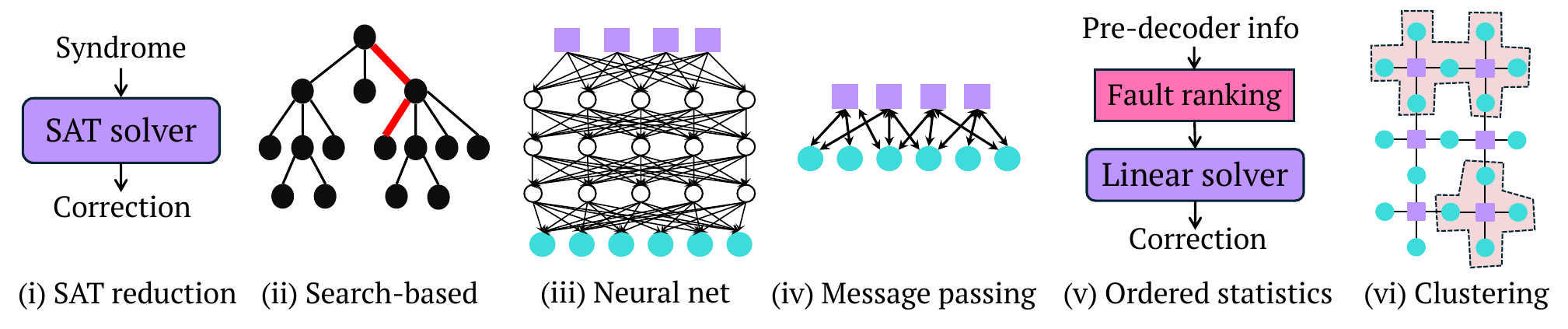}
    \caption{\textbf{Classes of qLDPC decoder.}
    Among these, classes (iv)–(vi) are identified as the most suitable for real-time decoding in FPGAs and are evaluated in this work.
    }
    \label{fig:decoder-classes}
\end{figure}

Existing decoders for general qLDPC codes can be grouped into six classes (\fig{decoder-classes}): (i) SAT-reduction, (ii) search-based, (iii) neural-network, (iv) message-passing, (v) ordered-statistics, and (vi) clustering.
SAT-reduction/SMT-based~\cite{Berent_2024,noormandipour2024maxsatdecodersarbitrarycss,shutty2022decoding,Zhou2025} and search-based~\cite{Ott2025,beni2025tesseract,wu2025minimum} decoders have variants that guarantee minimum-weight corrections, but they remain orders of magnitude slower than other approaches and rely on complex data structures and conditional logic that are inefficient on FPGAs~\cite{besta2019graphprocessingfpgastaxonomy,Halstead2011,Luo2017}.
Neural-network decoders~\cite{hu2025efficient,blue2025machine,Gong2024,Ninkovic2024,maan2025decoding} appear too slow for microsecond QEC cycles, and are better suited to GPUs than to FPGAs.
We therefore focus on the remaining three classes: message passing, ordered-statistics decoding (OSD), and clustering.
Our primary technical contributions are:
\begin{enumerate}[noitemsep]
\item An FPGA-tailored \emph{filtered-OSD}, avoiding standard OSD's costly explicit matrix inversion.
\item An FPGA-tailored \emph{cluster decoder} based on a generalized Union–Find decoder~\cite{delfosse2022toward}.
\item A \emph{systolic-array solver} for rank-deficient linear systems, used in both of the above decoders.
\item A comparison of logical error rates for these decoders against Relay message-passing~\cite{muller2025improved} under fixed FPGA cycle 
budgets to identify the most promising class for real-time decoding.

\end{enumerate}

Our FPGA-tailored algorithms use basic operations organized for efficient FPGA execution, yielding runtime and resource costs that differ sharply from CPU implementations of the same logic.
We model decoding via a binary \emph{decoding matrix} $H \in \mathbb{F}_2^{M\times N}$ that maps fault vectors 
$x \in \mathbb{F}_2^N$ to syndromes $\sigma \in \mathbb{F}_2^M$ through $Hx = \sigma$. 
Given $\sigma$, the decoder seeks a likely correction $y$ with $Hy = \sigma$.

\paragraph{Message-passing decoders.}
These algorithms exchange information iteratively between check and fault nodes in a graphical representation of the decoding matrix, with local update rules that map naturally onto parallel FPGA architectures~\cite{pamuk2011fpga,valls2021syndrome}.
The standard example is Min-Sum belief propagation (BP), which often fails to converge for quantum LDPC codes \cite{poulin2008on,maan2025decoding} in part due to loops and symmetries in the decoding graph.
Recent advances address these issues through decimation to break loops \cite{du2022stabilizer,gong2024toward,yao2024belief,tsubouchi2025degeneracy,du2024check,yin2024symbreak,demarti2024closed,demarti2024almost}, incorporating memory to damp message oscillations \cite{chytas2025enhanced,murphy2013loopy,muller2025improved}, and decoder ensembling \cite{koutsioumpas2025automorphism}, while others learn modified message-update rules by pretraining~\cite{liu2019neural}.
The recently introduced Relay decoder~\cite{muller2025improved} performs a sequence of BP runs (“legs”) with varying memory-strength parameters, progressively breaking trapping sets to improve convergence and accuracy.
We use Relay as a representative benchmark for the message-passing class, drawing on published FPGA performance analyses~\cite{muller2025improved,maurer2025realtimedecoding}.

\paragraph{Ordered-statistics decoders.}
OSD algorithms~\cite{panteleev2021degenerate,fossorier1995soft,fossorier2002iterative} leverage an unconverged pre-decoder’s output by reprocessing bits according to their estimated reliability.
Following Ref.~\cite{roffe2020}, the standard OSD selects a set $S$ of linearly independent columns of $H$, ordered by likelihood, with $|S|=\operatorname{rank}(H)$.
The corresponding submatrix $H_S \in \mathbb{F}_2^{M\times |S|}$ is inverted to obtain a correction $x = H_S^{-1}\sigma$ supported on $S$.
However, standard OSD is impractical on FPGAs~\cite{valls2021syndrome} because detecting independence and inverting the large $\operatorname{rank}(H)\!\times\!\operatorname{rank}(H)$ matrix dominate runtime, even when optimized~\cite{bascones2025exploring}.

Our first technical contribution is an FPGA-tailored variant, \emph{filtered-OSD}, which avoids explicit matrix inversion by solving
$H_{S'}x=\sigma$ directly on a smaller, potentially rank-deficient submatrix $H_{S'}$ (similar to Alg.2 in Ref~\cite{panteleev2021degenerate}).
We obtain $S'$ from a pre-decoder by discarding low-probability columns. 
Because $|S'| \ll \operatorname{rank}(H)$, column sorting and linear solving are both substantially accelerated.
Sparsity further reduces cost because many rows of $H_{S'}$ are zero and can be skipped by the solver.
This approach requires an FPGA-tailored solver for rank-deficient binary systems, described next.

\paragraph{Systolic rank-deficient solver.}
We provide a new linear-time systolic-array algorithm that solves $Ax=b$ for any matrix $A \in \mathbb{F}_2^{m\times n}$ and vector $b \in \mathbb{F}_2^{m}$.
\emph{Systolic arrays} are a class of highly parallel algorithms that implement computations on grids of processing elements and pass data locally each cycle, making them particularly well suited to FPGAs.
Existing FPGA solvers typically require $A$ has either full column or row rank (or both)~\cite{gentleman1982matrix,Hochet1989,cosnard1986matching,robert1985resolution,wang1993systolic,parkinson1984compact,Rupp2006,rupp2011hardware,jasinski2010gf2,scholl2013hardware,scholl2014hardware,Shoufan2010,wang2017fpga,bascones2025exploring}, while the only prior solver for arbitrary $A$~\cite{hu2024universal} employs relatively complex processing elements.
Our solver is simpler and may be of independent interest; it forms the computational core of our filtered-OSD and cluster decoder.

\paragraph{Cluster decoders.}
This class of decoders~\cite{kovalev2013fault,delfosse2021almost,delfosse2022toward,higgott2023improved,hillmann2024localized,wolanski2024ambiguity} group nearby detection events in the decoding graph that are likely caused by the same underlying faults, iteratively growing and merging clusters until each can be locally corrected.
Mathematically, by the end of the algorithm each cluster $C$ corresponds to a set of columns of $H$ with local corrections $x_C$ satisfying $H_C x_C=\sigma_C$, where $\sigma_C$ is the restriction of $\sigma$ to the neighborhood of $C$.
The local corrections are then combined to obtain a global $x$ obeying $Hx=\sigma$.
Initially, clusters may have no solution to $H_C x_C=\sigma_C$, indicating that they must be grown before becoming solvable.
Cluster-decoder variants differ in their growth and merge rules and may exploit code structure. 
For example, the Union–Find decoder~\cite{delfosse2021almost} avoids Gaussian elimination by leveraging surface-code structure.

We introduce an FPGA-tailored version of the algorithm in Ref.~\cite{delfosse2022toward}, 
which generalizes the Union–Find decoder~\cite{delfosse2017almost} to arbitrary qLDPC codes. 
This decoder uses the systolic solver to test cluster solvability and extract local corrections, and incorporates hardware-specific optimizations to maximize throughput.
As in filtered-OSD, it uses a pre-decoder: faults with high likelihoods are accepted immediately~\cite{caune2023belief}, 
while intermediate-likelihood faults are treated as erasures that seed the initial clusters together 
with unsatisfied syndrome nodes.

\begin{figure}[t]
    \centering
    \includegraphics[width=0.49\linewidth]{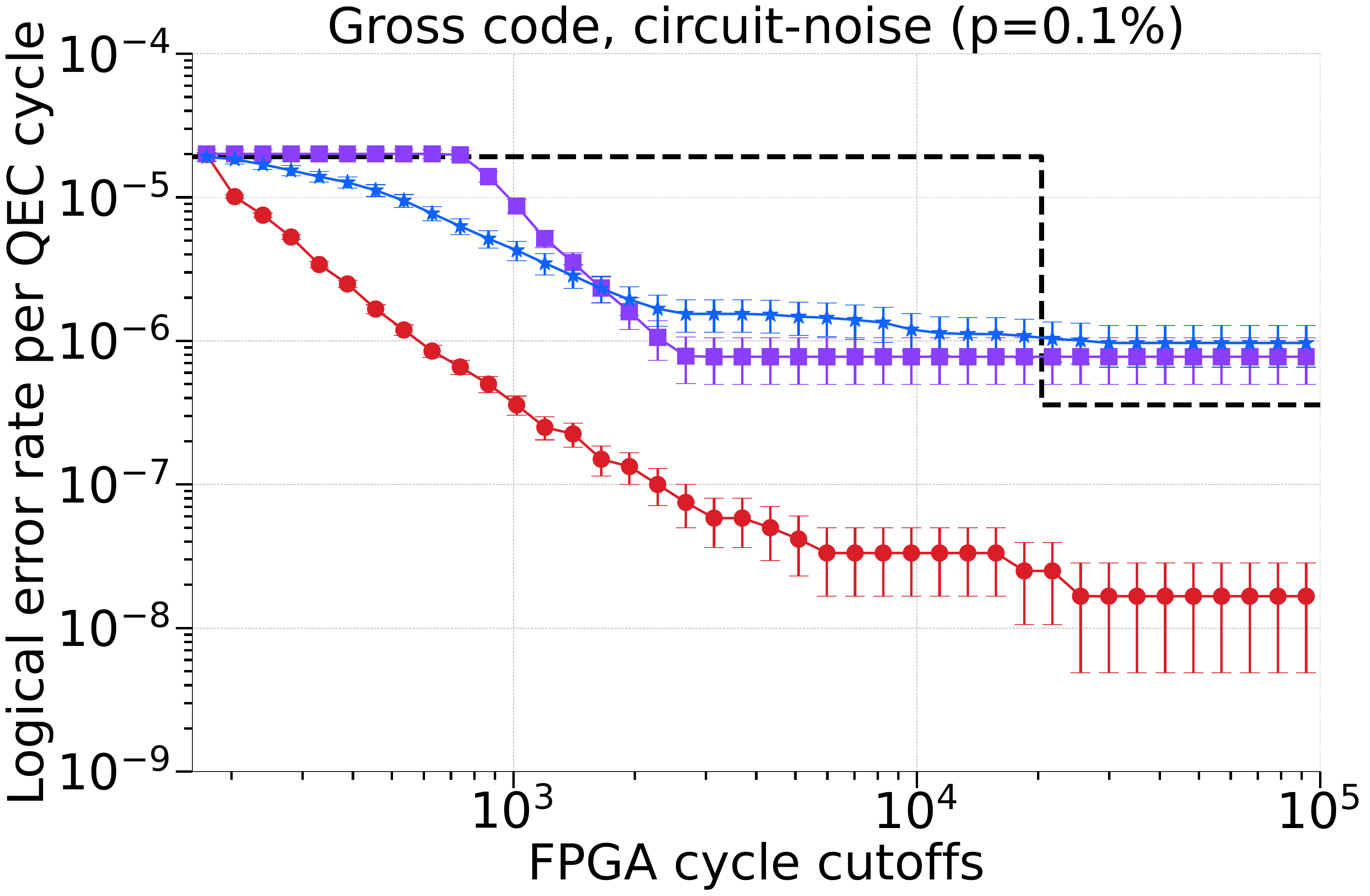}
    \hfill
    \includegraphics[width=0.49\linewidth]{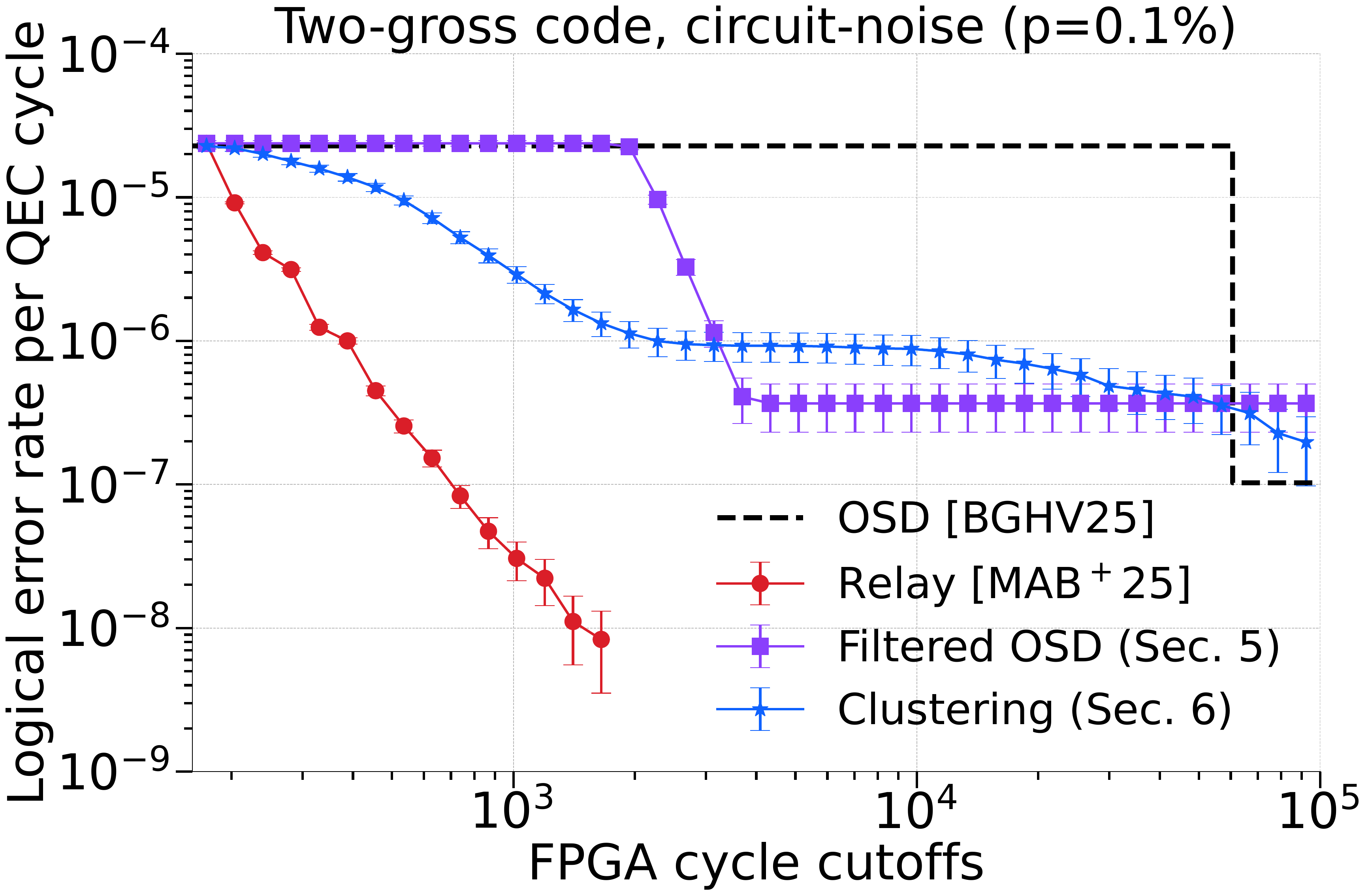}
    \caption{\textbf{Cutoff-time performance curves for FPGA-tailored decoders.} 
    Logical error rate versus FPGA cycle budget for Relay, filtered-OSD, cluster decoding, and standard OSD~\cite{bascones2025exploring}, with runs exceeding the cycle cutoff counted as failures.
    Relay achieves the lowest error rates across all tested budgets.
    Filtered-OSD starts up with significantly fewer cycles than standard OSD.
    }
    \label{fig:intro}
\end{figure}

\paragraph{Real-time decoding analysis.}
We compare the real-time feasibility of the aforementioned decoders in \fig{intro} using Monte Carlo simulations of the gross and two-gross qLDPC codes~\cite{bravyi2024high,yoder2025tour} to estimate logical error rates within a fixed FPGA cycle budget and to assess resource requirements on a representative commercial FPGA device (see~\sec{results}).
For a consistent comparison, we use the first leg of Relay as the pre-decoder for standard OSD, filtered-OSD, and cluster decoding, and all performance data in \fig{intro} are reported starting from the end of that leg.

Both our FPGA-tailored decoders achieve substantial runtime reductions relative to the standard OSD implementation of Ref.~\cite{bascones2025exploring}.
However, Relay reduces the logical error rate more rapidly with increasing cycle budget than the other three decoders, quickly reaching an order of magnitude lower rates for the gross code and two orders lower for the two-gross code.
Although \fig{intro} shows that Relay exhibits a long-tailed latency distribution, in \sec{real-time-decoding} we show that this tail does not lead to problematic backlog for our decoding examples, as they satisfy the latency-tail conditions we derive in \sec{real-time-background}.
These findings suggest that message-passing approaches, such as Relay, currently offer the most practical route to realizing real-time fault-tolerant decoding of large-scale quantum LDPC codes in FPGAs.

This overall conclusion appears unlikely to change without substantial new algorithmic advances in OSD or cluster decoding algorithms.
Although our cycle counts and resource estimates are obtained at the architectural level rather than from full Verilog implementations (see~\sec{results}), 
the observed performance gaps are simply too large to be bridged by modest clock-frequency variations. 
OSD could improve logical error rates by adding a combination-sweep (CS) stage~\cite{roffe2020}, but not enough to rival Relay~\cite{muller2025improved}.
Moreover, although CS adds negligible overhead on CPUs (where OSD’s elimination cost dominates over the matrix–vector multiplications in CS) it would cause a substantial slowdown on FPGAs, where elimination and matrix–vector multiplication have comparable linear parallel-time cost.
For similar reasons, although clustering could reuse partial elimination results~\cite{hillmann2024localized} for a speedup on CPUs, we expect little advantage on FPGAs where elimination is less of a bottleneck. 
Cluster decoders might also benefit from non-uniform, pre-decoder-biased growth~\cite{wolanski2024ambiguity}, but such extensions would introduce conditional logic and data-management overheads that reduce throughput.

\paragraph{Outline.}
In the remainder of this paper, we first provide background and definitions for QEC decoding and FPGA algorithms in \sec{qec-background} and \sec{fpga-background}.
Given the growing interest in using FPGAs for real-time decoding we take the opportunity to provide relatively self-contained technical introductions aimed at bridging the gaps between the communities.
We then present our systolic solver in \sec{gateware_lse} before presenting our FPGA-tailored decoders in \sec{fpga-osd} and \sec{fpga-cluster}, giving results and conclusions in \sec{results}.

\clearpage
\section{QEC decoding background and definitions}
\label{sec:qec-background}

This section provides a brief overview of quantum error correction (QEC), which protects quantum information by redundantly encoding logical qubits into many physical qubits and continuously monitoring for errors.
In this work, we consider a scenario where a dedicated circuit repeatedly measures the parity checks of a QEC code, producing classical \emph{syndrome} data that signals the occurrence of faults.
A \emph{decoder} is a classical algorithm that processes the syndrome stream to determine the required corrections.
For large-scale quantum computation, the decoder must operate in real time, processing syndrome data at the rate it is generated to prevent both stalling the quantum program and the accumulation of an unmanageable backlog~\cite{terhal2015quantum}.
This work focuses on implementing such real-time decoders in classical hardware.

In the rest of this section, we first review QEC codes, circuits, and the decoding process in \sec{qec-background-general}, followed by a discussion of the specific decoding algorithms relevant to this study.

\subsection{QEC codes, circuits, noise and decoding}
\label{sec:qec-background-general}

\begin{figure}[h]
\centering
\includegraphics[width=0.6\textwidth]{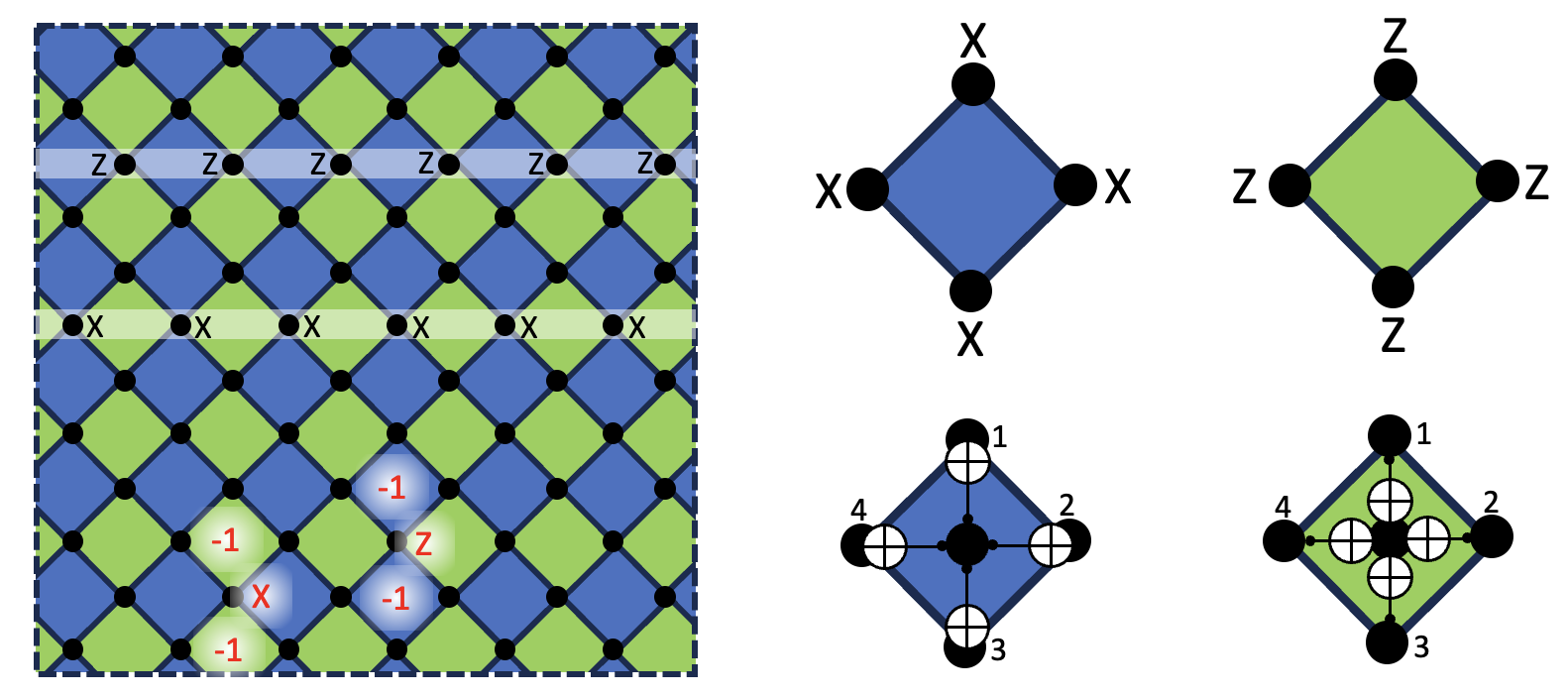}
\caption{
Distance-$6$ toric code on a square torus encoding $k=2$ logical qubits ($n=2d^2=72$ data qubits; black dots). 
Blue (green) faces denote $X$-type ($Z$-type) stabilizer generators, each acting on the four data qubits around a face. 
Representative $X$- and $Z$-type logical loops are highlighted. In the error-free case all stabilizer outcomes are $+1$; a single $Z$ ($X$) error flips the adjacent $X$-type ($Z$-type) checks (red). 
A standard circuit uses one ancilla per face: to measure $X$-checks, prepare $\ket{+}$, apply CNOTs with the ancilla as control to its neighbors, then measure in $X$; for $Z$-checks, prepare $\ket{0}$, use data qubits as controls into the ancilla, then measure in $Z$.
}
\label{fig:codes-and-circuits}
\end{figure}

\paragraph{Codes}
The most common class of quantum codes are \emph{stabilizer codes}.
A stabilizer code encodes $k$ logical qubits into $n$ physical qubits by defining the codespace as the joint $+1$ eigenspace of an abelian subgroup $S$ of the $n$-qubit Pauli group $\mathcal{P}_n$ that does not contain $-I$.
The subgroup $S$ is generated by $n-k$ independent \emph{stabilizer generators}, and its full group has size $2^{n-k}$.
Logical operators correspond to Pauli operators in the normalizer $N(S)$ that commute with all elements of $S$ but are not in $S$ (up to a phase).
The \emph{distance} $d$ of the code is the minimum weight (number of qubits acted upon) of such a logical operator and quantifies the number of errors the code can reliably correct.
We show an example in \fig{codes-and-circuits}.

In the simplest noise model, an unknown Pauli error $E$ acts on the system, after which the stabilizer generators are measured to produce a binary \emph{syndrome} vector whose $1$ entries correspond to $-1$ measurement outcomes.
Only generators that anticommute with $E$ produce $-1$ outcomes, allowing the syndrome to identify an error class.
A \emph{decoder} maps the syndrome to a Pauli correction $C$ with the same syndrome, returning the state to the codespace.
Up to a phase, the net operator $C E$ either belongs to $S$, in which case error correction succeeds, or to $N(S)\setminus S$, in which case a logical error has occurred.
Any error of weight strictly less than $d/2$ can be corrected.

\paragraph{Circuits}
In practice, stabilizer generators cannot be measured directly because quantum hardware only supports a limited set of physical operations.
Typical gate sets include single-qubit and two-qubit gates along with single-qubit measurements, and all stabilizer measurements must ultimately be decomposed into sequences of these elementary operations.
A standard approach is to introduce ancilla qubits that interact with the data qubits via two-qubit gates so that measuring the ancilla indirectly measures the desired stabilizer generator (see \fig{codes-and-circuits}).

Physical operations are subject to noise, which is more accurately modeled at the circuit level.
Because these noisy circuits produce unreliable stabilizer measurement outcomes, a standard strategy is to repeatedly execute the stabilizer measurement circuit.
Syndrome bits are obtained by taking the parity of consecutive outcomes for each generator, detecting faults that affect either the data qubits or the measurement process itself.

\paragraph{Noise and decoding}
In this work, we adopt the standard \emph{circuit noise model}, in which each gate, measurement, or idle period (a qubit not involved in any operation) has the same duration and fails independently with probability~$p$.
A fault is modeled as the ideal operation followed by a random non-trivial Pauli error on its qubits, and in the case of a measurement, a flip of the reported outcome. 

The complete mathematical description of the circuit decoding problem is specified by:
a binary decoding matrix $H \in \mathbb{F}_2^{M \times N}$,
a binary logical action matrix $A \in \mathbb{F}_2^{K \times N}$,
and a fault probability vector $\mathbf{p} \in [0,1]^N$.
There are $M$ detectors, which each take a value $0$ or $1$ during a noisy run of the circuit, forming a bitstring called the \emph{syndrome}, and $N$ distinct faults which occur independently, where the $j$th fault occurs with probability $p_j$. 
When the $j$th fault alone occurs, the observed syndrome is the $j$th column of $H$. 
When a set of faults occur together, their effect on the detectors is linear over $\mathbb{F}_2$.
Let the bitstring $F \in \mathbb{F}_2^{N}$ specify a set of faults (where $F_j = 1$ if and only if the $j$th fault occurs). 
The probability of $F$ is
\[
\Pr(F) = \prod_{j=1}^N (1-p_j) \left(\frac{p_j}{1-p_j} \right)^{F_j}.
\]
The syndrome $\sigma(F)$ is then $\sigma(F) = H F$,
where arithmetic (throughout this work, unless explicitly stated) is performed over $\mathbb{F}_2$.
The logical action of $F$ is
$a(F) = A F$.

A run of the noisy error correction circuit is modeled as follows:
The circuit is executed, and a set of (unknown) faults $F$ occurs with probability $\Pr(F)$. 
The syndrome bitstring $\sigma(F) = H F$ is known, even though $F$ is not. 
The syndrome is passed to a decoder $\mathcal{D}$, which is a classical algorithm with knowledge of $H$, $A$, and $\mathbf{p}$.
The decoder outputs a correction $\hat{F} = \mathcal{D}(\sigma(F))$.
If $a(F) + a(\hat{F}) = A(F+\hat{F}) = 0$, then we say that the decoder succeeds; otherwise, it fails.
Note that it is not necessary that $\hat{F} = F$ in order for the decoder to succeed.

We have not specified here exactly how to construct the decoding objects $H$, $A$ and $p$ from a specific circuit, or how to extract the syndrome $\sigma$ from measurement outcomes of the circuit in a given run, since the details of how this is done depends on the noise model.
We address this in more detail in our paragraph on decoding examples at the end of this subsection.

\paragraph{Stochastic and erasure faults}
It can be useful to separate faults into two classes: \emph{stochastic faults} (which occur with small probability in unknown locations) and \emph{erasure faults} (which occur with high probability, but in known locations).
In practice, these can arise due to quite different physical processes, with stochastic faults arising, for example, from dephasing errors, and erasure faults arising, for example, from a qubit that goes missing (with our knowledge) and is then replaced.
In the mathematical model we have presented, these two scenarios are simply covered by the probability of each fault --- with the probability $p_j$ of a stochastic fault being some small value, and the probability of an erasure fault being $p_j = 1/2$.

We can define a Tanner graph $\mathcal{G}$ for the decoding problem. 
$\mathcal{G}$ is a bipartite graph where each column of $H$ corresponds to a circular \emph{fault node}, and each row corresponds to a square \emph{check node}, and there is an edge connecting a pair of nodes if, and only if, the corresponding entry of $H$ is $1$.
Given the graph $\mathcal{G}$, we define the nodes adjacent to node $j$ as $\mathcal{N}(j)$.
We can think of the decoding problem in terms of the graph $\mathcal{G}$ as follows:
The set of faults $F$ that occur specifies a set of fault nodes in $\mathcal{G}$.
The syndrome $\sigma(F)$ is the set of check nodes of $\mathcal{G}$ that are connected to an odd number of fault nodes in $F$.
A correction $\hat{F}$ is a set of fault nodes such that each node in $\sigma(F)$ is connected to an odd number of fault nodes in $\hat{F}$.

We interchangeably use bitstring and set notation in a way that is standard in the QEC literature, using the same conventions as in Ref.~\cite{Ott2025}.
We use $j = 1,2,\dots,N$ to index fault nodes.
We then use $F$ to denote a set of fault nodes in two equivalent representations: as a set $F = \{j_1,j_2,\dots\}$, and as a bitstring $F \in \mathbb{F}_2^{N}$ with $F_j = 1$ if, and only if, $j \in \{j_1,j_2,\dots\}$.
Similarly, we use $\sigma$ to denote a set of check nodes both as a set $\sigma = \{i_1,i_2,\dots\}$ and as a bitstring $\sigma \in \mathbb{F}_2^{M}$ with $\sigma_i = 1$ if, and only if, $i \in \{i_1,i_2,\dots\}$.
When $U$ and $V$ are interpreted as sets, we use $U + V$ to denote their symmetric difference, consistent with $U + V$ being addition modulo two when $U$ and $V$ are interpreted as bitstrings.
When $U$ and $V$ are interpreted as bitstrings, we use $U \cup V$ to denote the (non-exclusive) OR, consistent with $U \cup V$ being the union when $U$ and $V$ are interpreted as sets.
Lastly, for $U$ and $V$ interpreted as sets, when $U \subseteq V$ we sometimes write $V - U$ or $V \setminus U$ to specify set difference (which is also equivalent to $U + V$ since arithmetic is over $\mathbb{F}_2$). 
The syndrome $\sigma$ of a fault set $F = \{j_1,j_2,\dots\}$ is $
\sigma = \mathcal{N}(j_1) + \mathcal{N}(j_2) + \dots = H F$, where arithmetic is over $\mathbb{F}_2$.

\begin{figure}[h]
\centering
\includegraphics[width=0.9\textwidth]{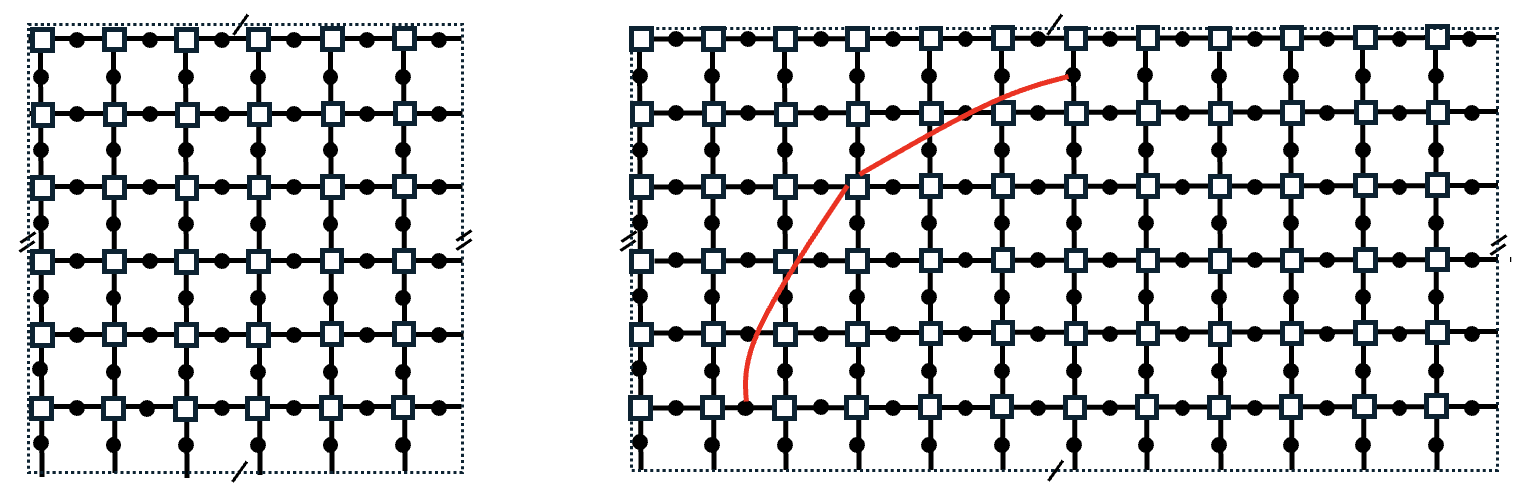}
\caption{Tanner graphs for a simple noise model where the circuit consists of the measurement of each check of the code (toric code on the left and gross code on the right), and the set of faults consists only of an $X$ error on each qubit prior to the check measurements.
Note that for the case of the gross code, we only draw the long-range edges (red) from one check, but all checks have similar edges which can be obtained by translation of those shown on the Torus.
}
\label{fig:Tanner-graphs}
\end{figure}

\paragraph{QEC circuit decoding examples.}
We benchmark our decoders on circuits for the $d=12$ (aka “gross”) and $d=18$ (aka “two-gross”) bivariate-bicycle codes of Refs.~\cite{bravyi2024high,bravyi2024github}, using the same setup. 
These decoding examples consist of $d$ noisy QEC rounds (implemented with an explicit circuit that measures all code checks) followed by one noiseless round to project to the codespace and test for a logical error.

We assume the standard circuit-level noise model: preparations and measurements fail with probability $p$ (state flip or outcome flip, respectively); idle gates experience $X$, $Y$, or $Z$ with probability $p/3$ each; CNOTs suffer one of the $15$ nontrivial two-qubit Pauli errors with probability $p/15$ each.
During each round, each check of the code is measured by the noisy circuit: we form a detector from the parity of a check's consecutive measurement (corresponding to a row of $H$), and each distinct failure event corresponds to a column of $H$. 
The probability $p_j$ is set to the value of the probability of fault $j$ occuring (for example, $p_j=p/15$ if fault $j$ is an $XY$ error on a CNOT). 
We set $H_{ij}=1$ if fault $j$ flips detector $i$, and set $H_{ij}=0$ otherwise.

Each row of $A$ corresponds to a logical operator generator (and as such there are $2k$ rows for a code encoding $k$ logical qubits).
Each fault produces a residual Pauli error on the data qubits, and we set $A_{ij}=1$ if the residual Pauli of fault $j$ anticommutes with logical operator generator $i$, and set $A_{ij}=0$ otherwise.
Because these codes are CSS~\cite{calderbank1996good,steane1996multiple}, $X$ and $Z$ decoding can be decoupled; here we consider $Z$-type decoding, with $X$-type decoding analogous.
This is done by separating rows of the decoding matrix $H$ into detectors formed from $X$-type check outcomes, and retaining only columns with support on those rows. 
The same set of columns are used to select a submatrix of $A$ and a subvector of $\mathbf{p}$.  
The resulting $Z$-type decoding problem is represented by a binary decoding matrix $H_Z \in \mathbb{F}_2^{M \times N}$,
a binary logical action matrix $A_Z \in \mathbb{F}_2^{K \times N}$,
and a fault probability vector $\mathbf{p}_Z \in [0,1]^N$.
The decoding matrices that arise in our two examples are $936 \times 8784$ and $2736 \times 26208$ for the gross and two-gross codes respectively.

\subsection{Real-time decoding and latency-tail conditions}
\label{sec:real-time-background}

A large-scale quantum computation may involve a sequence of more than $10^{10}$ QEC cycles which implement a sequence of fault-tolerant operations (`operations' for short). 
Here we discuss constraints that arise to ensure decoding keeps pace with the continuous stream of syndrome data produced by this sequence of QEC cycles.

In practice, decoding of the stream of syndrome data is broken into many smaller decoding problems, each involving a small overlapping batch of $\sim d$ consecutive cycles supported on a small number of code blocks.
For presentation clarity, we make the simplifying assumption that each decoding problem is disjoint: it corresponds to a single operation consisting of a contiguous sequence of QEC cycles on one code block, and different operations consist of non-overlapping sequences of cycles.  
This assumption lets us present the core ideas of real-time decoding without the added complexity of decoding problems that span overlapping windows of cycles and code blocks.

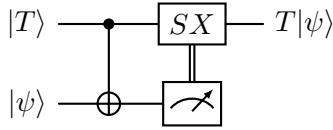
\begin{figure}[h]
    \centering
    \begin{quantikz}
    \lstick{$|T\rangle$} &     \ctrl{1} & \gate{SX} &  \rstick{$T|\psi\rangle$}\\
    \lstick{$|\psi\rangle$}&           \targ{} & \meter{} \wire[u][1]{c} 
    \end{quantikz}
    \caption{
    \textbf{Computational slowdown from decoding latency.}
    Many approaches to universal fault-tolerant quantum computation implement non-Clifford gates by injecting a $T$ state, as shown.
    This introduces a classically controlled feed-forward step: the $SX$ correction depends on a logical measurement outcome determined by the decoder.  
    As noted in Ref.~\cite{terhal2015quantum}, subsequent logical gates remain unspecified until this outcome is known, so decoding latency can slow the computation.
    }
    \label{fig:branching-and-latency}
\end{figure}

\paragraph{Computational slowdown from decoding latency.}
Many fault-tolerant architectures, including surface-code~\cite{fowler2012surface} and qLDPC-based~\cite{yoder2025tour} schemes, implement non-Clifford operations via state injection.
This introduces branching points in the computation, where the next operation depends on the decoder output from a previous operation (see \fig{branching-and-latency}). 
If the decoder requires $T_{\text{dec}}$ QEC cycles to return a result (for now we assume the decoder always takes the same amount of time for every decoding instance), the computation cannot proceed past the branch until that time. 

We distinguish two types of operations and their associated decoding problems.  
\emph{Logical operations} depend on earlier decoding outcomes, and decoding them produces results that may be needed to specify later logical operations, whereas  
\emph{idle operations} neither depend on earlier decoding nor produce outcomes required by subsequent logical operations. 
We define the resulting \emph{slowdown} as the ratio of the total duration of the computation as executed (including delays) to the duration of the `target duration' it would have if decoder latency were zero.
The decoding latency $T_{\text{dec}}$ sets a fundamental lower bound on the induced computational slowdown.

\paragraph{Decoding backlog.}
As new QEC cycles are streamed, the decoder must process each newly generated decoding problem every $T_{\text{gen}}$ cycles.
This immediately raises a scheduling question: what if the decoder has not finished the previous problem when the next one is generated?
If $T_{\text{dec}} > T_{\text{gen}}$, unresolved decoding tasks accumulate and form a backlog.  
A small backlog can be absorbed by inserting idle operations into the computation on the fly to postpone branching decisions until decoding completes.  
However, these idle cycles also require decoding and each takes longer to decode than to produce, so the backlog grows without progressing the computation causing a diverging slowdown.
This is known as the \emph{backlog problem}~\cite{terhal2015quantum}.

To avoid this compounding backlog when $T_{\text{dec}} > T_{\text{gen}}$, one can use multiple decoders per block and process subsequent time-windows in parallel to maintain throughput as the system scales~\cite{skoric2023parallel,tan2023scalable}, requiring coordinated scheduling~\cite{Maurya2024,khalid2025impacts}. 
This is only possible if the output of earlier time windows is not required in order to specify the decoding problem of later time windows~\cite{skoric2023parallel}.
In what follows, however, we assume one dedicated decoder per code block with no pooling across blocks. 
We also neglect all
latency sources other than decoding and the logical and idle operations themselves.

\paragraph{Variable decoding latency}
Many decoders do not have a fixed latency $T_{\text{dec}}$ but instead exhibit a distribution of latencies~$\tau$ that depend on the decoding instance.  
Variable decoding latency has been studied extensively for surface codes using numerical methods~\cite{Maurya2024,kurman2025benchmarking}; here we develop a general analytic approach that parallels the tail-latency criteria in Ref.~\cite{toshio2025decoder} while avoiding several of its technical complexities.

For decoders with worst-case and mean latencies $T_{\text{max}}$ and $T_{\text{avg}}$, there are three important cases:
\begin{enumerate}
    \item $T_{\text{max}} \le T_{\text{gen}}$,
    \item $T_{\text{avg}} < T_{\text{gen}}$ but $T_{\text{max}} > T_{\text{gen}}$,
    \item $T_{\text{avg}} > T_{\text{gen}}$.
\end{enumerate}
Case~1 is preferred, as the decoder is always faster than the syndrome-generation rate and no backlog will form.  
Case~3 must be avoided since it inevitably leads to a diverging backlog. 
Case~2 is more subtle; in the following we show that as long as the tails are not too long and do not occur with sufficiently low probability, the backlog problem can be avoided.

\begin{figure}[t]
    \centering
    (a)\includegraphics[width=0.52\linewidth]{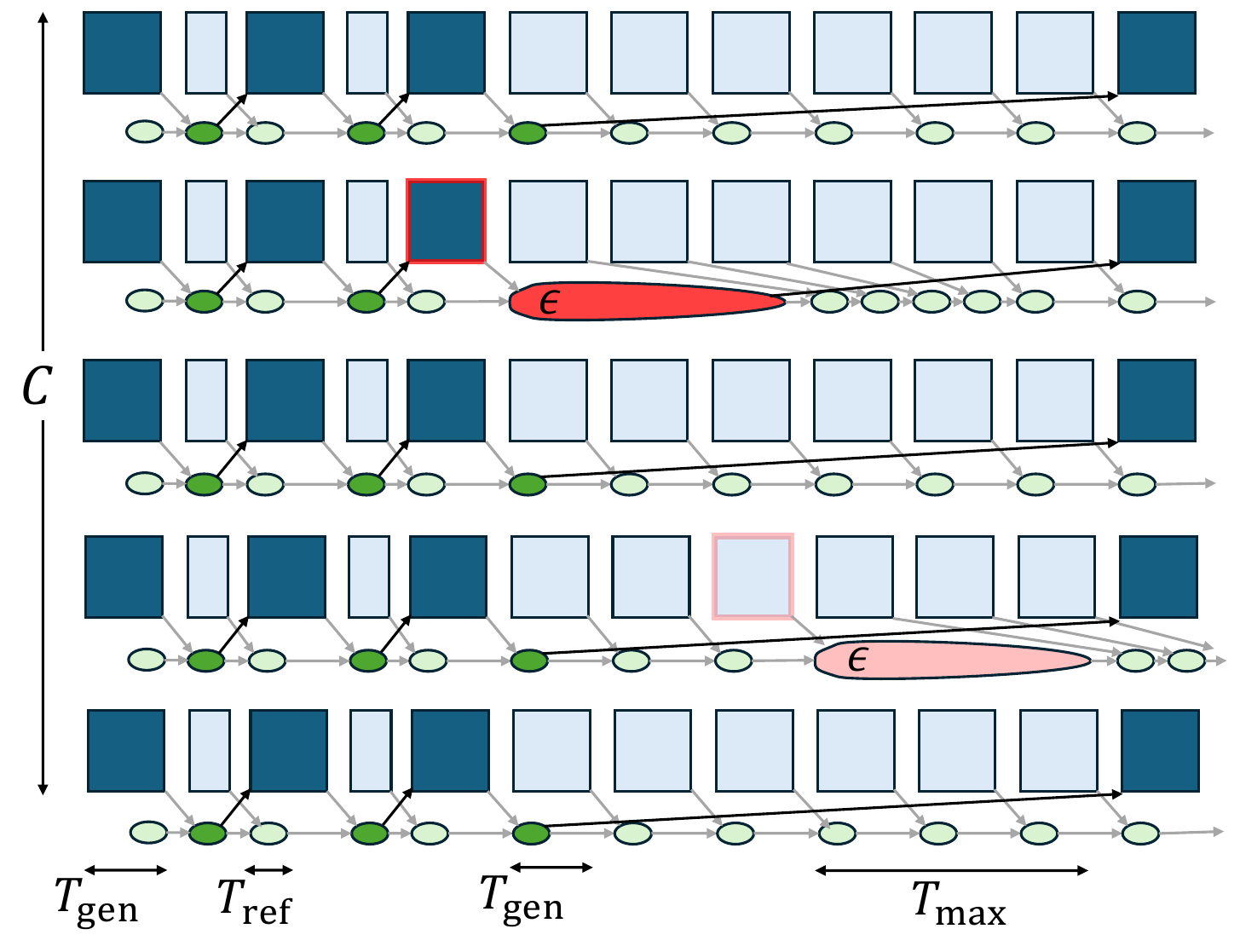}
    (b)\includegraphics[width=0.38\linewidth]{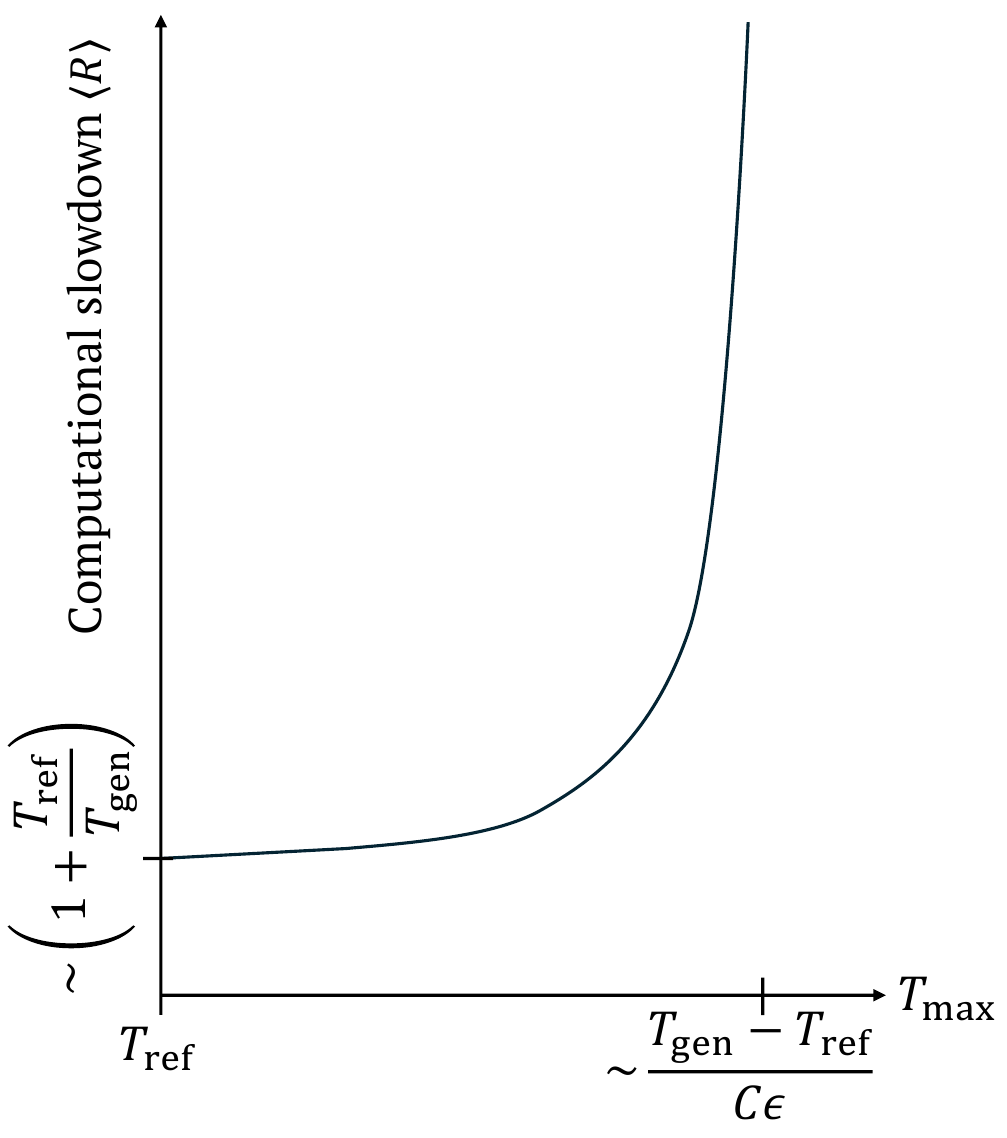}
    \caption{
    \textbf{Slowdown from variable decoding latency.}
    (a) Example decoding timeline for a computation with $C=5$ code blocks where each decoding task takes time $T_{\text{ref}} = T_{\text{gen}}/2$ with probability $1-\epsilon$, or $T_{\text{max}} = 3T_{\text{gen}}$ with probability $\epsilon$ (elongated ovals).  
    Dark squares indicate logical operations, light squares indicate idle operations and arrows indicate information passed to and from decoders. A slow decoding instance delays the computation by $T_{\text{max}}$, creating idle periods that themselves require decoding.  
    These delays can cascade if further slow decoding instances occur within the idle intervals introduced by earlier ones. 
    (b) The mean slowdown $\langle R \rangle$ grows slowly for small $T_{\text{max}}$, but increases sharply once $T_{\text{max}}$ approaches values near $(T_{\text{gen}} - T_{\text{ref}})/(C\epsilon)$, where cascading delays become likely.
    }
    \label{fig:tail_latencies}
\end{figure}

\paragraph{Latency-tail conditions.}
Consider a computation involving $C$ code blocks, each with a dedicated variable latency decoder with maximum decoding time $T_{\text{max}}$. 
Choose a reference time $T_{\text{ref}} < T_{\text{gen}}$ and let $\epsilon$ be the probability that a decoding run exceeds $T_{\text{ref}}$.  
If the following conditions are satisfied:
\begin{align}
\frac{T_{\text{ref}}}{T_{\text{gen}}} &< 1, \nonumber\\
\epsilon C\left\lceil
  \frac{T_{\text{max}} - T_{\text{ref}}}{T_{\text{gen}} - T_{\text{ref}}}
\right\rceil &< 1,
\label{eq:latency-tail-requirement}
\end{align}

then, the mean value $\langle R \rangle$ of the relative computational slowdown $R$ satisfies:
\begin{eqnarray}
\langle R \rangle 
&<& 
1 + \frac{T_{\text{ref}}}{T_{\text{gen}}}
   + \frac{2\gamma}{1 - \gamma}, \quad \text{where }
\gamma = \epsilon C \left\lceil
  \frac{T_{\text{max}} - T_{\text{ref}}}{T_{\text{gen}} - T_{\text{ref}}}
\right\rceil, \nonumber \\
&<&  1 + \frac{T_{\text{ref}}}{T_{\text{gen}}} + 2\epsilon C\left\lceil
  \frac{T_{\text{max}} - T_{\text{ref}}}{T_{\text{gen}} - T_{\text{ref}}}
\right\rceil +O(\gamma^2).
\label{eq:latency-tail-slowdown}
\end{eqnarray}

For a given decoder there is flexibility in how these constants are chosen, and
the bound tends to be tightest for small $T_{\text{ref}}/T_{\text{gen}}$ (e.g. $ \lesssim 0.5$),
moderate $T_{\text{max}}/T_{\text{gen}}$ (e.g. $ \lesssim 10$), and tiny $\epsilon$ (e.g. $\lesssim 10^{-4}$).  
One can freely choose $T_{\text{ref}}$ (which sets $\epsilon$).
If $T_{\text{max}}$ is very large and the logical error rate is well below the required target, one may impose a cutoff and declare failure when decoding exceeds a reduced $T_{\text{max}}$, thereby reducing the worst-case latency at the cost of a higher logical error rate.

\paragraph{Derivation.}
To show~\eqref{eq:latency-tail-slowdown}, we consider a worst-case target
computation consisting entirely of logical-operation layers, each of duration
$T_{\text{gen}}$ QEC cycles, and we upper bound all sources of latency.
We first specify the pessimistic modeling assumptions.  
Any decoding run that finishes within $T_{\text{ref}}$ is treated as taking
exactly $T_{\text{ref}}$, and any longer run is treated as taking
$T_{\text{max}}$.  
Between successive logical layers, when no slow decoding occurs, we assume
there is a single \emph{baseline} idle layer whose operations each last
$T_{\text{ref}}$ cycles.  
Whenever a slow decoding run occurs, a backlog is introduced: we extend the idle layer to $T_{\text{gen}}$ cycles, and continue inserting idle layers each of length $T_{\text{gen}}$ until the backlog has been eliminated. 
Finally, we assume that all decoding problems (whether arising from logical or long or short idle operations) have the same latency distribution.

Consider a single logical-operation layer.  
Associated with this layer are $C$ decoding runs for its logical operations and $C$ decoding runs for the idle layer immediately preceding it, for a total of $2C$ decoding jobs.  
By construction, all prior decoder jobs must have finished before this logical layer begins.  
If none of these $2C$ runs is slow, which occurs with probability $(1-\epsilon)^{2C}$, then only the baseline idle layer of duration $T_{\text{ref}}$ is inserted
before the next logical layer.  
In this case, the slowdown factor for this layer is $R_{\mathrm{fast}} = 1 + \frac{T_{\text{ref}}}{T_{\text{gen}}}$.

If at least one of the $2C$ decoder runs is slow, which occurs with probability $1-(1-\epsilon)^{2C}$, then $L$ idle layers (each lasting $T_{\text{gen}}$ QEC cycles) are inserted before the next logical layer begins.  
To ensure that the slow decoder run finishes and that the decoder can process the first $L-1$ inserted idle layers before the following logical layer starts (the final inserted idle layer begins decoding during the following logical layer), we require $LT_{\text{gen}} \ge (L-1) T_{\text{ref}} + T_{\text{max}}$.  
A sufficient choice is therefore $L = \lceil (T_{\text{max}} - T_{\text{ref}})/(T_{\text{gen}} - T_{\text{ref}}) \rceil$, provided none of the inserted idle decoding runs are slow (we address slow idle decoding runs below).  
If all $(L-1)C$ of these idle decoding runs are fast, which occurs with probability $(1-\epsilon)^{(L-1)C}$, then the slowdown for this layer is $R = 1 + L$.

We next consider the case where, after a slow decoding run in the first $2C$ decoding problems introduces $L$ idle layers, another slow run occurs within the first $L-1$ of these layers.  
To analyze this cascade, we partition the inserted idle layers into disjoint subsets called segments, labeled $s = 1,2,\dots,S$.  
Segment $1$ consists of the initial $L-1$ layers.  
Segment $s+1$ is created only if segment $s$ contains a slow decoding run, in which case all idle layers introduced by any slow runs in segment $s$ form segment $s+1$.  
A segment containing $l$ layers has at least one slow decoding run in its first $l-1$ layers with probability $1-(1-\epsilon)^{C(l-1)} \leq \epsilon C (l-1)$, using the bound
\begin{equation}
1-(1-x)^n \le xn \qquad \text{for all } 0<x<1,\; n\ge1.
\label{eq:prob-bound}
\end{equation}
Each segment contains at most $l \leq L-1$ layers, so the probability that a given segment has at least one slow decoding run is at most $1-(1-\epsilon)^{C(L-1)} < \epsilon C (L-1)$ (using \eq{prob-bound}).  
Conditioned on a slow decoding event occurring in the first $2C$ decoding runs, the expected number of idle layers is therefore at most
\[L+ (L-1) \sum_{s=1}^\infty (\epsilon C (L-1))^s = 1+ (L-1)/(1-\epsilon C (L-1)),\]
provided $\epsilon C (L-1) < 1$. 
Hence the mean relative slowdown satisfies
\begin{eqnarray}
\langle R \rangle &\le& [(1-\epsilon)^{2C}](1 + T_{\text{ref}}/T_{\text{gen}})
   +[1-(1-\epsilon)^{2C}] \left( 1 + \frac{L}{1-\epsilon C(L-1)} \right),\nonumber\\
   &\le& 1 + T_{\text{ref}}/T_{\text{gen}} + \frac{2\epsilon C L}{1-\epsilon C (L-1)} < 1 + T_{\text{ref}}/T_{\text{gen}} + \frac{2\gamma}{1-\gamma},\nonumber
\end{eqnarray}
using \eqref{eq:prob-bound} and the fact that $\gamma = \epsilon C L$.  
Since each layer of the computation can be analyzed similarly, this bounds the overall slowdown.

\subsection{Message passing decoding}
\label{sec:bp-decoding}

Message passing decoders iteratively pass messages between neighboring nodes in the decoding graph $\mathcal{G}$ that are used to make local guesses of which errors occurred until consistency with the syndrome is achieved.  
We describe two such message passing decoding algorithms.

\paragraph{Belief propagation decoding}
There are many versions of belief propagation (BP) applied to decoding. 
Here we present the MinSum variant described in Ref.~\cite{roffe2020}.

\emph{Initialization.}  
For all edges in $\mathcal{G}$, initial messages are set to
\begin{eqnarray}
m_{j \rightarrow i}(0) = \log{\left(\frac{1-p_j}{p_j}\right)}. \nonumber
\end{eqnarray}

\emph{Message updates.}  
In iteration $t$, messages are exchanged along edges of $\mathcal{G}$ in two steps:
\begin{align}
m_{i \rightarrow j}(t) &=(-1)^{\sigma_i}
\!\!\prod_{j' \in \mathcal{N}(i)\setminus\{j\}}\!\!\mathrm{sign}(m_{j'\rightarrow i}(t-1))
\;\times\!
\min_{j' \in \mathcal{N}(i)\setminus\{j\}}\!|m_{j'\rightarrow i}(t-1)|,
\label{eq:bp-check-to-error}\\[4pt]
m_{j \rightarrow i}(t) &= \log{\left(\frac{1-p_j}{p_j}\right)} + \sum_{i' \in \mathcal{N}(j)\setminus\{i\}} m_{i'\rightarrow j}(t).
\label{eq:bp-error-to-check}
\end{align}

\emph{Marginals and decision.}  
The marginal log-likelihood ratio (LLR) for fault $j$ and the corresponding hard decision are
\begin{eqnarray}
\Lambda_j(t) &=& \log{\left(\frac{1-p_j}{p_j}\right)} + \sum_{i' \in \mathcal{N}(j)} m_{i'\rightarrow j}(t)
\label{eq:llr},\\
\hat{F}_j(t)&=&\begin{cases}
1, & \text{if $\Lambda_j(t) > 0$,}\\
0, & \text{otherwise.}
\end{cases}\nonumber
\end{eqnarray}

\emph{Convergence criterion.}  
If $H\hat{F}(t) = \sigma$, the decoder halts and outputs $\hat{F}$; otherwise, iterations continue up to a maximum limit, after which the decoder is declared to have failed to converge.

\paragraph{Relay decoding}
Relay decoding generalizes BP by introducing local memory into the message updates to mitigate oscillations and trapping effects.  
Each decoding \emph{leg} runs a modified BP procedure, \emph{Disordered Memory Belief Propagation} (DMem-BP), which passes real-valued messages between check nodes $i$ and variable (fault) nodes $j$ on the Tanner graph $G$.

\emph{Initialization.}  
At the start of leg~1, for all edges $(j,i)$ in $\mathcal{G}$, messages and parameters are set as
\begin{equation}
\nu_{j\to i}(0) =\lambda_j(0)=\Lambda_j(0)=\log\!\left(\frac{1-p_j}{p_j}\right).
\nonumber
\end{equation}

\emph{Message updates.}  
In iteration $t$ in a leg, messages are exchanged along edges of $\mathcal{G}$ in two steps:
\begin{align}
\mu_{i\to j}(t)
&= (-1)^{\sigma_i} \!\!\prod_{j' \in \mathcal{N}(i)\setminus\{j\}}\!\!\mathrm{sign}(\nu_{j'\to i}(t-1))
\;\times\! \min_{j'\in\mathcal{N}(i)\setminus\{j\}}\!|\nu_{j'\to i}(t-1)|,
\nonumber\\[3pt]
\nu_{j\to i}(t)
&= \lambda_j(t) + \!\!\sum_{i'\in\mathcal{N}(j)\setminus\{i\}}\!\!\mu_{i'\to j}(t).
\nonumber
\end{align}

\emph{Marginals and decision rule.}  
The marginal for fault $j$ and the corresponding hard decision are
\begin{eqnarray}
\Lambda_j(t) &=& \lambda_j(t) + \sum_{i'\in\mathcal{N}(j)} \mu_{i'\to j}(t),
\nonumber\\
\hat{F}_j(t)&=&\begin{cases}
1, & \text{if $\Lambda_j(t) < 0$,}\\
0, & \text{otherwise.}
\end{cases}
\nonumber
\end{eqnarray}

\emph{Convergence criterion.} If $H\hat{F}(t)=\sigma$, the decoder halts and outputs $\hat{F}(t)$.

\emph{Memory update.}  
The local bias term evolves as
\begin{equation}
\lambda_j(t)
= (1-\gamma_j)\lambda_j(0) + \gamma_j\,\Lambda_j(t-1),
\qquad \gamma_j\in[0,1],
\nonumber
\end{equation}
with $\gamma_j$ controlling the degree of memory; $\gamma_j=0$ recovers MinSum BP.

\emph{Relay chaining.}  
Relay executes up to $R$ legs, each parameterized by $(\bm{\Gamma}_r,T_r)$, where $\bm{\Gamma}_r$ is the list of memory strengths $(\gamma_1, \gamma_2,\dots, \gamma_N)$ for leg $r$, and $T_r$ is the max iterations of leg $r$.  
For $r\ge2$, leg~$r$ is initialized using the previous leg’s terminal marginals:
\begin{equation}
\Lambda_j^{(r)}(0)=\Lambda_j^{(r-1)}(T_{r-1}), \quad
\lambda_j^{(r)}(0)=\Lambda_j^{(r)}(0), \quad
\nu_{j\to i}^{(r)}(0)=\lambda_j^{(r)}(0),
\nonumber
\end{equation} 
Each leg terminates when a valid $\hat{F}$ is found or after $T_r$ iterations.
There is also an option in Relay to continue until $S$ solutions are obtained, but in this work, we assume $S=1$.

\subsection{Ordered statistics decoding}
\label{sec:osd-decoding}

Ordered statistics decoding (OSD) obtains a valid correction by Guassian elimination, prioritizing faults that it has been told are most likely to have occurred~\cite{panteleev2021degenerate,fossorier1995soft,fossorier2002iterative}. 
Following the description in Ref.~\cite{roffe2020}:

\paragraph{Algorithm input.}
The algorithm takes as input the binary decoding matrix $H \in \mathbb{F}_2^{M \times N}$, the observed syndrome $\sigma \in \mathbb{F}_2^{M}$, and a score $q_j$ for each column $j \in 1,\dots,N$, where low scores correspond to more likely faults.

\paragraph{Fault ranking.}
Obtain a list $R$ of fault indices (columns) ranked from lowest to highest score.

\paragraph{Correction in highest-ranked independent columns.}

Proceed in two stages:

\begin{enumerate}
    \item \textbf{Linearly independent columns.}
    Find the set $S$, where 
    $$S = \{\text{first $\operatorname{rank}(H)$ linearly independent columns of $H$ in the order $R$}\}.$$
     \item \textbf{Inverse of sub-matrix.}
    Find the inverse $X$ of $H_S$ via Gauss–Jordan elimination, where $H_S$ is the submatrix formed from the columns of $H$ indexed by $S$.
    The inverse exists because $H_S$ has full column rank, with $|S| = \operatorname{rank}(H)$.  
\end{enumerate}

Applying $x = X \sigma$ then yields $x \in \mathbb{F}_2^{|S|}$ satisfying $H_S x = \sigma$ by construction. 

\paragraph{Algorithm output.} 
The entries of $x$ are placed at the positions in $S$ (zeros elsewhere) to form the full correction $F \in \mathbb{F}_2^{N}$.

\paragraph{Combination sweep.}
A standard extension performs a \emph{combination sweep}~\cite{roffe2020,fossorier1995} forming an algorithm referred to as OSD-$k$ with $k \ge 1$.
This makes use of the inverse $X$ computed in OSD to efficiently solve multiple related systems.  
Loosely, in combination sweep a small set of faults on indices outside $S$ is toggled, and $X$ is used to determine the complementary indices inside $S$ that combine with them to form a valid correction for $\sigma$.  

\paragraph{BP–OSD decoding.}
The ranking $R$ is typically obtained by running belief-propagation (BP) decoding for up to a maximum number of iterations.  
If BP converges, its output correction is returned; otherwise, $R$ is formed by sorting indices $j$ in decreasing order of the log-likelihood ratio $\Lambda_j$ from the final BP iteration (see~\eq{llr}).  
OSD (or OSD-$k$) is then applied as a post-processing step.

\subsection{Cluster decoding}
\label{sec:union-find-algo}

Cluster decoders~\cite{kovalev2013fault,delfosse2021almost,delfosse2022toward} iteratively grow connected subgraphs of the Tanner graph $\mathcal{G}$ (called \emph{clusters}) until each cluster can be corrected independently.  
After each growth step, a correction is sought that is supported entirely within the current clusters; if no such correction exists for a given cluster, that cluster is grown further.  
Conceptually cluster decoders are similar to OSD (which solves a linear system on a sub-matrix of faults prioritized by a pre-decoder) but they decompose this sub-matrix into small blocks corresponding to clusters in the decoding graph.

The Union–Find (UF) decoder of Ref.~\cite{delfosse2021almost} is a well-known clustering algorithm specialized to surface codes (and other codes whose Tanner graphs have maximum variable-node degree two), equipped with an efficient subroutine for finding a correction.
Here we use a generalization of UF to quantum LDPC codes, similar to that in Ref.~\cite{delfosse2022toward}.
We consider a noise model with stochastic and erasure faults, where faults in 
$\mathcal{E} \subset \{1,\dots,N\}$ are erasures (probability $1/2$), and all others occur independently with probability $p_j$.

\paragraph{Algorithm input.} 
The algorithm takes as input the binary decoding matrix $H \in \mathbb{F}_2^{M \times N}$, the observed syndrome $\sigma \subseteq \{1,\dots,M \}$, and an erasure set $\mathcal{E} \subseteq \{1,\dots,N\}$ (subsets of row and column indices of $H$ respectively).

\paragraph{Cluster initialization.}
Each cluster is a connected subgraph of the Tanner graph.
We initialize the cluster set as
\[ \mathcal{C} = \{ G^{(1)}, \dots, G^{(|\sigma|+|\mathcal{E}|)} \}, \]
where each subgraph consists of a single vertex in $\sigma \,\cup\, \mathcal{E}$.

For a cluster $G \in \mathcal{C}$ with vertex set $V$, the algorithm makes use of the following subroutines:
\begin{enumerate}
    \item \textbf{Cluster validity:}  
    Let $S$ be the set of check nodes in $G$, and $E$ the set of fault nodes in $G$.  
    Let $\tilde{E} \subset E$ be the \emph{interior fault nodes} of $G$, which is the subset of fault nodes which have all neighbors are in $G$.
    The \emph{cluster syndrome} is $\sigma|_{G} = \sigma \cap S$.  
    Let $H|_G$ denote the restriction of the decoding matrix $H$ to rows $S$ and columns $\tilde{E}$.  
    The cluster is \emph{valid} if $\sigma|_G \in \operatorname{Im}(H|_G)$, where $\operatorname{Im}(H|_G)$ is the image of $H|_G$.
    This is equivalent to there existing an $\tilde{F} \subseteq \tilde{E}$ such that
    \[
    H|_G \, \tilde{F} = \sigma|_G ,
    \]
    with all arithmetic over $\mathbb{F}_2$.
    This is precisely the \emph{solution existence problem} defined later in \sec{linear-system-problems}.
    
    \item \textbf{Cluster growth:}  
    The grown cluster is
    \[
    \mathrm{grow}(G) = \mathcal{G}\big[ V \cup \mathcal{N}(V) \big],
    \]
    the subgraph induced by $V$ and its neighbors.  
    If $G$ is connected, so is $\mathrm{grow}(G)$.
    
    \item \textbf{Cluster merging:}  
    Given two clusters $G$ and $G'$ with vertex sets $V$ and $V'$, define
    \[
    \mathrm{merge}(G, G') = \mathcal{G}\big[ V \cup V' \big].
    \]
    If each cluster is connected and $V \cap V' \neq \emptyset$, the merged cluster is connected.
\end{enumerate}

\paragraph{Cluster update loop.}  
After the cluster set $\mathcal{C}$ has been initialized, the validity of each of the clusters is tested before applying the main loop of the algorithm.
While there exists at least one invalid cluster in $\mathcal{C}$:
\begin{enumerate}[label=(\roman*)]
    \item Remove an invalid cluster $G$ from $\mathcal{C}$.
    \item Form the grown cluster $G' = \mathrm{grow}(G)$, with vertex set $V'$.
    \item Remove all clusters from $\mathcal{C}$ whose vertex sets intersect $V'$ and merge them sequentially into $G'$ using $\mathrm{merge}(\cdot,\cdot)$.
    \item Test if the merged cluster $G'$ is valid and insert it back into $\mathcal{C}$.
\end{enumerate}

When the cluster update loop terminates, every cluster in $\mathcal{C}$ is valid.  
For each cluster $G^{(l)}$, a correction $\tilde{F}^{(l)} \in \mathbb{F}_2^{N}$ is obtained by solving the \emph{find solution problem} (\sec{linear-system-problems}). 

\paragraph{Algorithm output.}
The overall correction is then computed and output:
\[
\hat{F} = \bigcup_{l} \tilde{F}^{(l)} .
\]

\begin{figure}[ht!]
    \centering
    \begin{subfigure}[b]{0.48\linewidth}
        \centering
        \includegraphics[width=\linewidth]{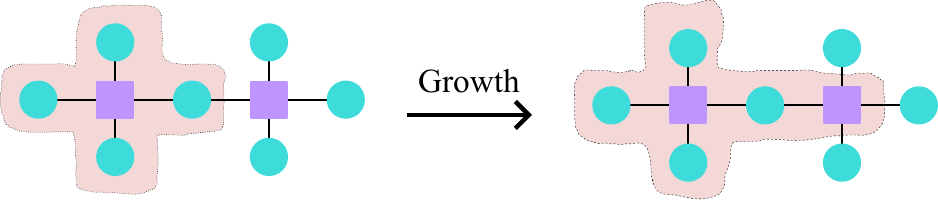}
        \caption{Cluster growth}
        \label{subfig:cluster_growth}
    \end{subfigure}
    \begin{subfigure}[b]{0.48\linewidth}
        \centering
        \includegraphics[width=\linewidth]{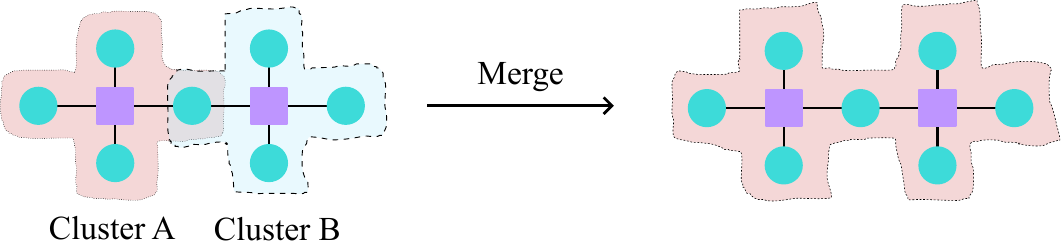}
        \caption{Cluster merging}
        \label{subfig:cluster_merge}
    \end{subfigure}
    \caption{Illustrations of the cluster growth and merge subroutines used in Union–Find decoding.}
    \label{fig:union_find}
\end{figure}

Note that by construction, every cluster has a boundary consisting entirely of check nodes or entirely of fault nodes. 
Recall that $\mathcal{G}$ is bipartite, with disjoint fault and check node sets.  
Initially, each cluster’s boundary contains only check nodes.  
A growth step extends $G$ by one edge from every boundary node, flipping the boundary type (check $\leftrightarrow$ fault).  
When two clusters merge, it is because the grown boundary of one has contacted the boundary of another; since both boundaries are of the same type, the merged cluster inherits that type.

\paragraph{BP–UF decoding.}
The assignment of erasures has not yet been specified. 
In this work, we use BP as a pre-decoder for up to a maximum number of iterations.
We specify two values $\Lambda_\text{accept}$ and $\Lambda_\text{suspicious}$.
If BP converges, the resulting correction is output. 
Otherwise, faults with log-likelihood ratios $\Lambda_j \le \Lambda_\text{accept}$ are applied directly, those with $\Lambda_\text{accept} < \Lambda_j \le \Lambda_\text{suspicious}$ are marked as erasures, and UF decoding is applied to the updated syndrome.

\clearpage
\section{FPGA background and definitions}
\label{sec:fpga-background}

Field Programmable Gate Arrays (FPGAs) provide a compelling platform for the realization of real-time QEC decoders. 
With the QEC community in mind, this section introduces FPGA algorithms.
In \sec{design-graphs} we introduce the simple computational model of \emph{computation graphs}, which can be thought of as pseudocode-level descriptions of FPGA algorithms. 
In \sec{examples-designs} we present illustrative examples of computation graph algorithms, and in \sec{fpgas} we briefly review FPGA architectures and how computation graphs map to them.

\subsection{Computation graphs (FPGA-oriented pseudocode)}
\label{sec:design-graphs}

\begin{figure}[ht!]
    \centering
    (a) \includegraphics[height=0.32\textwidth]{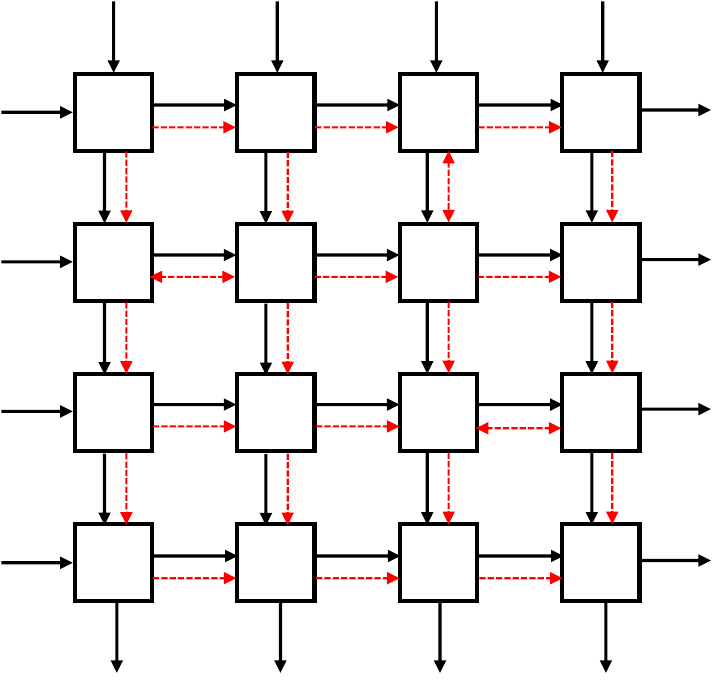}
    \hspace{1.0cm}
    (b) \includegraphics[height=0.32\textwidth]{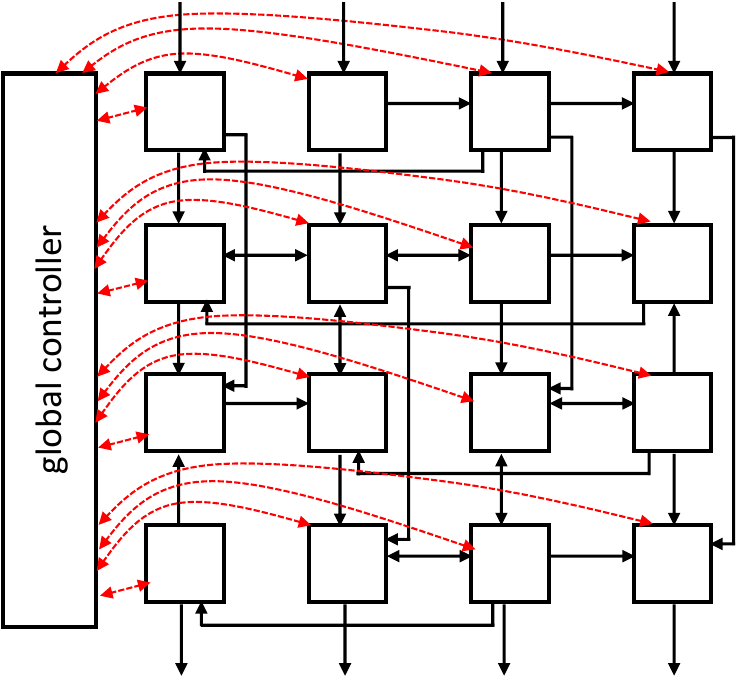}
    \caption{
    Examples of computation graphs. 
    Control links are shown as red dashed edges and double-headed arrows denote two oppositely directed edges.
    (a) Data enters via I/O ports on the top and left, traverses processing nodes left-to-right and top-to-bottom, and exits via I/O ports on the right and bottom. 
    Data flow is local in a 2D grid.
    (b) The I/O ports are as before, but the data flow is no-longer local in a 2D grid, and a global controller is connected to all nodes. 
    }
    \label{fig:example-designs}
\end{figure}

Here we define a simple graphical computational model that abstractly captures how FPGAs compute.
We define a \textit{computation graph} $\Gamma$, where the nodes are simple processing units and the directed edges represent data transmission lines between nodes\footnote{Technically, instead of a graph, we should call this a “multi-digrapgh with independent identities” or a “quiver”, defined as tuple $(V,E,s,t)$ where $V$ is a vertex set, $E$ is an edge set, $s$ and $t$ are the source and target maps $s,t:E\rightarrow V$.}. 
Each node has a function $f$ which maps incoming and currently stored data to outgoing and updated stored data,
We allow $\Gamma$ to have some dangling edges (ports) to allow data inputs and outputs of the algorithm. 
The algorithm executes as a sequence of iterations: the computation graph and per-node behaviors are fixed, but data and local memories evolve in such a way that performs the desired computation. 
If each node’s action may be any gate from a universal logic set, the model is Turing-complete. 
\fig{example-designs} illustrates simple computation graphs; for clarity, node memories and action maps are omitted.

\paragraph{Algorithms from computation graphs:}
An algorithm on a computation graph executes over a sequence of iterations, where in each iteration, every node/processor follows two steps:
\begin{enumerate}
    \item[(i)] \emph{Message passing:} Control and data messages (computed in the previous iteration) are sent along edges.
    \item[(ii)] \emph{Computation:} The node updates its memory data and generates outgoing data based on the control signal, using both stored and incoming data as inputs.
\end{enumerate}
\noindent Computation graphs have no universal termination rule. 
Each algorithm specifies its own criterion, for example halting after a fixed number of iterations or when a \texttt{terminate} signal is emitted on a designated port.
The resulting description, consisting of the computation graph, initialization, termination conditions, control specifications, etc., serves as a graphical pseudo code.

\paragraph{Local versus non-local control}
Local control as in \fig{example-designs}(a) can be specified at each node just as for logic functions that process data. 
Sometimes a computation-graph algorithm requires a global controller that broadcasts signals to all nodes. 
Global control is typically expressed as a \emph{finite state machine} (FSM) that cycles through predefined states to issue the appropriate control signals as an algorithm proceeds.
Although the controller can itself be implemented as a computation graph, for visualization we often depict it as a single compound node within the computation graph, as in \fig{example-designs}(b). 

A computation graph is called a \emph{systolic network} if it has bounded degree (independent of problem size) and all control is local with respect to the computation graph. 
Two important special cases are the \emph{systolic array} (such as the example in \fig{example-designs}(a)), based on a 2D grid, and the \emph{systolic line}, based on a 1D chain. 
It is often seen as desirable to find a systolic array or systolic line version of an algorithm (see \sec{fpgas}).

\paragraph{Algorithm cost:}
The abstract cost of a computation-graph algorithm is characterized by the size of the graph and the number of iterations to termination, rather than the bit-operation counts typically used for CPU algorithms.
These quantities serve as proxies for implementation costs such as FPGA footprint and cycle count.
However, another important feature of an FPGA implementation not captured by this is the clock frequency, which we discuss in \sec{fpgas}.

\subsection{Examples of computation-graph algorithms}
\label{sec:examples-designs}

\begin{figure}[ht]
    \centering
    \includegraphics[width=0.6\textwidth]{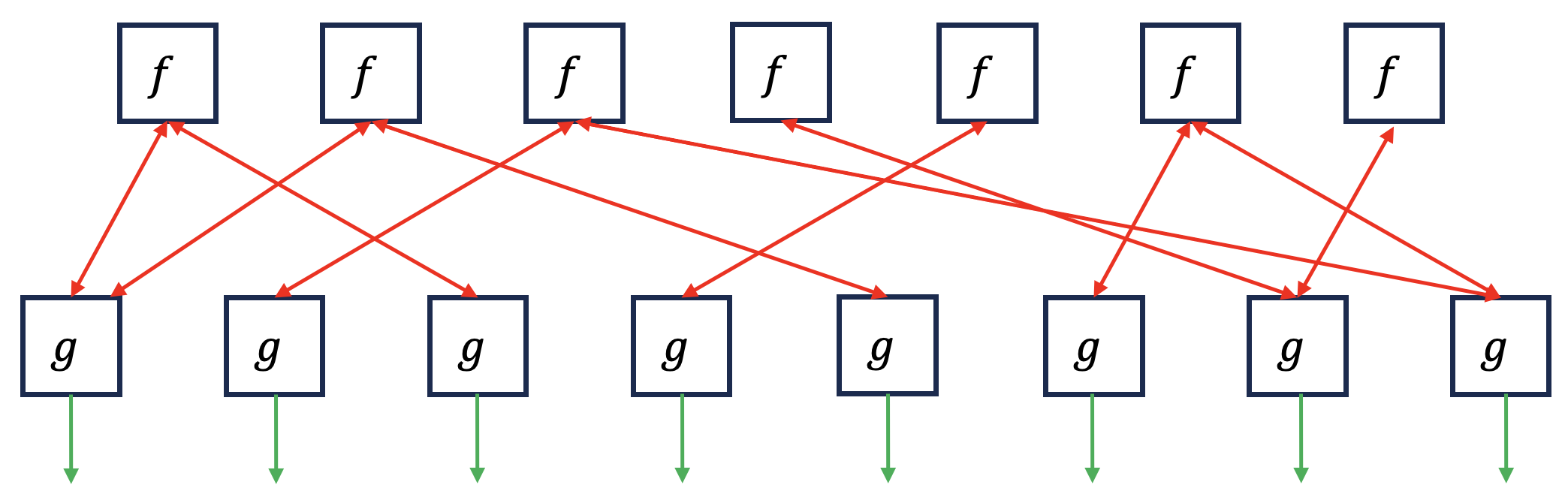}
    \caption{
    Illustration of part of a computation graph for belief propagation (BP), with a row of check nodes ($f$ from \eq{bp-check-to-error}) and a row of variable nodes ($g$ from \eq{bp-error-to-check}).
    Each iteration performs one BP iteration: nodes exchange messages with neighbors (red arrows), variable nodes output hard decisions (green arrows), and all nodes update messages for the next iteration.
    The components that check syndrome satisfaction, initializations, etc. are omitted.
    }
    \label{fig:design-graph-bp}
\end{figure}

A natural example of a computation graph algorithm is belief propagation (BP) as shown in \fig{design-graph-bp}. 
The computation graph comprises a row of check nodes applying $f$ from \eq{bp-check-to-error} and a row of variable nodes applying $g$ from \eq{bp-error-to-check}. 
Each iteration exchanges messages along red links, emits hard decisions from variable nodes along green links, and updates all node messages for the next round.

\begin{figure}[ht!]
    \centering
    (a)\includegraphics[width=0.45\textwidth]{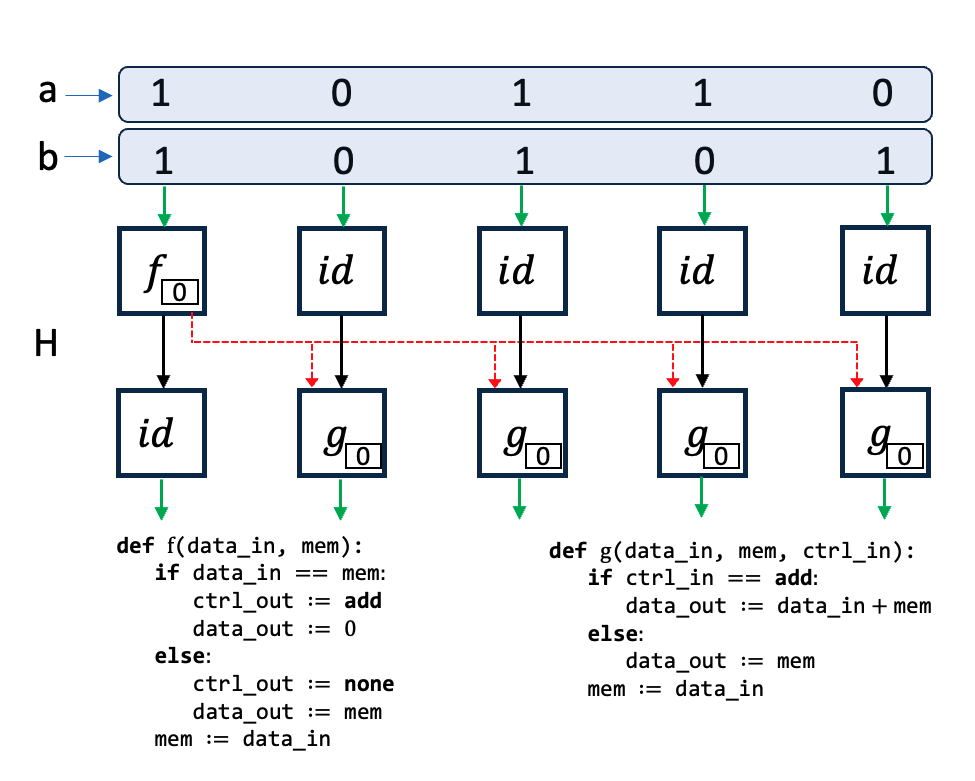}
    (b)\includegraphics[width=0.45\textwidth]{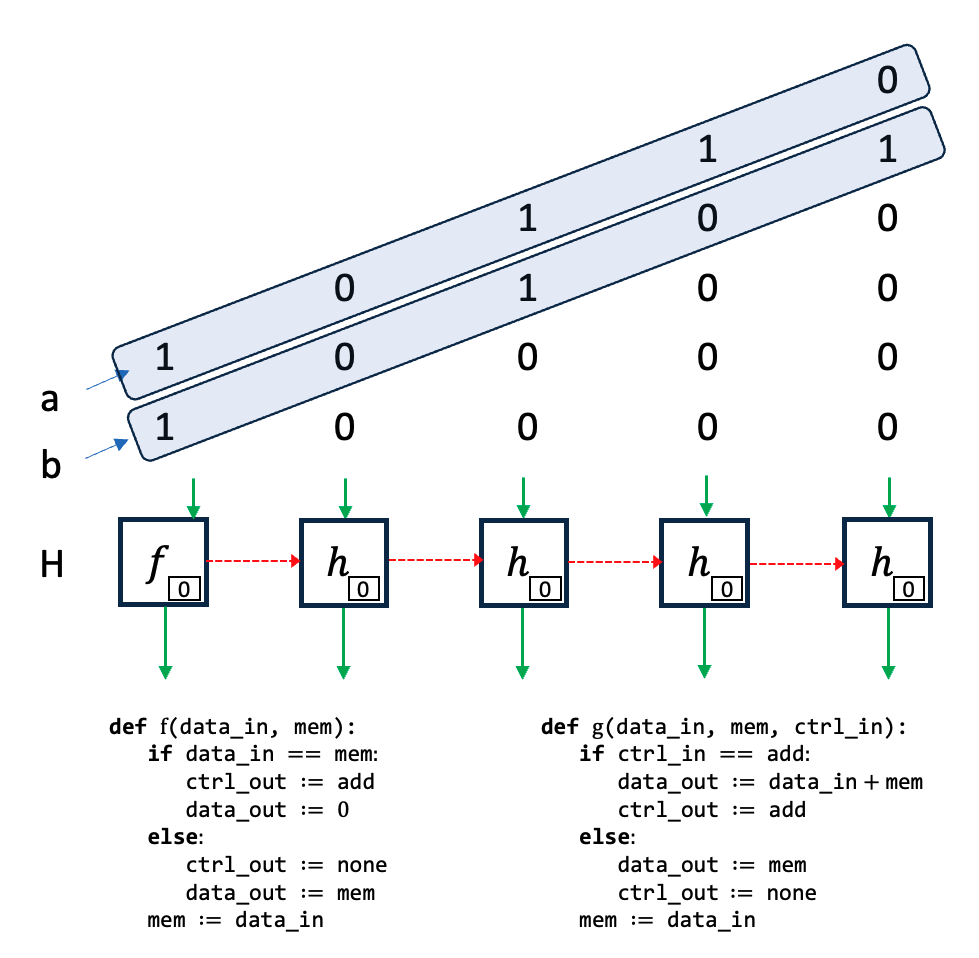}
    (c)\includegraphics[width=0.45\textwidth]{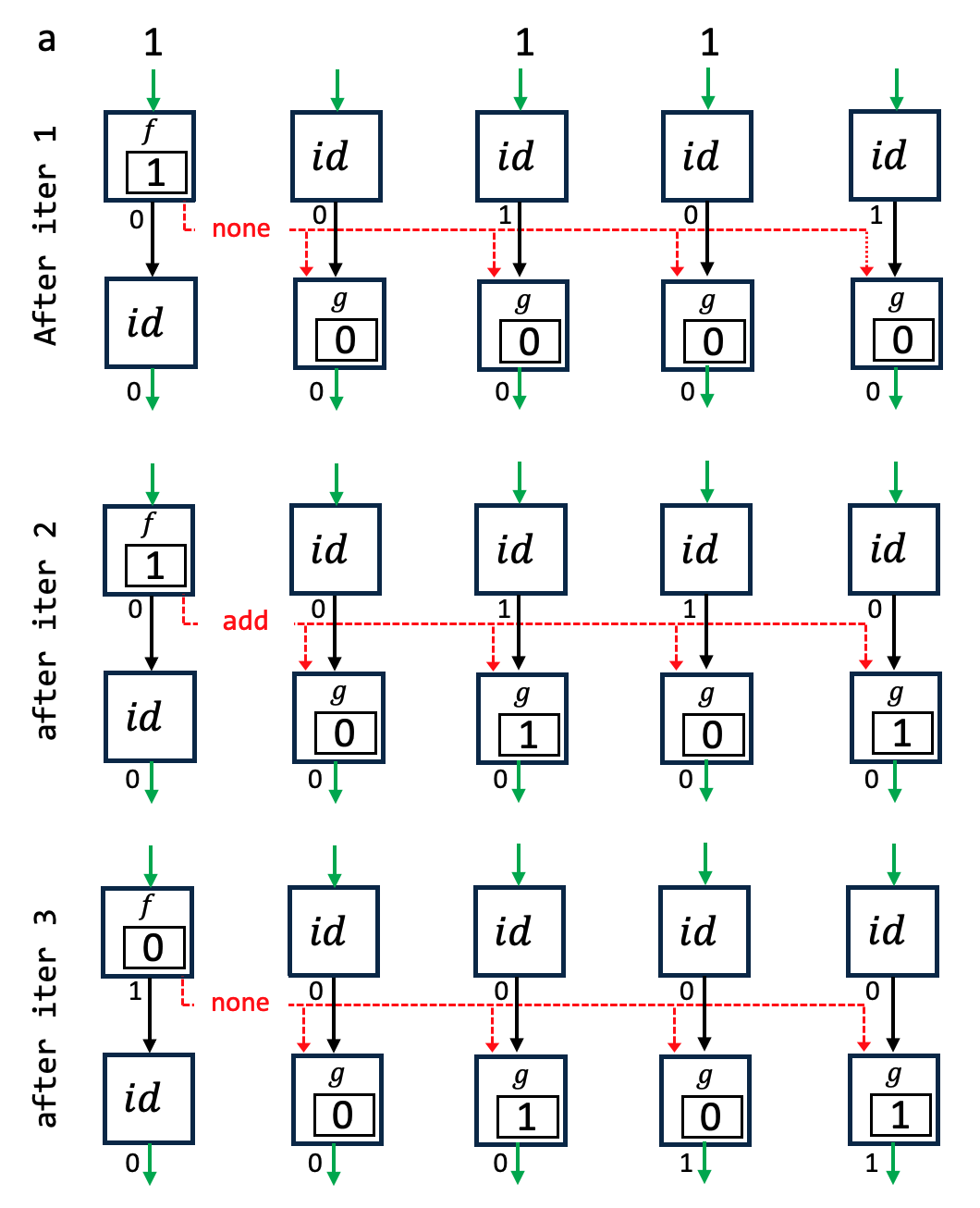}
    (d)\includegraphics[width=0.45\textwidth]{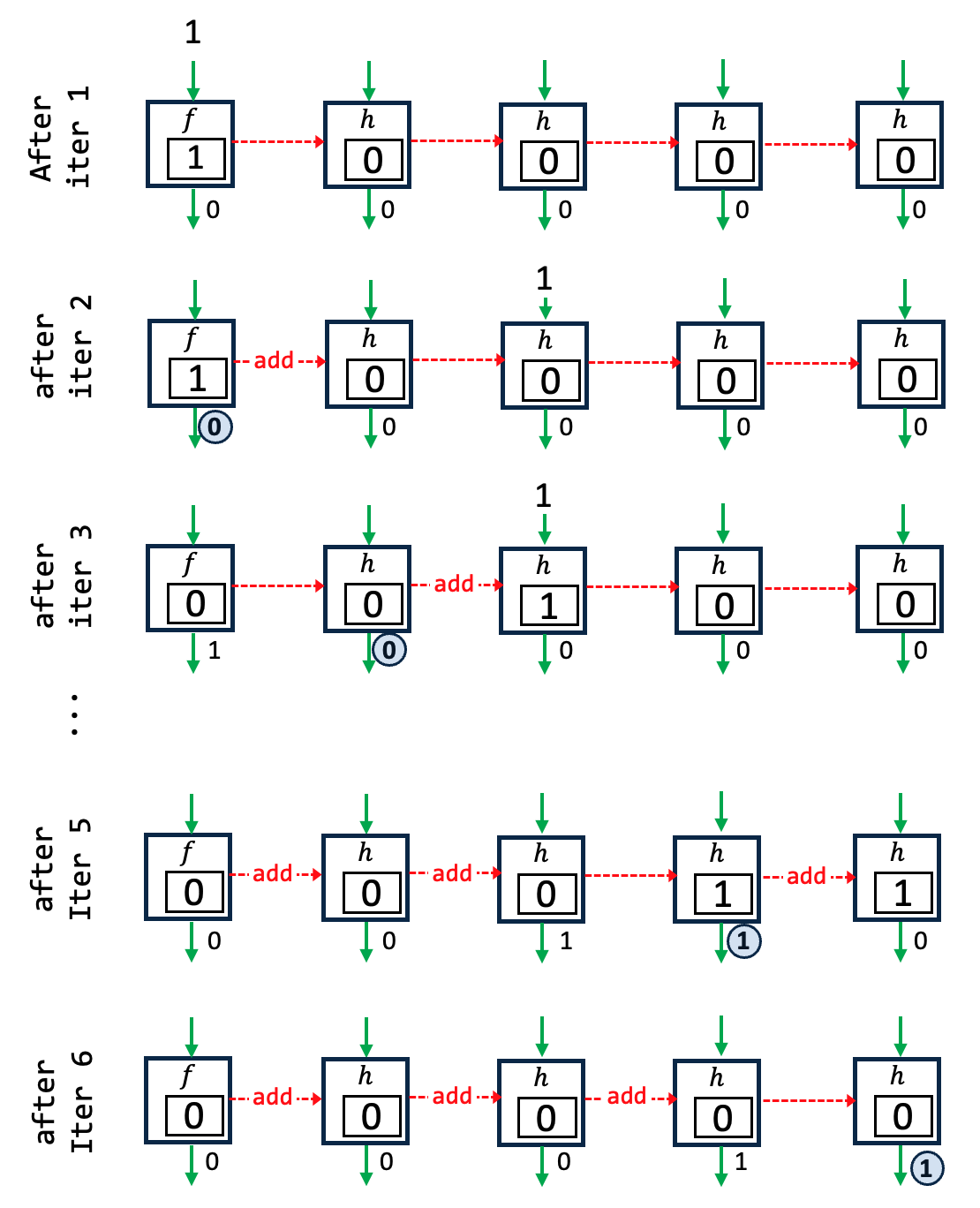}
    \caption{Two computation graph algorithms to compute $H(a,b)$ on length-$n$ bit strings (shown for $n=5$ with $a=10110$, $b=10101$, yielding $H(a,b)=00011$). 
    (a) A non-systolic version where a control decision is made about which case is required in the left-most node and signals are sent to all $n-1$ remaining nodes. 
    (b) A systolic-line version with only local communication. 
    Inputs (and outputs) must be staggered to give the control signal time to flow along the line. 
    (c) and (d) illustrate the state after each iteration in the two versions. 
    }
    \label{fig:design-examples}
\end{figure}

Let us consider another example in more detail. 
Let $H$ be the map from two length-$n$ bitstrings to one length-$n$ bitstring:
\[
H(a,b) = \begin{cases}
    a+b, &\text{ if } a_{1}=b_{1}, \\
    b &\text{ otherwise }.
\end{cases}
\]
\noindent where bitstrings from $\mathbb{F}_2^n$ are written as $a=a_{1}a_{2} \dots a_n$. 
As elsewhere in this work, addition of bit strings is over $\mathbb{F}_2$ (XOR). 
\fig{design-examples} presents two computation-graph implementations of $H$.
The first uses a broadcast controller that, for each pair, signals whether to output $a+b$ or $b$. 
The second is a systolic line; it avoids non-local control but requires staggered inputs and produces staggered outputs. 
Both computation graph algorithms can operate on a stream of bit strings $x^{(1)},x^{(2)},\ldots$, producing $H(x^{(1)},x^{(2)}), H(x^{(2)},x^{(3)}), \ldots$. 
Both initialize internal memories and messages at $0$ or \textbf{none} and both emit results as computed.
Other computation graph algorithms can retain outputs in the memory of nodes until explicitly read out.

\clearpage
\subsection{FPGAs and other computing platforms}
\label{sec:fpgas}

\begin{figure}[h!]
    \centering
    \includegraphics[width=0.8\textwidth]{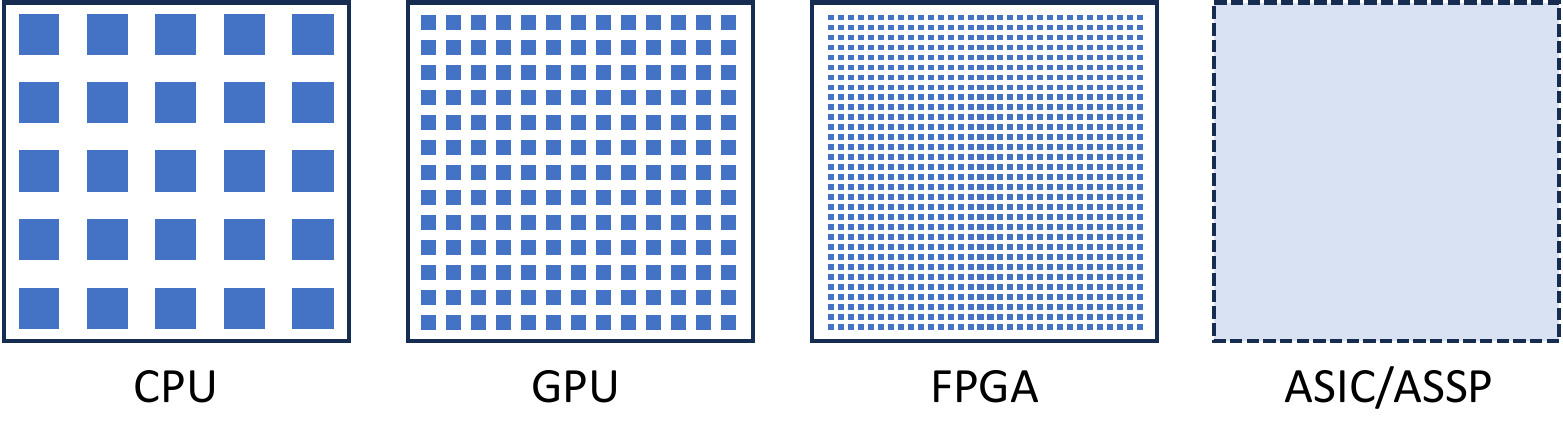}
    \caption{
    Granularity of parallelism across common computing platforms (approximate current scales). 
    CPUs: tens of general-purpose cores (coarse-grained). 
    GPUs: thousands of smaller, specialized cores (medium-grained parallelism). 
    FPGAs: millions of tiny logic blocks (fine-grained, large-scale parallelism). 
    ASICs/ASSPs: task-specific designs whose on-chip organization is tailored to the application rather than following a standardized template. 
    }
    \label{fig:CPU-GPU-FPGA}
\end{figure}

\begin{figure}[h!]
    \centering
    \includegraphics[width=0.9\textwidth]{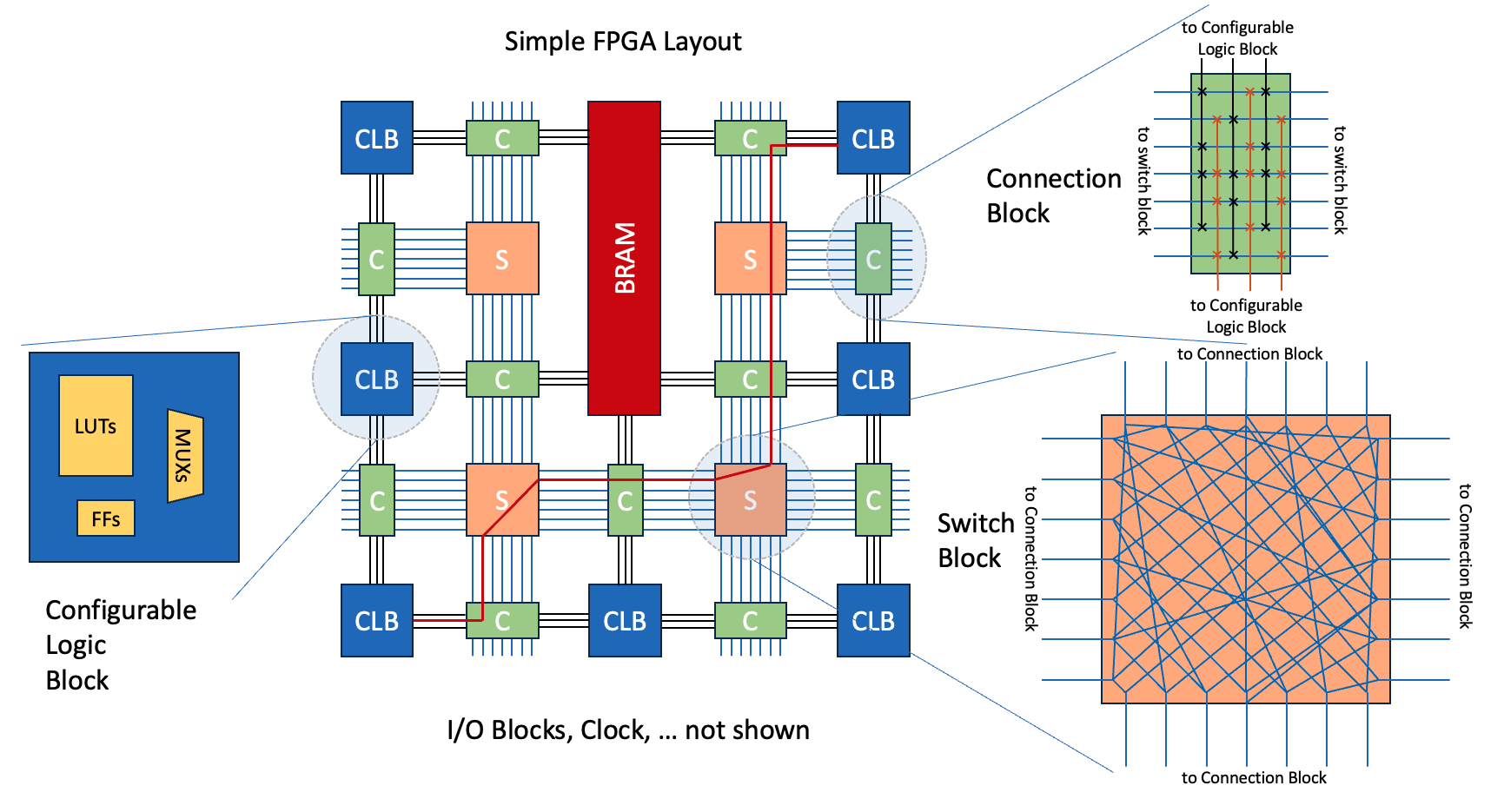}
    \caption{ 
    Island-style FPGA layout, where configurable logic blocks (CLB)s are placed as discrete “islands” on a 2D grid, separated by programmable routing channels. Connection blocks attach CLB pins to these channels (via programmable switches such as pass transistors or multiplexers), and switch blocks at channel intersections steer signals between tracks.  The red path illustrates a route from the bottom-left CLB to the top-right CLB. 
    RAM blocks can also be included in the layout with a block RAM (BRAM) being shown in this example. Other types of specialized components may also be includes such as Digital Signal Processors (DSPs). To avoid clutter I/O blocks, the clock and other important components are not included in the above layout.
    }
    \label{fig:FPGAFabric}
\end{figure}

Nearly all classical computing today is performed with integrated circuits. 
Differing configurations of this common substrate give rise to a range of platforms that span a broad spectrum of flexibility (see \fig{CPU-GPU-FPGA}).
Among the most flexible are central processing units (CPUs) and graphics processing units (GPUs).
CPUs and GPUs come as prefabricated, general-purpose chips that execute instruction streams and therefore support many algorithms, while exposing little user control over how on-chip resources are allocated to specific parts of the algorithm.
At the opposite extreme, application-specific integrated circuits (ASICs) and application-specific standard products (ASSPs) are fixed-function devices: they are hardwired at fabrication to implement a specific algorithm and cannot be repurposed afterward.
Field-programmable gate arrays (FPGAs) occupy the middle ground: prefabricated chips whose logic and interconnect can be reconfigured to implement different algorithms, giving users more control over on-chip resource use. 
While FPGA device details vary, the following are some key common elements:

\begin{enumerate}
    \item \emph{Clock.} 
    FPGA algorithms proceed over a sequence of cycles, and a periodic clock is used to synchronize updates across the device. 
    To ensure operations complete before the next clock cycle starts, the clock period is algorithm specific and must exceed the worst-case delay (i.e., the longest path through the device during once cycle that could be executed by the algorithm)~\cite{HOLDSWORTH2002163}. 
    Typical clock frequencies are $\sim\!100$–$500\,\mathrm{MHz}$, depending on the FPGA device and algorithm.
 
    \item \emph{Configurable logic blocks (CLBs).} CLBs provide the FPGA’s basic compute and state. 
    Each CLB contains lookup tables (LUTs) to realize arbitrary Boolean functions of their inputs, flip-flops\footnote{A flip-flop is a small memory component that can store a single bit.} (FFs) for single-bit storage, and multiplexers\footnote{A multiplexer is a gadget that chooses which one of several inputs to pass through to a single output, according to simple control bits.} (MUXes) for selection. 
    Programming a LUT’s truth table lets a CLB implement, for example, an AND gate or a small combinational subcircuit, with FFs registering results for synchronous pipelines. 
    Many devices also include fast carry chains and related primitives to accelerate adders and counters.
    
    \item \emph{Configurable interconnect.} A programmable routing network that links CLBs into larger circuits. 
    We illustrate an island-style layout in \fig{FPGAFabric}. For an accessible overview of FPGA routing, see \cite{FPGARouting}.
    
    \item \emph{I/O blocks.} Interface the FPGA with external hardware, setting input/output direction, matching signal properties and timing to ensure reliable data transfer to devices such as analog-to-digital converters, memories, or neighboring FPGAs.
    
    \item \emph{Configuration/boot logic.} Handles power-on reset and device bring-up, loads the configuration for the algorithm, enforces security and integrity checks, and provides interfaces for monitoring and debugging.
    
\end{enumerate}

Modern FPGAs also integrate specialized blocks, including digital signal processing slices (fixed, high-speed multiplier–accumulator units for arithmetic), and dedicated static random-access memory (RAM) for bulk storage~\cite{IBMFPGA}.
Many devices provide multiple RAM classes, including numerous small, distributed block RAMs (BRAMs) intended as local buffer memory along with fewer, larger UltraRAMs (URAMs) suited to large tables or deep queues with less strictly local access. 

\paragraph{Connection to computation graphs.}
The computation graphs of \sec{design-graphs} are abstractions of FPGA algorithms. 
Node functions become computations on configurable LUTs, while graph edges become data and control links realized via the configurable interconnect.
In the simplest case, each node fits in a CLB and, per cycle, consumes inputs, updates its local state, and emits outputs.
In this setting, one computation graph iteration corresponds to one FPGA clock cycle. 
If a node spans multiple CLBs on the other hand, a single graph iteration may take several cycles, but the step-by-step semantics are preserved. 
Together with initialization, termination, and control policies, the computation graph serves as graphical pseudo-code from which a concrete FPGA configuration can be implemented in a hardware description language (e.g., Verilog or VHDL).

Systolic lines and arrays map to FPGAs with regular, local links and bounded per-node work, so per-step routing and compute delay is approximately constant as problem sizes grow. 
Consequently, for systolic lines and arrays the computation graph’s iteration count is a good predictor of runtime at a nearly size-independent clock frequency that can be measured once for a given implementation. 
By contrast, in non-systolic algorithms the maximum clock frequency typically decreases with size (due to longer critical paths and routing congestion) and performance can be characterized empirically across problem sizes.

\paragraph{Practical considerations: Avoiding expensive computations.}
Some functions and structures are expensive to implement in FPGAs. 
Expensive in this context refers to either the amount of space on the FPGA needed (footprint) or a high cost in either latency or computational time. 
The following functions are expensive~\cite{intel_avoid_expensive_functions_2023} in FPGAs as examples:
\begin{enumerate}
\item Arithmetic operations (e.g. division and modulo (remainder) operations) except addition, multiplication, absolute value, and comparison operations.
\item Atomic operations (updates to shared state, for example incrementing a counter used by many modules). 
\end{enumerate}
\noindent What may be inexpensive on a CPU can be costly on an FPGA. 
For instance, an if/else conditional is an atomic operation which on a CPU typically incurs the cost of evaluating the condition and executing only the selected branch. 
In contrast, FPGAs compute both branches in parallel, and the final result is selected using a multiplexer. 
This parallelism can lead to increased resource usage and latency, making such constructs more expensive in FPGA implementations. 
Inexpensive functions have minimal effects on performance, and their implementation consumes minimal hardware. 
The following functions are inexpensive in FPGAs:
\begin{enumerate}
    \item \textbf{Binary logic:} Element-wise logical operations such as \texttt{AND}, \texttt{NAND}, \texttt{OR}, \texttt{NOR}, \texttt{XOR}, and \texttt{XNOR}.
    \item \textbf{Shift by constant:} Bit-string shifts by a fixed number of positions, with zero padding on the vacated slots.
    \item \textbf{Scaled integer arithmetic:} Multiplication and division of an integer by powers of two (realized as left and right shifts respectively).
\end{enumerate}

\paragraph{FPGAs vs.\ ASICs}
FPGAs can be compiled, loaded, and tested quickly (often within hours), making them well suited to rapid iteration once a fairly explicit algorithm (say at the computation-graph level) is in hand. 
ASICs require a full design and fabrication cycle with substantial development cost (often years and millions of USD).
However, for a fixed algorithm ASICs typically deliver higher clock rates (up to a factor-five speedup is common), the achievable sizes can be larger (about an order of magnitude is common), and their per-unit cost can be lower at volume as development cost is amortized. 
In practice, FPGAs are used to prototype and validate; ASICs are chosen when the design is stable and expected volume or performance justifies the investment.
A key constraint for both platforms is finite device size. 
High-end FPGAs provide on the order of millions of LUTs and flip-flops; ASICs are ultimately limited by silicon wafer size, often by cost well before that. 
Computation-graph algorithms must therefore either fit on a single device or be designed as distributed algorithms that run across multiple devices with explicit off-chip communication.
\clearpage
\section{Computation graph algorithms for linear systems}
\label{sec:gateware_lse}

In this section we 
(i) formalize the linear-system problems required by our decoders (\sec{linear-system-problems}); 
(ii) survey computation-graph algorithms for solving linear systems, noting that most assume full rank and therefore exclude the rank-deficient systems over $\mathbb{F}_2$ that arise in our setting (\sec{lit-review-gea}); and 
(iii) present a systolic-array computation-graph algorithm that solves arbitrary-rank systems over $\mathbb{F}_2$ (\sec{systolic-gaussian-elimination-algorithm}).
Our algorithm is simpler than the only other systolic array approach we are aware of for this task~\cite{hu2024universal}.
To explain it, we introduce a ``lifted'' version of Gauss-Jordan elimination that places pivots on the diagonal and retains non-pivot rows while zeroing their entries.

\subsection{Linear system problems and Gaussian elimination algorithms}
\label{sec:linear-system-problems}

First, we define three key linear algebra problems relevant to the Clustering and OSD algorithms.
Given a binary matrix $A \in \mathbb{F}_2^{m \times n}$, consider the following problems:
\begin{enumerate}
    \def\enumgap{0.7ex}
    \item[(i)] \textbf{Solution existence problem:} Given $y \in \mathbb{F}_2^{m}$, is $y\in  \operatorname{Im}(A)$? \\[\enumgap]
    This is used in the Clustering algorithm to verify cluster validity.
    
    \item[(ii)] \textbf{Find solution problem:} Given $y \in  \operatorname{Im}(A)$, find any $x \in \mathbb{F}_2^{n}$ satisfying $Ax = y$. \\[\enumgap]
    This is used at the end of the Clustering algorithm to determine a correction for valid clusters and in our filtered-OSD algorithm described in \sec{filtered-osd-decoding}.
    
    \item[(iii)] \textbf{Find generalized inverse problem:} Find generalized inverse $X \in \mathbb{F}_2^{n \times m}$ of $A$. \\[\enumgap]
    We define the \textit{generalized inverse} as any matrix $X$ such that $AXA=A$. 
    In the special case where $A$ has full column rank, this is equivalent to finding the left-inverse of $A$, which is used in the standard OSD algorithm in \sec{osd-decoding}.  
\end{enumerate}
For each of these problems, the binary matrix $A$ may be non-square ($m \neq n$) and rank-deficient ($\text{rank}(A) < \min(m, n)$). 
Regardless of the rank properties of $A$ we see that a generalized inverse exists and this property guarantees that for any $y \in \operatorname{Im}(A)$, one can find a solution $x$ to $A x = y$.  
Indeed, if $y = A x'$ for some $x'$, then setting $x = X y$ gives $A x = A X y = A X A x' = A x' = y$,
so $x$ reproduces $y$ exactly.

Gaussian elimination solves linear systems by transforming a matrix into row-echelon form using row operations (row swaps or bitwise XOR additions over $\mathbb{F}_2$). 
In row-echelon form, each row’s leading $1$ (pivot) appears in a strictly increasing column index, with no $1$s below it. 
Gauss–Jordan elimination continues to reduced row-echelon form, where each pivot column contains exactly one $1$.
The standard algorithm \alg{gauss_elim_double_pass} performs this in two passes: first to row-echelon form, then to reduced row-echelon form.

\begin{algorithm}
\SetAlgoLined
\KwInput{$A \in \mathbb{F}_2^{m \times n}$}
\KwOutput{$A$ in reduced row–echelon form}

\textbf{Forward pass: find pivots and clear below}\; Set $\text{pivotRow}\gets 1$\;
\For{each column index $p$ from $1$ to $n$}{
    Find first row $r \ge \text{pivotRow}$ with $A[r,p]=1$\;
    \If{such $r$ exists}{
        Swap rows $r$ and $\text{pivotRow}$\;
        Clear all $1$s below in col $p$ by adding the pivot row\;
        Move pivot row index down by $1$\;
    }
}

\textbf{Backward pass: clear above pivots}\;
\For{each pivot row from bottom to top}{
    Let $p$ be the index of the leftmost $1$ in $A[r,:]$\;
    Clear all $1$s above it by adding the pivot row\;
}
\caption{Gauss–Jordan over $\mathbb{F}_2$}
\label{alg:gauss_elim_double_pass}
\end{algorithm}

\paragraph{Solving linear system problems via Gauss-Jordan elimination}
Given an augmented matrix $[A \,|\, B]$ with 
$A \in \mathbb{F}_2^{m \times n}$ and 
$B \in \mathbb{F}_2^{m \times l}$, 
we perform Gauss-Jordan elimination on $A$ while applying the same row operations to $B$, obtaining $[A' \,|\, B']$ with 
$A' \in \mathbb{F}_2^{m \times n}$ and 
$B' \in \mathbb{F}_2^{m \times l}$. 
By taking an appropriate choice of $B$, this procedure solves each of the problems (i)–(iii).
Let $r = \operatorname{rank}(A)$. 
The \emph{pivot positions} of $A'$ are $P = \{(i, j_i) \mid i = 1, \dots, r\}$, where $(i, j_i)$ is the position of the leading $1$ in the $i$-th nonzero row of $A'$. 
In reduced row-epsilon form, the pivot columns satisfy 
$j_1 < j_2 < \dots < j_r$.
One can solve the three problems as follows:
\begin{itemize}
    \item \emph{Solution existence}: Set $B = y$ and check whether any $B'_{i} = 1$ for a non-pivot row $i$, which occurs if and only if no solution exists. This is equivalent to the required condition $\operatorname{rank}(A) = \operatorname{rank}(A | y)$ which is the Rouché–Capelli theorem. This only requires the forward pass of \alg{gauss_elim_double_pass} to evaluate.

    \item \emph{Find solution}: Set $B = y$. then to find a solution $x \in \mathbb{F}_2^n$ satisfying $Ax = y$, we define $x_j = B'_{i}$ for each $(i,j) \in P$ and $x_j = 0$ for all other $j$. 

    \item \emph{Find generalized inverse}: Set $B = I_m$ then to find a solution $X \in \mathbb{F}_2^{n \times m}$ satisfying $AXA = A$ we define $X_{j,*} = B'_{i,*}$ for all $(i,j) \in P$ and $X_{j,*}=0^m$ for all other $j$.

\end{itemize}

Therefore, if Gauss–Jordan elimination can be performed on a binary matrix $A$ with $B$ as a spectator, any of the problems (i)–(iii) can be solved.  
However, as discussed in the next subsection, most existing algorithms designed for FPGAs can not perform Gauss–Jordan elimination on arbitrary binary matrices.
We later introduce a modified version, lifted Gauss–Jordan elimination~\alg{lifted_gauss_elim}, which is the basis of our systolic-array computation graph algorithms for solving each of the aforementioned problems for arbitrary matrices over $\mathbb{F}_2$.

\subsection{Literature review of Gaussian elimination algorithms with computation graphs}
\label{sec:lit-review-gea}

As discussed in \sec{linear-system-problems}, our decoding applications require performing Gaussian elimination on a binary matrix $A \in \mathbb{F}_2^{m \times n}$, which may be non-square and rank-deficient, along with an auxiliary matrix $B \in \mathbb{F}_2^{m \times l}$ subjected to the same row operations. Here we briefly review computation graph algorithms for Gaussian elimination. 
Most known computation graph algorithms for Gaussian elimination fall into two classes, which we will refer to as “systolic” and “shifting” approaches respectively.

\textbf{Systolic approaches:}
In 1981 Gentleman and Kung~\cite{gentleman1982matrix} introduced the first systolic array for the triangulation, via Gaussian elimination of a $m\times n$ matrix $X$ over the reals $\mathbb{R}$ with $n\leq m$. 
Gentleman and Kung noted that the standard pivot-selection strategy in Gaussian elimination, which identifies pivots through a global search of the matrix, is incompatible with systolic algorithms because it requires global communication. 
Instead, they used a neighbor-pivoting strategy, which they point out was known at least since the 1960s for unrelated reasons~\cite{national1961modern}. 

Later Hochet et. al. in 1989~\cite{Hochet1989} adapted the neighbor pivoting technique of~\cite{gentleman1982matrix} to matrices over $\mathbb{F}_p$ and extended it to Gauss-Jordon elimination by adapting the systolic arrays for real matrices of~\cite{cosnard1986matching} and~\cite{robert1985resolution}. 
The final algorithm of Hochet et. al. was designed for dense non-singular square $n\times n$ matrices over $\mathbb{F}_p$ and aborted otherwise. 
Later, Wang and Lin~\cite{wang1993systolic} use a similar but different (uses multiple control signals) systolic array to solve a system of $n=2q-1$ equations over $\mathbb{F}_2$ to find inverses in $\mathbb{F}_{2^q}$. 
Both approaches use $O(n^2)$ nodes and complete the first solution in $O(n)$ iterations which is a runtime improvement over sequential CPU-based Gaussian elimination, which requires $\Omega(n^3)$ bit operations in the worst case~\footnote{Faster approaches exist in the CPU-model for solving linear systems that do not use Gaussian elimination, with runtime $O(n^\omega)$ bit operations (with the best exponent~\cite{vassilevska2023omega} known to date $\omega\approx 2.37$), still much slower than the $O(n)$ time for the parallel computation graph model.}. 

\textbf{Shifting approaches:}
Systolic arrays are attractive due to their simple control mechanisms and constant cycle time per operation. 
However, non-systolic array architectures can outperform them in certain regimes by reducing the number of cycles or requiring a smaller footprint, even if their cycle time scales with problem size. 
The shifting approaches for Gaussian elimination over~$\mathbb{F}_2$ store the matrix in a two-dimensional array of nodes, with each of the first row of nodes having connections to all other nodes in the same column. 
The matrix rows are processed over time, with the pivot row of the matrix being shifted to the first row of nodes for parallel elimination. 
This method was introduced abstractly by \cite{parkinson1984compact} and later described in more detail by \cite{Rupp2006}. 
Both achieve a runtime of~$\mathcal{O}(n^2)$ iterations using~$\mathcal{O}(n^2)$ processing elements for an $n \times n$ matrix.

Later improvements, such as those in \cite{rupp2011hardware, jasinski2010gf2}, reduce this to $O(n)$ iterations while maintaining $O(n^2)$ nodes. 
While the non-systolic algorithms in \cite{rupp2011hardware} and \cite{jasinski2010gf2} are similar to each other, their reported runtime prefactors differ: \cite{rupp2011hardware} requires $3n$ iterations for matrices over $\mathbb{F}_2$, while \cite{jasinski2010gf2} claims $n$ iterations suffice (but lacks details on its processing element design).
In contrast to the systolic approaches, these non-systolic approaches with $O(n)$ iteration runtime are based on a single-pass version of Gaussian elimination rather than double-pass Gaussian elimination version in \alg{gauss_elim_double_pass}, and operate on an in-place array rather than a skewed flow.

\textbf{Reduced footprint approaches:}
The increasing demand for cryptographic applications involving large matrices (too large to fit on available FPGAs) has driven the development of Gaussian elimination algorithms with smaller footprints or distributed implementations.  
A linear systolic array, explored in \cite{rupp2011hardware}, sequentializes the systolic design, acting on just one row of the matrix at a time, with other rows stored outside the array, reducing node count to $O(n)$ at the cost of $O(n^2)$ iterations.
Distributed approaches, such as \cite{geiselmann2003hardware, wang2016solving}, leverage sparse matrix representations to reduce communication overhead but significantly increase iteration cost.

\textbf{Extensions to row-rank-deficient rectangular matrices:}
For binary matrices all the aforementioned references consider cases where the input matrix is square and full rank or at least they include error signals when a matrix is not full rank then terminate. 
For decoding applications the matrices are in general neither square nor full rank (see \sec{qec-background} and \sec{linear-system-problems}). 
Refs.~\cite{scholl2013hardware, scholl2014hardware} consider Gaussian elimination algorithms for rectangular matrices in the context of decoding, assuming a fixed, linearly independent column set is provided (an assumption that does not hold for our decoding problems). Refs.~\cite{Shoufan2010,wang2017fpga} state the algorithms they use support rectangular matrices, but either require full rank or abort if this is not the case. 
Ref.~\cite{hu2024universal} is the only work we are aware of that explicitly considers arbitrary binary matrices, including row rank-deficient and rectangular cases. 
Their approach uses the same array for the entire Gauss-Jordon elimination process where processing elements have a switch that dynamically reassigns rows as pivots or non-pivots during execution. 

In \sec{systolic-gaussian-elimination-algorithm}, we provide a modified version of the algorithm from Ref.~\cite{Hochet1989}, which can handle arbitrary binary matrices without requiring more complicated processing elements.

\subsection{Systolic array for lifted Gauss-Jordan elimination of arbitrary binary matrices}
\label{sec:systolic-gaussian-elimination-algorithm}

Here we modify the algorithm presented in Ref.~\cite{Hochet1989} to form a systolic array computation graph algorithm on an arbitrary binary matrix $A \in \mathbb{F}_2^{m \times n}$ with a spectator matrix $B \in \mathbb{F}_2^{m \times l}$.  
To specify this algorithm, we first define the lifted version of row-echelon and reduced row echelon forms:

\begin{dfn}[Lifted (reduced) row-echelon matrix]
Let $A \in \mathbb{F}_2^{m \times n}$, and let $A' \in \mathbb{F}_2^{m \times n}$ be the reduced row-echelon form of $A$.  
The \emph{lifted reduced row-echelon matrix} $A^{\uparrow} \in \mathbb{F}_2^{n \times n}$ is obtained by:
\begin{enumerate}[label=(\roman*)]
    \item For each pivot column $j$ of $A'$, set row $j$ of $A^{\uparrow}$ to be the pivot row of $A'$ containing that pivot.
    \item For all non-pivot columns $j$, row $j$ of $A^{\uparrow}$ is the zero vector.
\end{enumerate}
The \emph{lifted row-echelon matrix} is defined similarly, taking $A'$ to be the (non-reduced) row-echelon form of $A$ in place of the reduced row-echelon form.
\end{dfn}

When $A$ is a full-rank square matrix, the lifted and standard row–echelon and reduced row–echelon forms are identical.  
We modify the Gauss–Jordan algorithm slightly to obtain \alg{lifted_gauss_elim}, which computes the lifted reduced row–echelon form of $A$.  

\begin{algorithm}[h]
\SetAlgoLined
\KwInput{$A \in \mathbb{F}_2^{m \times n}$}
\KwOutput{Lifted reduced row–echelon form $A^{\uparrow} \in \mathbb{F}_2^{n \times n}$}

\textbf{Forward pass: stream rows, lock pivots}\;
Initialize $A^{\uparrow} \gets 0^{n \times n}$ and mark all columns unlocked\;
\For{each row $r$ of $A$}{
    \While{$r \neq 0^{n}$}{
        Let $p$ be the index of the leftmost $1$ in $r$\;
        \If{column $p$ is unlocked}{
            Store $r$ in row $p$ of $A^{\uparrow}$; lock column $p$; \textbf{break}\;
        }
        \Else{
            Set $r \gets r \oplus A^{\uparrow}[p,:]$ to clear column $p$\;
        }
    }
}

\textbf{Backward pass: clear above pivots}\;
\For{each column index $p$ from $n$ down to $1$}{
    \For{each $q<p$ with $A^{\uparrow}[q,p]=1$}{
        Add pivot row $p$ to row $q$ in $A^{\uparrow}$\;
    }
}
\caption{Lifted Gauss–Jordan over $\mathbb{F}_2$}
\label{alg:lifted_gauss_elim}
\end{algorithm}

\begin{figure}[ht!]
    \centering
    \includegraphics[width=0.5\textwidth]{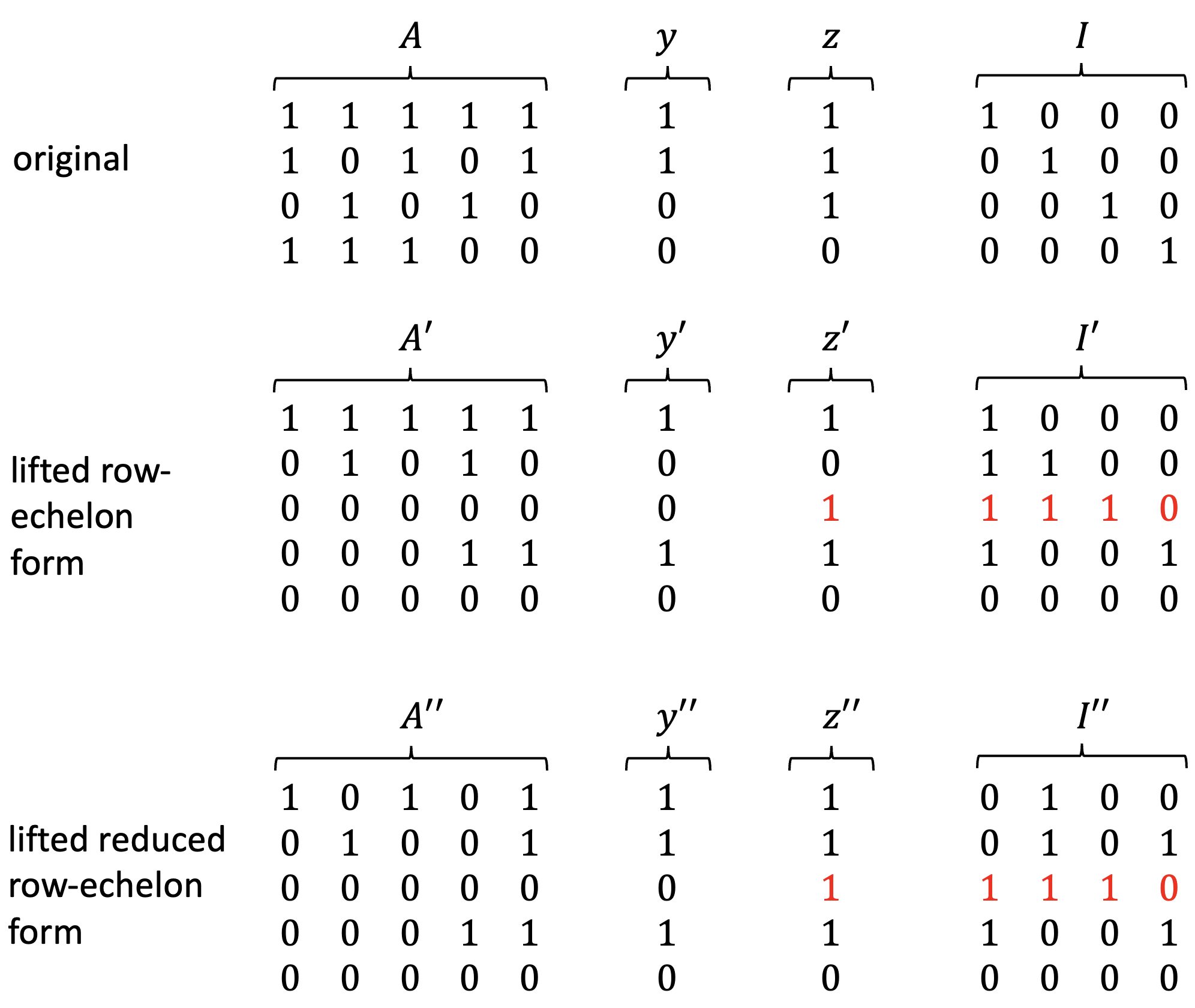}
    \captionsetup{aboveskip=0.5ex} 
    \caption{
    \textbf{Lifted reduced row–echelon form.}  
    Example of transforming a matrix $A$ into lifted reduced row–echelon form using \alg{lifted_gauss_elim}, shown with three spectator arrays $B$ (vectors $y$ and $z$ and the identity matrix $I$).  
    After the first pass, $A$ is in lifted row–echelon form with no pivots in rows three or five.  
    The target $z$ is unattainable because $z'$ has a $1$ in non–pivot row three, whereas $y$ does not, implying that a solution $x$ to $Ax = y$ exists.  
    After the second pass, $x = y''$ is obtained, and the identity has been transformed into the pseudo-inverse of $A$, satisfying $A I'' A = A$.
    }
    \label{fig:lifted-gauss-jordan}
\end{figure}

Lifted Gauss–Jordan elimination handles arbitrary matrices in a straightforward manner, and when applied to a spectator matrix $B$ in along with $A$, the spectator array is transformed directly into the solution of the corresponding problem; see \fig{lifted-gauss-jordan}.  
For an $m \times n$ binary matrix $A$, a length-$m$ spectator vector $y$ is mapped to a length-$n$ vector $x$ satisfying $Ax = y$ (when a solution exists), while an $m \times m$ identity spectator is mapped to an $n \times m$ pseudo-inverse $X$ satisfying $AXA = A$.

\begin{figure}[ht!]
    \centering
    (a)\includegraphics[width=0.95\textwidth]{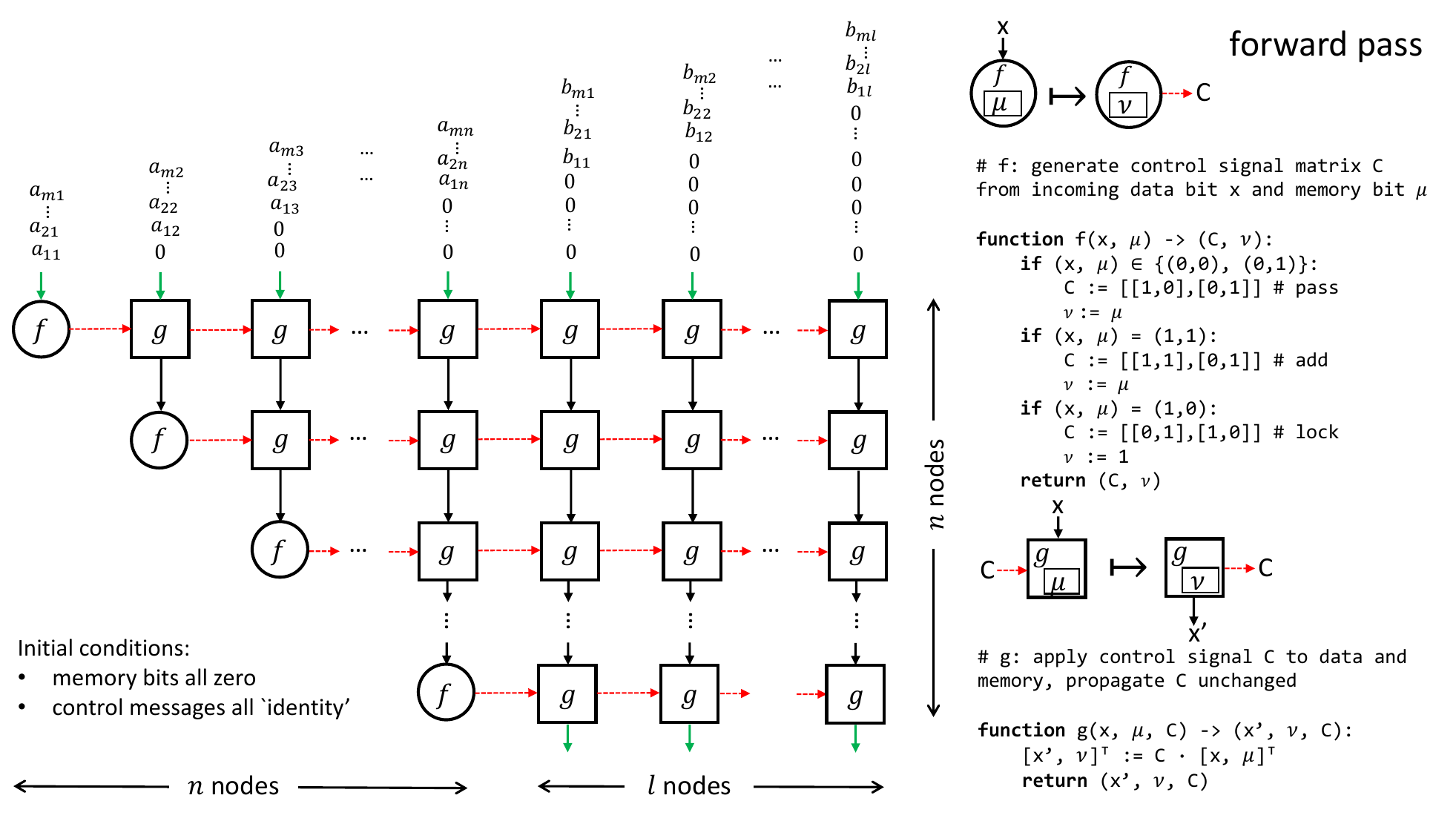}

    \vspace{2ex} 

    (b)\includegraphics[width=0.95\textwidth]{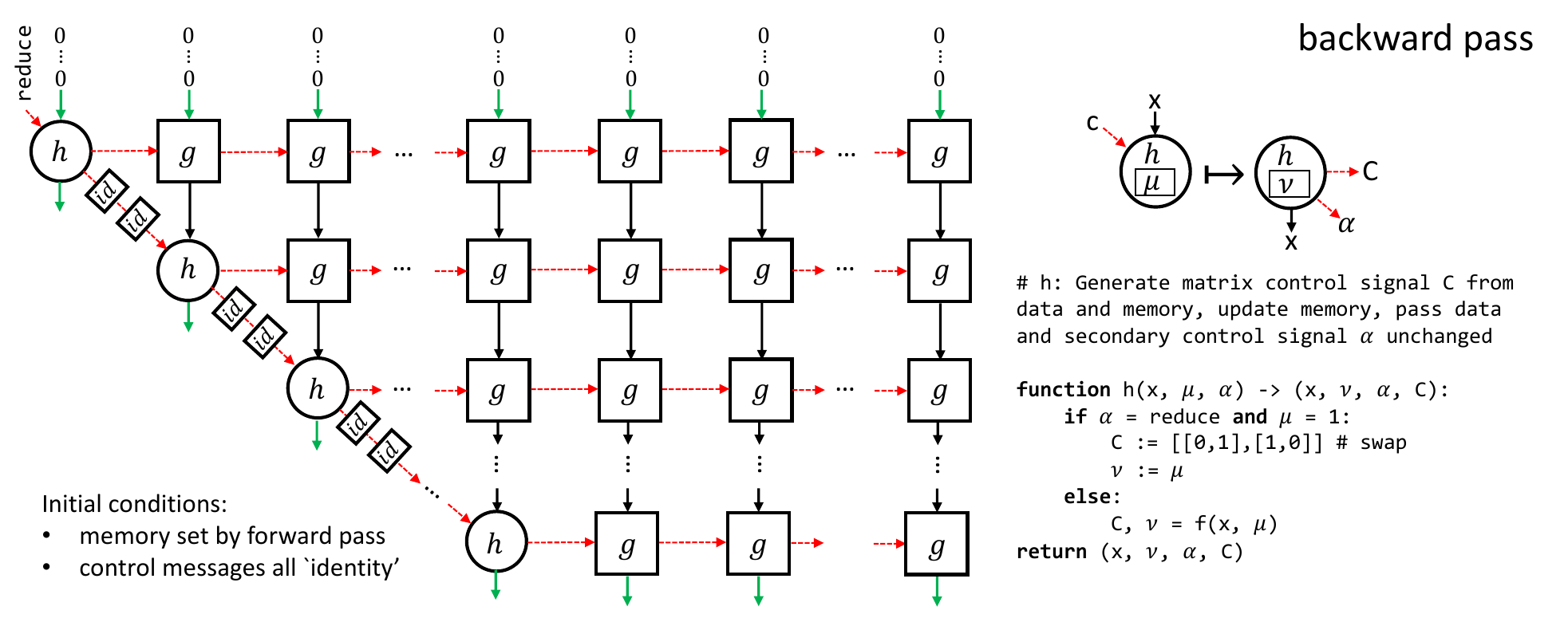}

    \captionsetup{aboveskip=0.5ex} 
    \caption{
    \textbf{Systolic array for lifted Gauss–Jordan elimination.}
    Adapted from Hochet, Quinton, and Robert~\cite{Hochet1989}.
    (a) Forward pass: rows of $[A|B]$ enter from the top with staggered timing and propagate down the array. 
    Circular (diagonal) nodes detect and lock pivots, store the corresponding row, and send $2\times 2$ control matrices to the right, directing square nodes to \emph{pass}, \emph{add}, or \emph{lock} rows. 
    For $l=1$ with $B=y$, a nonzero output at the bottom of column $n+1$ signals that $Ax = y$ has no solution.
    (b) Backward pass: locked rows flow downward from the top, eliminating $1$s they have in pivot columns of lower rows. 
    This array reduces to the forward-pass array if the “reduce” control is omitted.
    }
    \label{fig:general-systolic-gauss}
\end{figure}

We implement our computation graph algorithm for transforming a matrix $A$ with spectator $B$ into lifted reduced row–echelon form, as shown in \fig{general-systolic-gauss}.
Our systolic array computation graph algorithm only involves minor changes to the Ref.~\cite{Hochet1989} algorithm (in fact, some of our changes from the original algorithm streamline the associated computation graph a little even though it can be applied to more general matrices).
In the forward pass, rows of the augmented matrix $[A|B]$ stream downward through the systolic array with staggered input.
Our design extends the array of Ref.~\cite{Hochet1989} to the right for $l>1$ and initializes with zeros, removing the original “init” control signal.
Whereas the original assumed a full–rank square $n\times n$ input (and declared failure if singular), we observe that their forward-stage design processes any $A\in\mathbb{F}_2^{m\times n}$ into lifted row–echelon form.

The $j$th diagonal (circular) node records whether a $1$ has been encountered in column $j$.
Each iteration, it sends a $2\times 2$ binary control matrix along its row\footnote{In practice, only a two–bit identifier is needed, since there are three possible matrices.}, which directs square nodes on how to compute and update their bits from the incoming and stored values.
When a circular node first encounters a $1$, the column is locked: its memory bit is set to $1$, and the corresponding row is stored locally instead of passing downward.
Unlocked rows continue downward; if they encounter a locked row, they either pass unchanged or, if they have a $1$ in the locked row’s pivot column, are updated by adding (XORing) the locked row.

After exactly $2n+m+l-1$ iterations, all rows have passed through the array.  
Rows that were never locked exit at the bottom, implying their $A$–portion is zero (a non–pivot row) and their spectator portion can be read out.  
The locked rows remain stored in the array’s memory bits, forming the remaining rows of the $n\times (n+l)$ matrix $[A'|B']$.  
At this stage $A'$ is in lifted row–echelon form, though not yet in lifted reduced row–echelon form.

For the solution–existence problem (i), set $B = y$ and perform only the forward pass.  
If $A$ has full row rank ($r = m$), all rows are locked and a solution exists for any $y$.  
If $r < m$, then $m - r$ rows of $A$ and their corresponding bits of $y$ pass through the array, undergoing row operations.  
A solution $x$ exists if, and only if, no $1$s appear at the bottom node during the $2n + m + l - 1$ forward–pass iterations.

To obtain the lifted reduced row–echelon form $[A^\uparrow|B'']$ from the lifted row–echelon form $[A'|B']$ stored at the end of the forward pass, we use the modified computation graph algorithm in \fig{general-systolic-gauss}(b).  
We use a slight modification of the backward pass computation graph from Ref.~\cite{Hochet1989} with less complex nodes and control.
Rows are processed sequentially from top to bottom: each row flows downward through the array, adding any locked row it encounters if it has a $1$ in that row’s pivot column.  
When a row exits the bottom, all its $1$s in any other pivot column have been cleared, yielding the corresponding row of $[A^\uparrow|B'']$.  
This process repeats for each row, starting the next row’s descent as soon as the previous one has cleared the array, until all rows of $A^\uparrow$ have been collected.

Let us consider how many iterations the backward pass takes. 
The right-most entry of the first row only starts moving down after $n+l$ iterations since it takes that long to receive the signal, and then it takes it $n-1$ iterations to be output at the bottom of the array. 
After that, the right-most element of each of the remaining $n-1$ rows passes through one at a time, taking another $n-1$ iterations, so that the total number of iterations is $3n+l-2$. Alternatively, it takes $3(n-1)+1$ iterations for the \texttt{reduce} signal to reach the last $h$ node and another $l$ iterations to finish the task. 

Now lets consider the scenario in which we wish to run the forward pass and then the backward pass right after. 
First note that the computation graph for the backward pass actually recovers that for the forward pass if we do not input the “reduce” control signal in the top left port -- since in this case $h$ reduces to $f$, and the function $g$ in the remaining nodes is unchanged.
Hence, we can run the full algorithm (forward then backward pass) on the computation graph for the backward pass, simply by feeding zeros in the top after the arrays have been passed in, along with feeding “reduce” into the leftmost control port after $m$ iterations have elapsed. 
This begins the backward pass while the forward pass is still going on.
The total amount of time from start to finish is then $m+(3n+l-2)=3n + m + l -2$ iterations. 
We note here that after the forward pass has completed the result of the systolic calculation is stored in the internal memory of the $g$ nodes. 
Likewise, after the backwards pass results are contained in the $g$ nodes internal memory and do another $O(n)$ iterations are required to extract the result.

\paragraph{Resources summary}
The systolic-array algorithm presented here solves $Ax=b$ for any $A \in \mathbb{F}_2^{m\times n}$ and $b \in \mathbb{F}_2^{m}$ using $n(n+1)/2 + n$ processing elements and completing in $3n + m - 1$ cycles (or declaring unsolvability after $n + m - 1$ cycles).
Computing a generalized inverse of $A$ requires an extended array with $n(n+1)/2 + n^2$ processing elements, with the same $3n + m - 1$ cycle count.

A fixed array of size $n(n+1)/2 + n$ can also solve $Ax=b$ for any $A \in \mathbb{F}_2^{m\times n'}$ with $n' \leq n$.
Padding $A$ with $(n-n')$ zero columns yields a valid instance and incurs the full $3n + m - 1$ cycles.
However, we expect a minor modification to the control logic should allow the same hardware to solve the $m \times n'$ system in $3n' + m - 1$ cycles without padding, although we have not analyzed this variant in detail.
\clearpage
\section{Filtered ordered statistics decoding with FPGAs}
\label{sec:fpga-osd}

\begin{figure}[h]
    \centering
    \includegraphics[width=0.83\linewidth]{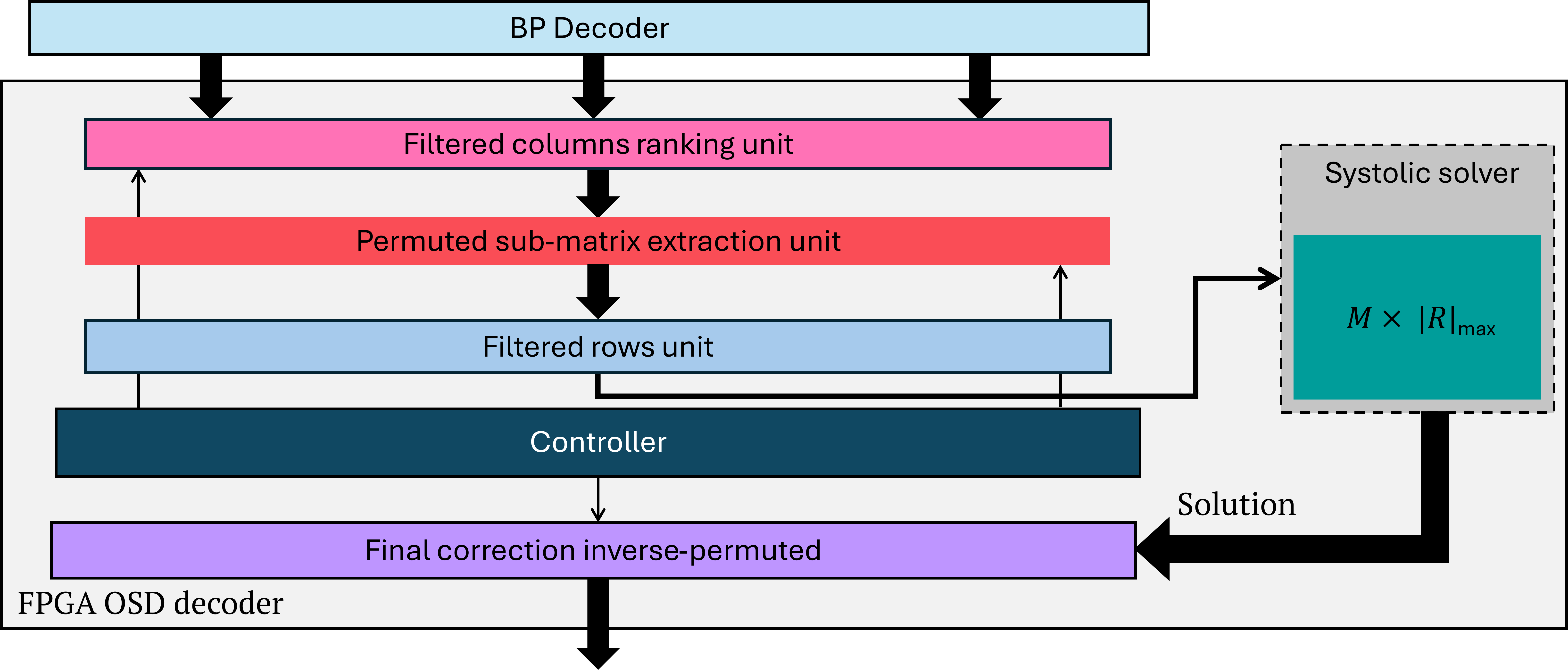}
    \caption{
    \textbf{FPGA architecture overview for filtered-OSD for quantum LDPC codes.}
    When the pre-decoder fails to converge, it forwards per-fault log-likelihood ratios (LLRs) to the FPGA OSD stage. 
    Fault indices with LLR values below $\Lambda_{\text{confident}}$ are collected and sorted in increasing LLR (most likely first). 
    The corresponding columns of the decoding matrix $H$ are extracted and permuted so that lower-LLR columns appear leftmost. 
    The resulting reduced, reordered matrix is sent to the systolic solver, which obtains a solution on the permuted subset of faults. 
    The resulting solution is then inverse-permuted to recover the final correction on the full set of faults in the original ordering.
    }
    \label{fig:osd-arch}
\end{figure}

In this section we present an FPGA implementation of a filtered version of the ordered-statistics decoding (OSD) algorithm for quantum LDPC codes~\cite{panteleev2021degenerate,fossorier1995soft,fossorier2002iterative,roffe2020}. 
We modify OSD from \sec{osd-decoding} to take advantage of our systolic solver to find a correction without first computing linearly independent columns or a matrix inverse.
The result is a filtered version of OSD which requires a much smaller linear solution as described in \sec{filtered-osd-decoding}.
We provide an FPGA-tailored design for filtered-OSD (summarized in \fig{osd-arch}) with \sec{osd-input}, \sec{fault-ranking}, \sec{submatrix-extraction}, \sec{systolic-solver}, and \sec{osd-output} detailing each component of the design. 
In \sec{osd-extensions} we outline how to extend our design in various ways, including how to implement the combination sweep to form OSD–$k$. 

Our implementation overcomes two significant obstacles for OSD in FPGAs.
First, by restricting to a filtered column set, the systolic solver acts on a rank-deficient matrix which is much narrower than that which is inverted in the standard OSD algorithm, reducing the FPGA footprint considerably (which otherwise is unlikely to fit on available FPGAs~\cite{valls2021syndrome}). 
Second, the lifted Gauss–Jordan elimination implemented by the systolic solver in \sec{systolic-gaussian-elimination-algorithm} removes linearly dependent columns on the fly, avoiding a costly pre-processing stage (this second obstacle is also overcome by the FPGA implementation of OSD in the recent work of Ref.~\cite{bascones2025exploring}).

\subsection{Filtered ordered statistics decoding}
\label{sec:filtered-osd-decoding}

We introduce a variant of ordered-statistics decoding that markedly reduces FPGA cost. 
Standard OSD selects an ordered column list $S$ from the predecoder with $|S|=\operatorname{rank}(H)$ and computes a correction supported in $S$ via a left inverse of $H_S$ (e.g., $x=H_S^{-1}\sigma$), which requires $H_S$ to have full row rank.
In practice, the true support is usually much smaller and concentrated near the front of $S$ (which the pre-decoder concluded to be most likely to have errors), so we restrict to a smaller prefix $R\subset S$ and solve $H_{R}y=\sigma$ with our systolic solver, avoiding explicit inverse construction. 
This targeted restriction and direct solve yield substantial savings in our FPGA implementation.
In what follows, we state the filtered version of OSD.

\paragraph{Algorithm input.}
The algorithm takes as input the binary decoding matrix $H \in \mathbb{F}_2^{M \times N}$, the observed syndrome $\sigma \in \mathbb{F}_2^{M}$, and a score $q_j$ for each column $j \in 1,\dots,N$, where low scores correspond to more likely faults, and a cut-off $q_\text{filter}$.

\paragraph{Filtered fault ranking.}
Obtain a list $R$ of fault indices (columns) ranked from lowest to highest score with $q_j<q_\text{filter}$. 

\paragraph{Correction in highest-ranked independent columns.}
Find the solution $x \in \mathbb{F}_2^{|R|}$ such that $H_R x = \sigma$.

\paragraph{Algorithm output.} 
The entries of $x$ are placed at the positions in $R$ (zeros elsewhere) to form the full correction $F \in \mathbb{F}_2^{N}$.

\subsection{Algorithm input}
\label{sec:osd-input}

As described in \sec{osd-decoding}, OSD takes as input the decoding matrix $H \in \mathbb{F}_2^{M \times N}$, the syndrome vector $\sigma \in \mathbb{F}_2^{M}$, and the log-likelihood ratio (LLR) vector $\mathbf{\Lambda} \in \mathbb{R}_2^{M}$ outputted by a pre-decoder. 

Let us consider the data structures used to store these objects.
As mentioned in \sec{fpgas}, the primary options for memory in FPGAs are flip-flops (FFs) and random access memories (RAMs).
The syndrome vector $\sigma$ is a bit-string and can be stored in FFs.
The LLRs are saved as floating-point/integer values making it inconvenient to store them in FFs, and so we instead store them in RAMs.
The decoding matrix $H$ requires fast lookup, and is therefore stored in RAM. 
We provide additional FPGA details in \sec{results} and \app{osd-resources} for our example decoding problems.

\subsection{Filtered column ranking}
\label{sec:fault-ranking}

To make use of the LLRs $\Lambda_j$, we build a filtered ranking of column (fault) indices. 
We first keep those with $\Lambda_j<\Lambda_{\text{confident}}$ (a threshold above which a fault is deemed unlikely), then sort the survivors in increasing $\Lambda_j$ to obtain $R=[j_1,j_2,\ldots, j_{|R|}]$, where typically $|R|$ is considerably less than $N$.
If without padding, $|R|>|R|_\text{max}$, we say that the decoding algorithm fails.
To avoid the complications that could arise from a variable number of survivors in an FPGA implementation, we pad the list $R$ with empty entries such that $|R|=|R|_\text{max}$ for a fixed length irrespective of the decoding instance.

\begin{figure}[h]
    \centering
    \includegraphics[width=0.75\linewidth]{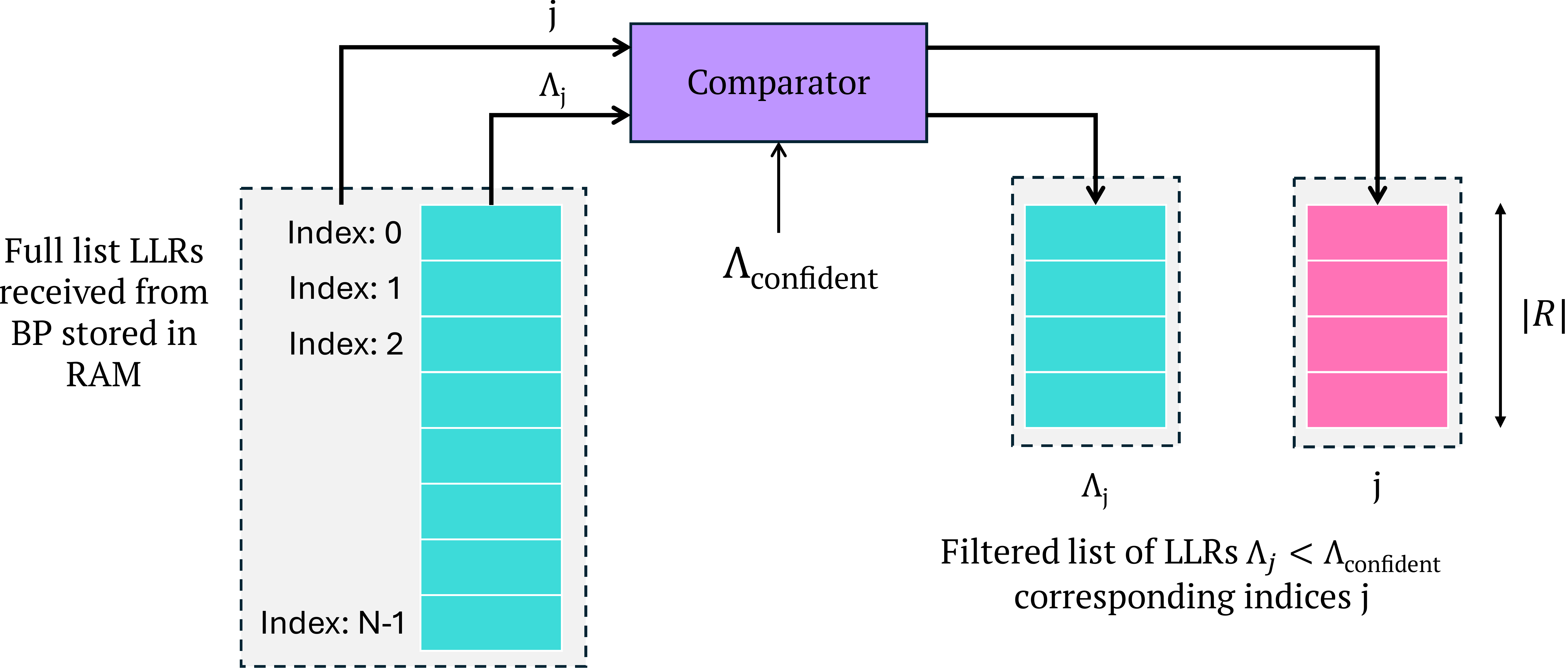}
    \caption{
    Filtering the list of LLRs to find all $\Lambda_j$ for which $\Lambda_j<\Lambda_{\text{confident}}$: 
    The LLRs received from the pre-decoder are stored in a RAM input register. 
    The input register is read sequentially, and the comparator passes the $(\Lambda_j, j)$ pairs for which $\Lambda_j<\Lambda_{\text{confident}}$ to the corresponding RAM output registers on the right.
    }
    \label{fig:truncating}
\end{figure}

\paragraph{Filtering the columns}
Since the LLRs $\Lambda_j$ are stored in RAM, we can read them sequentially and compare every value read from memory with $\Lambda_{\text{confident}}$. 
Successful comparisons are then pushed to a separate pair of RAMs, one each for $\Lambda_j$ and index $j$. 
The approach is described in \fig{truncating}.
In our decoding examples analyzed in \sec{results} and \app{cluster-resources}, we implement this in a banked manner which partially parallelizes the filtering stage of the algorithm.

\begin{figure}[h]
    \centering
    (a)\includegraphics[width=0.68\linewidth]{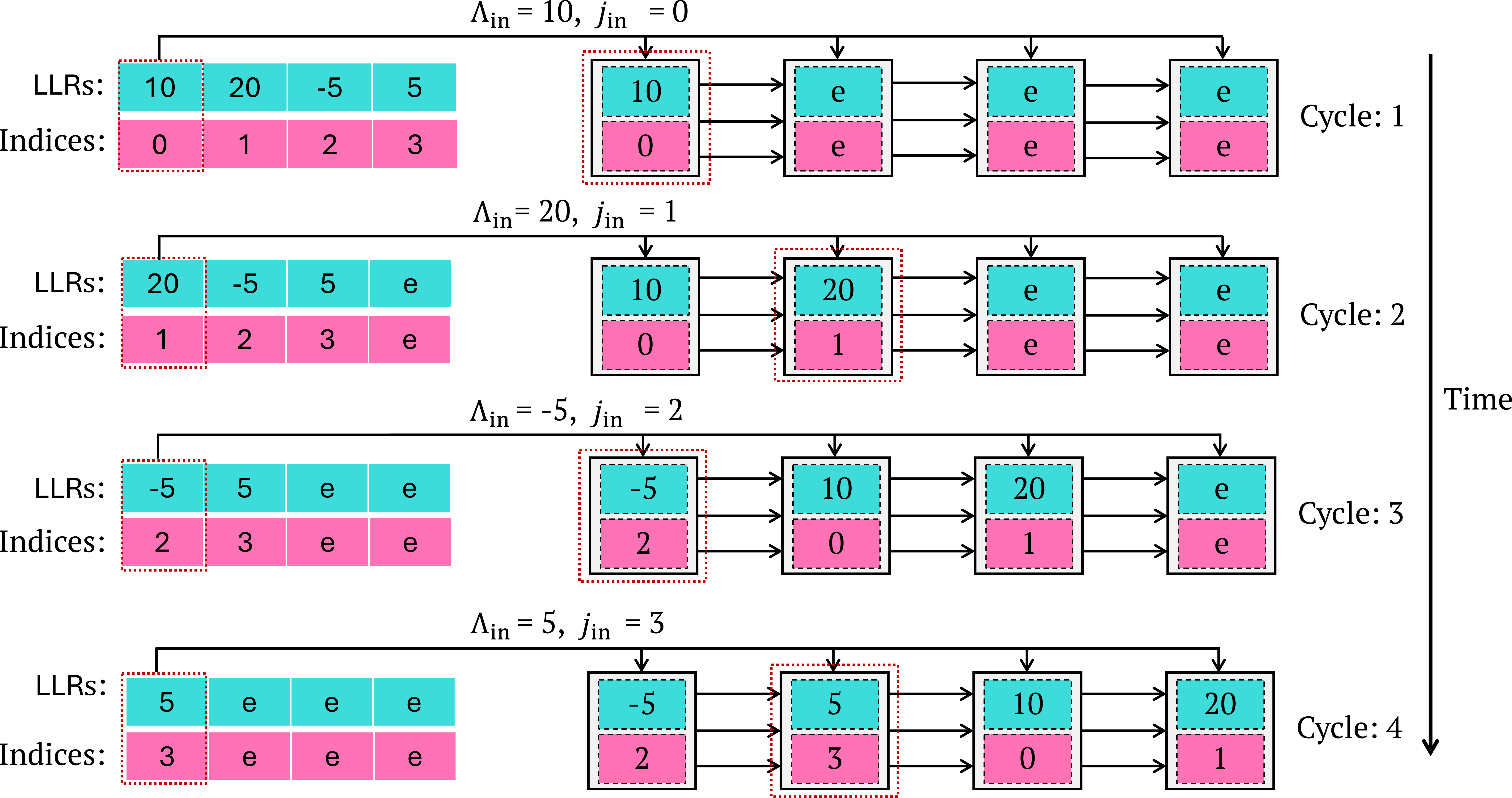}
    (b)\includegraphics[width=0.23\linewidth]{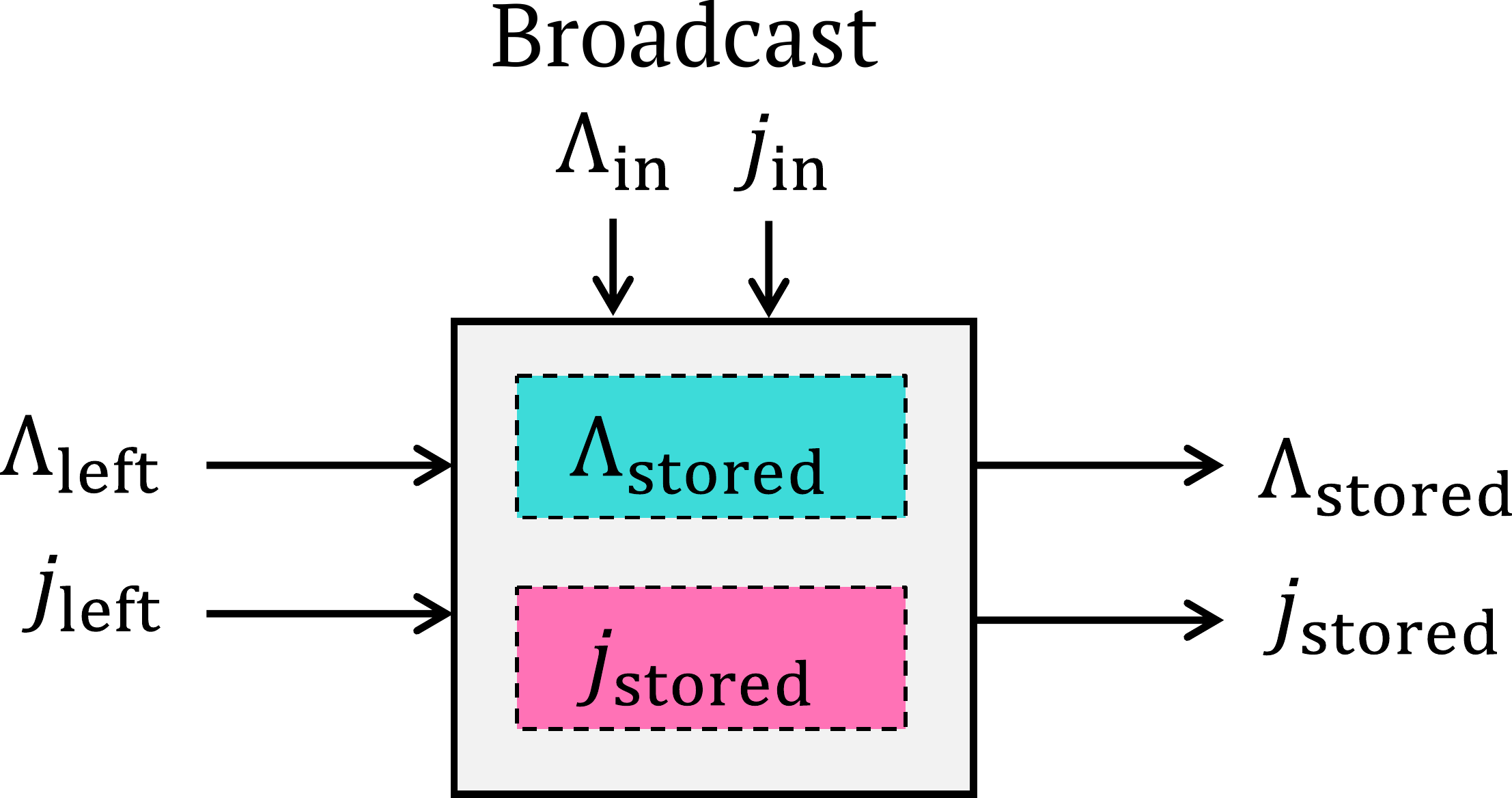}
    \caption{
    \textbf{Broadcast insertion sorter.} 
    This computation-graph algorithm sorts a filtered list of $|R|$ LLRs in $|R|$ cycles (example shown for $|R|=4$). 
    (a) Two length-$|R|$ registers are used: an input register initially holding the unsorted pairs and an (initially empty) output register. 
    Each cycle removes the front pair from the input, broadcasts it to all output nodes, inserts it at the correct position, and shifts the remaining input pairs forward by one slot. 
    Dotted red boxes indicate the output nodes that capture the broadcast value on that cycle.  
    (b) Each output-register node updates in parallel each cycle: it forwards its stored pair to the right, receives the broadcast candidate $(\Lambda_{\text{in}},j_{\text{in}})$, and compares $\Lambda_{\text{in}}$ with its stored $\Lambda_{\text{stored}}$ and the left neighbor’s $\Lambda_{\text{left}}$. 
    If $\Lambda_{\text{left}} \le \Lambda_{\text{in}} < \Lambda_{\text{stored}}$, the stored pair becomes $(\Lambda_{\text{in}},j_{\text{in}})$ (insertion); if $\Lambda_{\text{in}} < \Lambda_{\text{left}}$, it becomes $(\Lambda_{\text{left}},j_{\text{left}})$ (shift right); otherwise it is unchanged.
    Empty nodes (marked “e”) compare as $\Lambda_{\text{stored}}=+\infty$.
    After $|R|$ cycles the register stores $(\Lambda,j)$ in non-decreasing $\Lambda$, and the sorted indices are used by the submatrix extraction module.
    }
    \label{fig:osd_sorting}
\end{figure}

\paragraph{Sorting filtered columns.}
There are several FPGA-friendly ways to sort the pairs~\cite{papaphilippou2020adaptable,ortiz2011streaming,zuluaga2016streaming,li2020extended} $$(\Lambda_{j_1}, j_1), (\Lambda_{j_2}, j_2), \ldots, (\Lambda_{j_{|R|}}, j_{|R|}),$$ 
for example, bubble sort has a natural 1-D systolic-line implementation which completes in $2|R|$ cycles via alternating odd and even comparisons and swaps. 
In this work we instead use the broadcast–insertion sorter~\cite{lee1995shift} shown in \fig{osd_sorting}: although not systolic, it finishes in just $|R|$ cycles by broadcasting one input pair per cycle to all output-register nodes.
We note that since our filtered version of OSD requires the sorting of a much smaller list than the full list of columns, the contribution of the sorting step to the overall algorithm cost is relatively small.

\begin{figure}[h]
    \centering
    \includegraphics[width=0.95\linewidth]{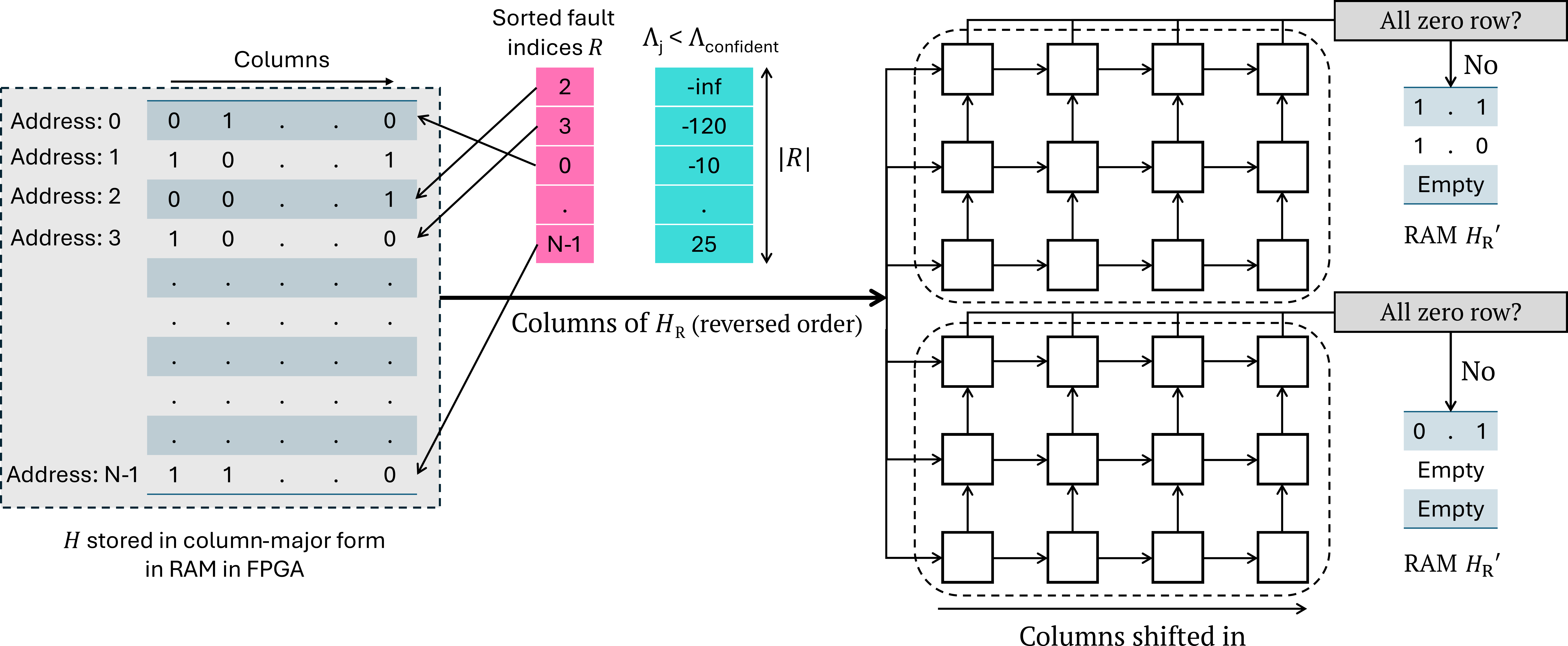}
    \caption{
    \textbf{Extracting permuted submatrix.} 
    $H$ (displayed transposed $H^T$) has one memory address per column. Storing $H$ in on-chip RAM enables fast column access with low FPGA routing cost.
    The sorted, filtered list of indices in $R$ are traversed from largest to smallest LLR, and the corresponding columns are read to assemble the permuted submatrix $H_\text{R}$.
    The order in which the LLRs are traversed is to ensure the columns are arranged in the right order in the intermediate storage (explained in the next step).
    \\
    \textbf{Filtering all zero rows from the extracted submatrix.} The columns of $H_\text{R}$ are shifted into an intermediate storage composed of an array of $M \times |R|_\text{max}$ flip-flops. 
    Once this extraction is completed, the array of flip-flops can be partitioned into multiple groups. Within each group, rows are shifted upwards and checked if they are all zero. If not, they are pushed into RAMs that are allocated per group of the array of flip-flops. These RAMs together store the filtered submatrix $H_\text{R}'$.
    Dividing the rows of the $M \times |R|_\text{max}$ array into multiple blocks allows faster filtering of all zero rows from the extracted submatrix.
    }
    \label{fig:filtered_rows}
\end{figure}

\subsection{Extracting submatrix and filtering out zero rows}
\label{sec:submatrix-extraction}

Once the filtered set of column indices are sorted, a submatrix $H_R$ of the original decoding matrix $H$ must be extracted, indexed by these column indices. 
Due to the sparsity of $H$, $H_R$ will contain many rows composed of only zeros. These rows are inconsequential for decoding, and can be ignored to yield a reduced submatrix $H_R'$. 
We now show how the submatrix can be extracted and then how all zero rows can be filtered out.

\paragraph{Column extraction}
From the fixed decoding matrix $H \in \mathbb{F}_2^{M \times N}$, select the columns indexed by $R$ to form the submatrix $H_R \in \mathbb{F}_2^{M \times |R|}$, consisting of columns of $H$ ordered by the filtered ranking $R$.
This is achieved as shown in \fig{filtered_rows}.

\paragraph{Filtering all zero rows}
We now exploit the sparsity of $H$.
Since $H_R$ consists of a relatively small subset of columns of the decoding matrix, many rows can be expected to be all zero (in \sec{results} when we consider example decoding problems we observe about 90\% of rows are zero in typical cases) and can be skipped (reducing the cycle count).
\fig{filtered_rows} shows how this additional filtering step can be done to get a filtered submatrix $H_\text{R}'$. 
During the initial extraction of the submatrix into the array of flip-flops, the right-most column of the array would contain the first column extracted.
To preserve the ordering of columns of the matrix, the submatrix should thus either be extracted last column first or reversed in the zero row filtering step.

\subsection{Correction in submatrix (systolic solver)} 
\label{sec:systolic-solver}

In \sec{osd-decoding}, the OSD-0 algorithm finds a correction within the highest-ranked independent columns using two stages.
First, the highest-ranked linearly independent set $S$ of columns is identified such that $|S|=\text{rank}(H)$.
Second, an inverse of the matrix $H_S$ is obtained.

Our FPGA implementation attains the same result (with high probability\footnote{Filtering can alter OSD in rare cases: in some runs (i) more than $|R|$ LLRs may fall below $\Lambda_{\text{confident}}$, exceeding the register capacity, or (ii) the retained indices may not admit a consistent correction; we treat either case as an OSD failure. 
In practice we choose $|R|$ and $\Lambda_{\text{confident}}$ so that this failure probability is negligible.}) but via a slightly different route. 
First, we use only $|R| \ll |S| = \operatorname{rank}(H)$ columns.
This is possible since $\operatorname{rank}(H)$ greatly exceeds the typically correction support, which is very likely contained in the lowest-LLR subset of $S$, and because the systolic solver never inverts a submatrix of $H$, it need not operate on a full-rank submatrix. 
Second, we do not pre-identify linearly independent columns: the systolic solver discovers them on the fly and sets their correction entries to $0$, which is equivalent to having excluded those columns from the outset.

We have already described in \sec{fault-ranking} how to identify the filtered column ranking $R$. Once the submatrix is extracted and all zero rows are filtered out, we use the systolic array algorithm from \sec{systolic-gaussian-elimination-algorithm} to determine  $x \in \mathbb{F}_2^{|R|}$, where $H_R x = \sigma$. 
Note that any time the systolic algorithm encounters a linearly dependent column, it is not identified as a pivot column and therefore will not be included in the correction support.

\subsection{Algorithm output}
\label{sec:osd-output}

\begin{figure}
    \centering
    \includegraphics[width=0.8\linewidth]{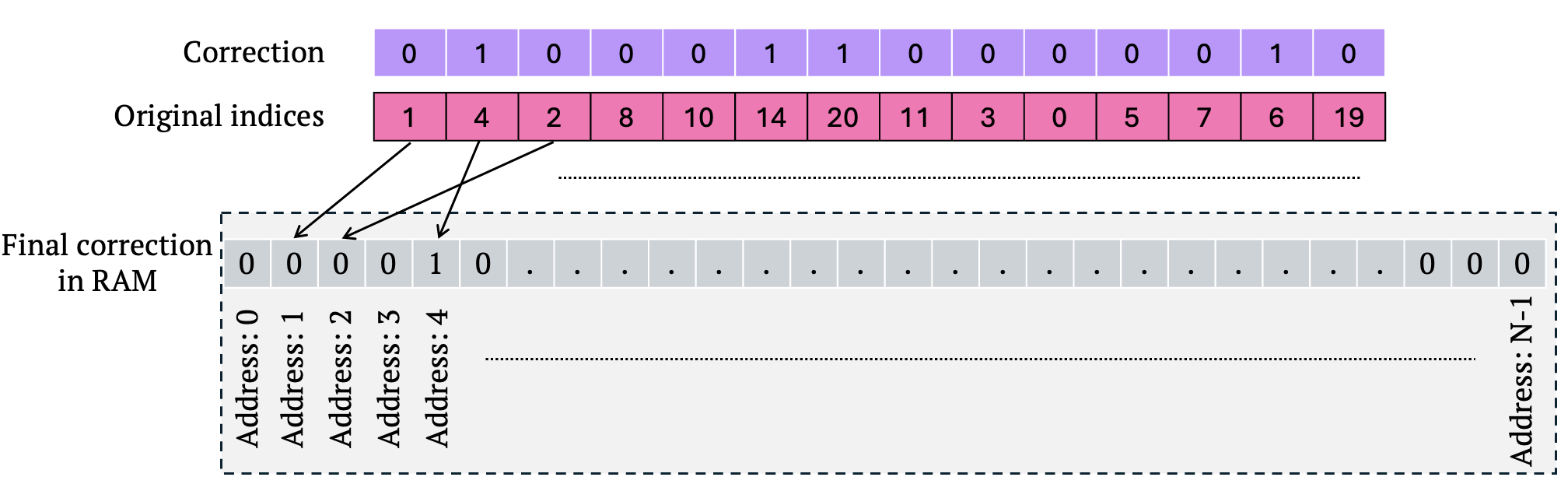}
    \caption{
    \textbf{Inverse permutation for final correction.} 
    The systolic solver outputs indices $R$ and bits $x \in \mathbb{F}_2^{|R|}$ as pairs $(R_j,x_j)$. 
    Each pair inserts $x_j$ at address $R_j$ in the length-$N$ correction vector held in on-chip RAM, initialized to $0^N$.
    }
    \label{fig:descramble}
\end{figure}

The correction $x \in \mathbb{F}_2^{|R|}$ we have obtained is supported on the faults in $R$ (and is expressed in the order of $R$).
To lift this to a correction $F \in \mathbb{F}_2^N$, one should set $F_j=1$ for each $j$ such that $x_{r} = 1$ for $R_r=j$.
To implement this in an FPGA, we use the approach described in \fig{descramble}. 
The final correction for all $N$ faults is initialized in RAM. 
Using the indices $R$ as the address for this RAM, the bits of the final correction can be set to the corresponding bit in the correction generated by the systolic solver. 

\subsection{Extensions}
\label{sec:osd-extensions}

We briefly outline natural extensions to the FPGA-tailored algorithm of filtered-OSD we have presented above.

First, to improve the run-time, one can retain as $R$ only those indices $j$ with $\Lambda_j<\Lambda_{\text{confident}}$, without padding to $|R|_\text{max}$. 
This introduces mild dynamism—sorting a variable-length list of LLRs and extracting a variable number of columns from the decoding matrix but reduces average runtime; the worst case (when $|R|=|R|_\text{max}$) is unchanged.

One could reduce memory requirements by storing columns of the decoding matrix $H$, which is sparse, as row-index lists (reducing the memory requirements). 
Another opportunity to reduce the memory requirements to store the decoding matrix could be to exploit its structure. 
In our decoding examples for instance, the circuit consists of many repeated syndrome-cycles which result in the decoding matrix containing many copies of the same submatrix. 
This could potentially be exploited to store the matrix more efficiently.
All of these approaches are likely to slow down the extraction of parts of the decoding matrix however.

Finally, one may wish to implement the combination-sweep extension (OSD–$k$), which can lower logical error rates at the cost of additional runtime. 
The simplest adaptation to achieve this is to iterate over small initial fault sets, update the syndrome for each candidate, and run the full OSD–0 pipeline to obtain the resulting net corrections, selecting the correction with the lowest weight. This multiplies runtime by the number of candidates and is less advantageous on FPGA than on CPU, where a single expensive-to-compute inverse can be reused cheaply.\footnote{One could compute the full inverse on FPGA using version (iii) of the systolic solver in \sec{gateware_lse}, but this offers little benefit in the FPGA setting: both inversion and application take time linear in the matrix dimension, whereas on a CPU inversion costs $O(n^\omega)$ bit operations (with the best exponent~\cite{vassilevska2023omega} known to date $\omega\approx 2.37$) while a matrix-vector multiply costs just $O(n^2)$ bit operations.}

\clearpage
\section{Cluster decoding with FPGAs}
\label{sec:fpga-cluster}

In this section, we present an FPGA-tailored algorithm of the cluster decoding algorithm for quantum LDPC codes~\cite{delfosse2021almost,delfosse2022toward} described in \sec{union-find-algo}.
Conceptually, the cluster decoder is similar to OSD, which solves a linear system on a sub-matrix of faults prioritized by a pre-decoder, but the cluster decoder solves a set of even smaller small blocks corresponding to clusters in the decoding graph.

\begin{figure}[h]
    \centering
    \includegraphics[width=0.83\linewidth]{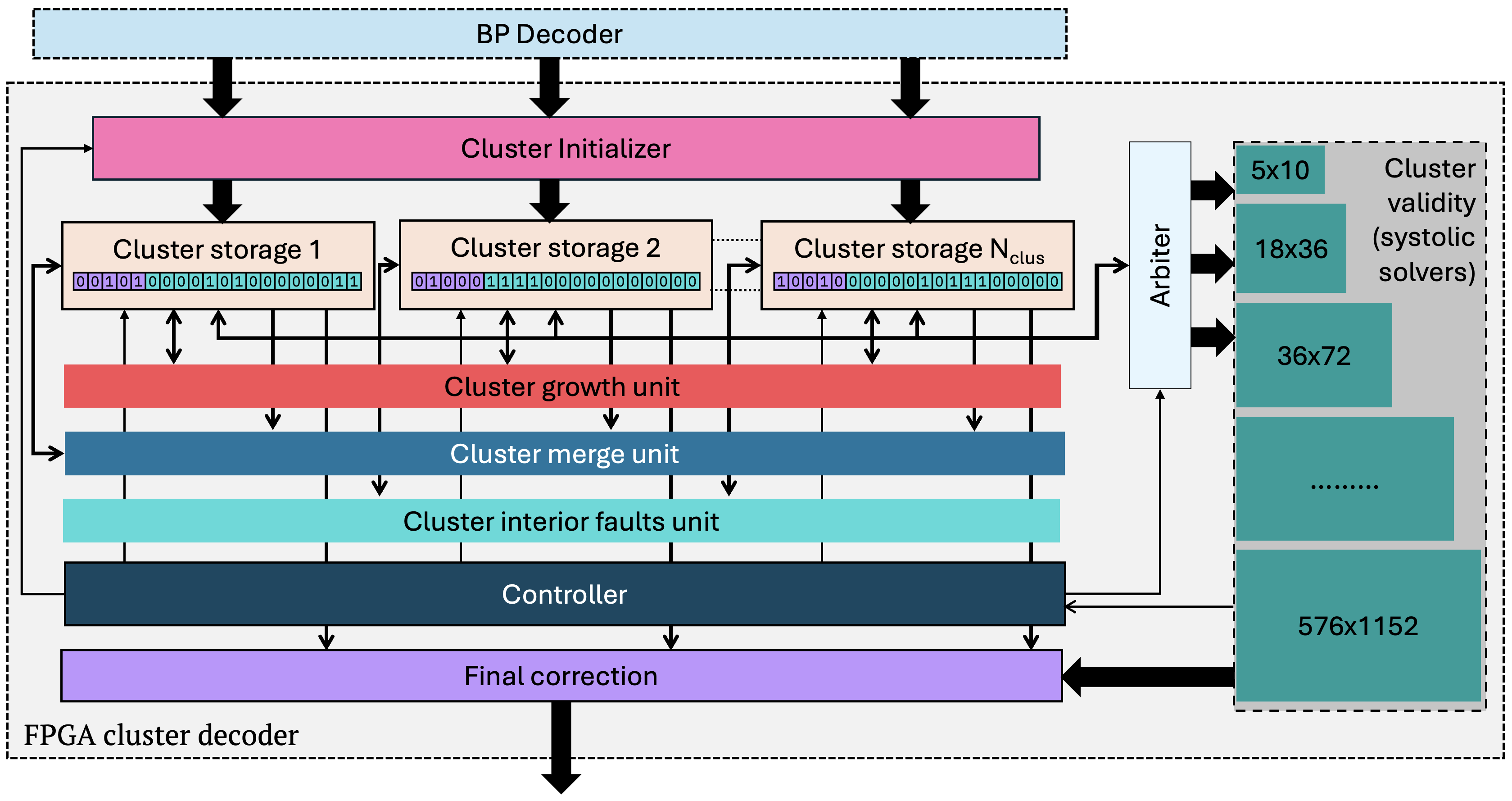}
    \caption{
    \textbf{FPGA architecture overview for cluster decoder for quantum LDPC codes.}
    A pre-decoder first processes the input syndrome. If no valid correction is found, it applies a partial correction, identifies an erasure set, and passes the residual syndrome and erasure set to the cluster decoder.
    The cluster initializer constructs the initial set of clusters, each stored as a bitmap in a dedicated cluster storage block.
    The cluster growth unit expands clusters by adding neighboring nodes, and the cluster merge unit detects and merges overlapping clusters.
    Cluster validity is checked using a pool of systolic-array linear system solvers from~\sec{systolic-gaussian-elimination-algorithm}, with multiple solver sizes available to match cluster size; an arbiter assigns solvers accordingly.
    A controller sequences the growth, merge, and validity stages, and the final correction unit aggregates corrections once all clusters are valid to produce the decoder’s output.
    }
    \label{fig:arch}
\end{figure}

The FPGA-tailored algorithm of cluster decoding we present here preserves the high-level algorithmic structure (cluster initialization, growth, merging, and validity checking) as specified in \sec{union-find-algo}, but realizes these steps in an FPGA-tailored architecture as summarized in \fig{arch}.
Our implementation addresses a key challenge for cluster decoding on FPGAs: incorporating pre-decoder information. 
Rather than using non-uniform weighted cluster growth~\cite{wolanski2024ambiguity}, which seems difficult to implement efficiently on FPGAs, we accept faults the pre-decoder classifies as high-confidence and initialize uncertain locations as erasures.
The following subsections describe in detail the FPGA-tailored algorithm of each aspect of the cluster decoder.

\subsection{Data structures and bitmap notation} 
\label{sec:data-structures}

First, we address the main FPGA-friendly data structures that we use to implement the algorithm.

\begin{figure}[ht!]
    \centering
    \includegraphics[width=0.4\linewidth]{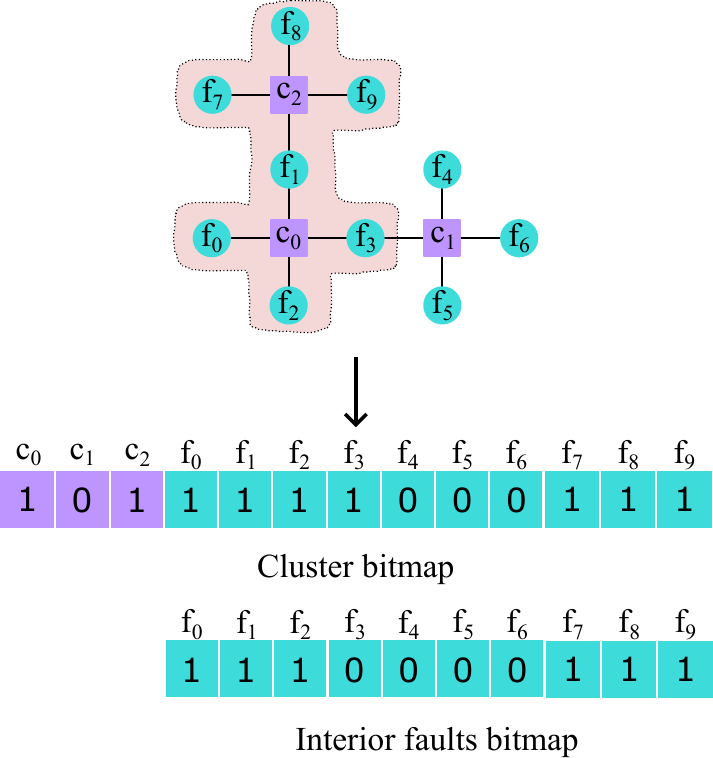}
    \caption{    
    The bitmap representation of a cluster assigns a bit to each node in the Tanner graph $\mathcal{G}$; the left bits correspond to syndrome nodes and the right bits to fault nodes.
    The shaded box highlights a cluster containing nodes $c_0, f_0, f_1, f_2, f_3$, which are assigned a value of $1$ in the cluster's bitmap, with all other nodes assigned $0$.
    Note that in practice the full decoding graph is much larger than this example.
    }
    \label{fig:cluster_repr}
\end{figure}

\paragraph{Cluster representation.}
The data structure used to represent clusters in an FPGA has a direct impact on the complexity of subsequent operations such as growth, merging, and validity checks.
A straightforward approach would be to assign each node in the fault graph a unique ID and store, for each cluster, the list of its node IDs, but this makes merge detection in an FPGA cumbersome.
Instead, as shown in \fig{cluster_repr}, we assign each node in the fault graph a fixed position in a \emph{cluster bitmap}, which is a bitstring $b \in \mathbb{F}_2^{M + N}$ in which a bit value of $1$ indicates that the corresponding node belongs to the cluster (in FPGA literature, such a representation is often called a \emph{bitmap}).
This is similar to the bitstring representation of errors and syndromes described in \sec{qec-background-general}, except that here a single length-$(M+N)$ bitstring represents an arbitrary (possibly mixed) set of check and fault nodes.
We assume $N_\text{clus}$ dedicated registers, which along with FFs to contain the bitmap, include a FF which marks whether or not the cluster is in use, and a FF which marks whether or not an in-use cluster is valid. 

As we will see later, we also use the \emph{interior faults bitmap} of a cluster,  
a length-$N$ bitmap indicating the fault nodes whose neighboring check nodes all lie within the cluster.

\paragraph{Decoding matrix representation}
We assume the decoding matrix $H \in \mathbb{F}_2^{M \times N}$ is stored in RAM since its entries need to be accessed quickly.
We also assume an `extended' adjacency matrix $H^\mathrm{ext} \in \mathbb{F}_2^{(M+N) \times (M+N)}$ by $H^\mathrm{ext}_{ij} = 1$ iff $i = j$ or vertices $i$ and $j$ are neighbors in $\mathcal{G}$.
Unlike $H$, the constant matrix $H^\mathrm{ext}$ does not require independent access of different columns or rows and can be embedded directly in an FPGA without using RAMs or FFs (see \app{cluster-resources} for more detail).

\paragraph{Bitmap notation.}
Let $\mathbf{b} \in \mathbb{F}_2^{n}$ denote a bit-vector (bitmap) stored in FFs in an FPGA. 
We use the following notation to describe standard bit-level operations:
\begin{enumerate}
    \item \textbf{$\mathbf{b} \ \&\ \mathbf{b'}$:} Element-wise logical \texttt{AND} between two bit-vectors.
    \item \textbf{$\mathbf{b} \ |\ \mathbf{b'}$:} Element-wise logical \texttt{OR} between two bit-vectors.
    \item \textbf{$\mathbf{b} + \mathbf{b'}$:} Element-wise logical \texttt{XOR} between two bit-vectors.
    \item \textbf{$\neg \mathbf{b}$:} Bit-wise logical \texttt{NOT} (complement) of $\mathbf{b}$.
    \item \textbf{$\mathrm{OR\_reduce}(\mathbf{b})$:} Reduction of $\mathbf{b}$ under logical \texttt{OR}, returning a single bit equal to~1 if any entry of $\mathbf{b}$ is~1. 
    \item \textbf{$\mathrm{AND\_reduce}(\mathbf{b})$:} Reduction of $\mathbf{b}$ under logical \texttt{AND}, returning~1 only if all entries of $\mathbf{b}$ are~1.
\end{enumerate}

Most of these operations can be done within a single FPGA clock cycle for reasonably sized inputs ($<1000$ bits), except the two reduction operations which can require multiple cycles for an input bitmap $\mathbf{b}$ with more than 1000 bits.
Large input sizes for can require multiple cycles since the reduction operation can be trivially decomposed into a tree-like manner, with each level of the tree requiring a single clock cycle. 
This decomposition is often necessary to help prevent the clock frequency from being severely degraded. 

\subsection{Algorithm input}
\label{sec:uf-input}

As described in \sec{union-find-algo}, the cluster decoder operates on an input consisting of the observed syndrome $\sigma$ and an erasure set $\mathcal{E}$, from which the initial clusters are constructed.  
In this work, we consider the scenario where the noise originates from a purely stochastic model, yielding an initial syndrome $\sigma_\text{original}$, and a pre-decoder has been run but failed to converge on a valid correction.  
We describe how the pre-decoder's outputs are used along with $\sigma_\text{original}$ to compute the inputs $\sigma$ and $\mathcal{E}$ for the cluster decoder, and how this computation can be implemented efficiently in FPGAs.

\begin{figure}[ht!]
    \centering
    \includegraphics[width=0.65\linewidth]{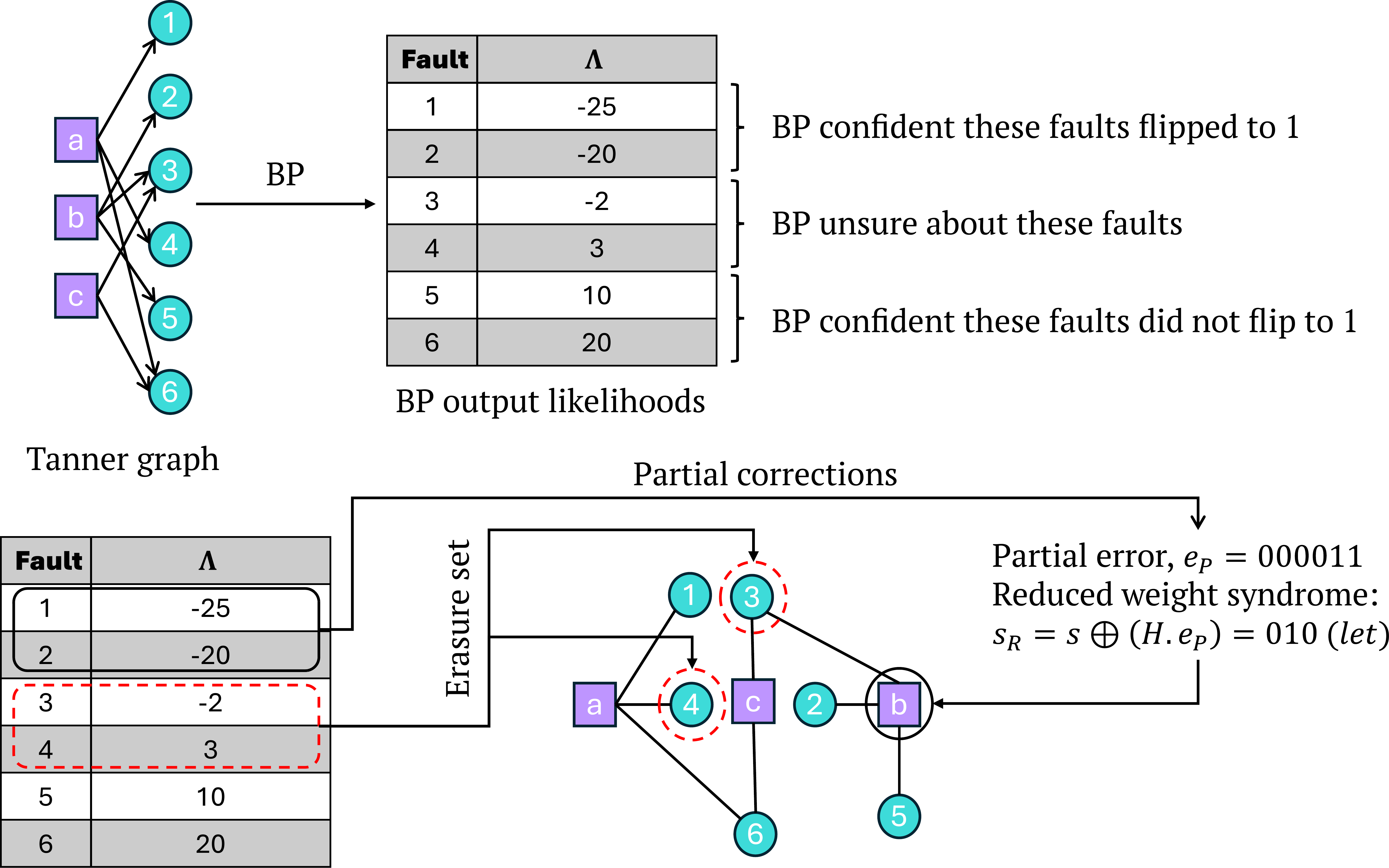}
    \caption{
    Mapping pre-decoder outputs to cluster decoder inputs.  
    Confident errors ($\Lambda_j \ge \Lambda_\text{accept}$) are flipped to form the partial correction $\tilde{F}_\text{pre}$, reducing the syndrome weight, while uncertain faults ($\Lambda_\text{erase} \le \Lambda_j < \Lambda^+_\text{accept}$) form the erasure set $\mathcal{E}$.
    }
    \label{fig:bpuf_flow}
\end{figure}

The pre-decoder produces log-likelihood ratios $\Lambda_j$ indicating confidence that fault $j$ occurred.  
Faults with $\Lambda_j \ge \Lambda_\text{accept}$ are treated as confident errors and flipped to form a \emph{partial correction} $\tilde{F}_\text{pre}$, which reduces the syndrome weight and thus the number of initial clusters.  
Faults with $\Lambda_\text{erase} \le \Lambda_j < \Lambda_\text{accept}$ are treated as uncertain and form the erasure set $\mathcal{E}$, initializing clusters over the corresponding fault nodes in the decoding graph (\fig{bpuf_flow}).  
The effect of varying the cutoffs $\Lambda_\text{accept}$ and $\Lambda_\text{erase}$ is analyzed in \app{cluster-resources}.

In the FPGA-tailored algorithm, the length-$N$ list of pre-decoder outputs is stored in a chain of $N$ nodes.  
Comparators threshold the pre-decoder outputs against the cutoffs to produce binary vectors for $\tilde{F}_\text{pre}$ and $\mathcal{E}$.  
Once $\tilde{F}_\text{pre}$ is obtained, the updated syndrome is computed via binary matrix–vector multiplication over $\mathbb{F}_2$ followed by entry-wise binary addition:
\[
\sigma \leftarrow \sigma_\text{original} + (H \tilde{F}_\text{pre}),
\]
both of which can be implemented efficiently in FPGAs.

\subsection{Cluster initialization}
\label{sec:uf-initialization}

At the beginning of the algorithm, clusters are initialized from the input syndrome $\sigma$ and the erasure set $\mathcal{E}$ obtained from the pre-decoder output (see \sec{union-find-algo}).
We represent $\sigma$ and $\mathcal{E}$ together as a length-$(M+N)$ bitmap $b_\text{init}$, where the first $M$ bits correspond to check nodes and the remaining $N$ bits correspond to fault nodes.
Each `1’ in $b_\text{init}$ seeds a unit-bitmap cluster; for example, $b_\text{init} = 10101$ produces unit bitmaps $10000$, $00100$, and $00001$.
For later use, we define the length-$(N+M)$ bitmap $u$ to be all zeros except for a single one in the right-most position.

To extract each unit-bitmap initialization seed from $b_\mathrm{init}$, we first set $b \gets b_\mathrm{init}$.
We then iteratively isolate and remove the right-most one in $b$ using the following loop:
\begin{enumerate}
    \item Compute $c = b + u = \neg(10101) + 00001 = 01011$.
    \item Compute $d = b \ \& \ c = 10101 \ \& \ 01011 = 00001$.
    \item The result $d$ is the desired unit-bitmap; store it.
    \item Update $b \gets b + d = 10101 + 00001 = 10100$ to clear the extracted bit.
    \item Continue while $\mathrm{OR\_reduce}(b)$ is nonzero.
\end{enumerate}

This procedure can be applied independently to the check-node and fault-node portions of $b_\text{init}$, allowing parallel initialization of syndrome and erasure clusters and achieving roughly a two-fold reduction in initialization time.

\subsection{Cluster growth}
\label{sec:uf-growth}

As described in \sec{union-find-algo}, the cluster decoder cluster growth subroutine expands each cluster by adding all vertices adjacent (in the decoding graph $\mathcal{G}$) to any vertex currently in the cluster, as illustrated in \fig{cluster_growth}(a).
In our FPGA-tailored design, we implement this operation using the extended adjacency matrix $H^\mathrm{ext}$ defined in \sec{data-structures}.

Given a cluster bitmap $b \in \mathbb{F}_2^{M+N}$, growth is performed as follows:
\begin{enumerate}
    \item For each column of $H^\mathrm{ext}$, perform a bitwise AND with $b$, yielding an intermediate matrix $B$ whose rows match those of $H^\mathrm{ext}$ where $b=1$ and are zero otherwise.
    \item Apply a column-wise OR to $B$ to obtain the grown bitmap $b'$.
\end{enumerate}

\begin{figure}[ht!]
    \centering
    (a)\includegraphics[width=0.45\linewidth]{figures/cluster_growth_graph.pdf}
    (b)\includegraphics[width=0.45\linewidth]{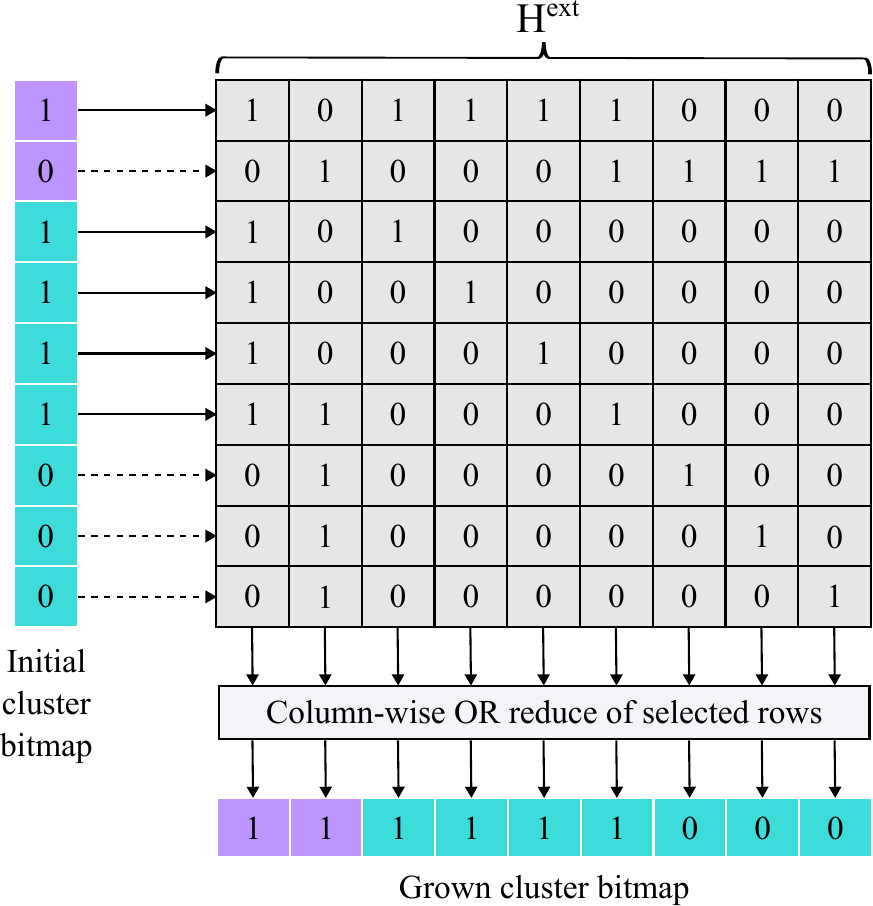}
    \caption{
    (a) Example of cluster growth: all vertices adjacent to any vertex in the current cluster (highlighted) are added to the cluster.
    (b) Gateware implementation: columns of $H^\mathrm{ext}$ corresponding to $1$ entries in $b_\text{init}$ are selected; a column-wise OR reduction over these columns yields $b'$.
    }
    \label{fig:cluster_growth}
\end{figure}

\subsection{Cluster merging}
\label{sec:uf-merge}

As described in \sec{union-find-algo}, clusters are merged whenever they share at least one vertex.  
In the bitmap representation introduced in \sec{data-structures}, this condition can be detected via a simple bitwise operation: if the bitwise \texttt{AND} of two cluster bitmaps contains any $1$, the clusters overlap and must be merged.
Given two cluster bitmaps $b$ and $b'$ of length $(M+N)$, the merge procedure is:
\begin{enumerate}
    \item \textbf{Merge predicate:} Compute $m = \mathrm{OR\_reduce}(b \ \& \ b')$.  
    If $m = 1$, the clusters share at least one vertex.
    \item \textbf{Merge operation:} If $m = 1$, set $b \leftarrow b \ | \ b'$ to obtain the merged bitmap, and assign the register that held cluster $b'$ as empty.
\end{enumerate}
Here \texttt{\&} denotes bitwise AND, \texttt{|} denotes bitwise OR, and $\mathrm{OR\_reduce}$ collapses all bits to a single value via OR.

In our cluster decoder implementation, one cluster is grown per update loop.  
Let $G^{(1)}$ be the grown cluster and $\{G^{(2)}, \dots, G^{(r)}\}$ the others.  
Before growth, each cluster represented a distinct connected component of $\mathcal{G}$ with disjoint bitmap support.  
After growth, $G^{(1)}$ may overlap with other clusters and must be merged with them.  
This is done by checking the merge predicate between $G^{(1)}$ and each of the $r-1$ remaining clusters in turn, merging into $G^{(1)}$ whenever required.

\begin{figure}[ht!]
    \centering
    (a)\includegraphics[width=0.3\textwidth]{figures/cluster_merge_graph.pdf}
    (b)\includegraphics[width=0.3\textwidth]{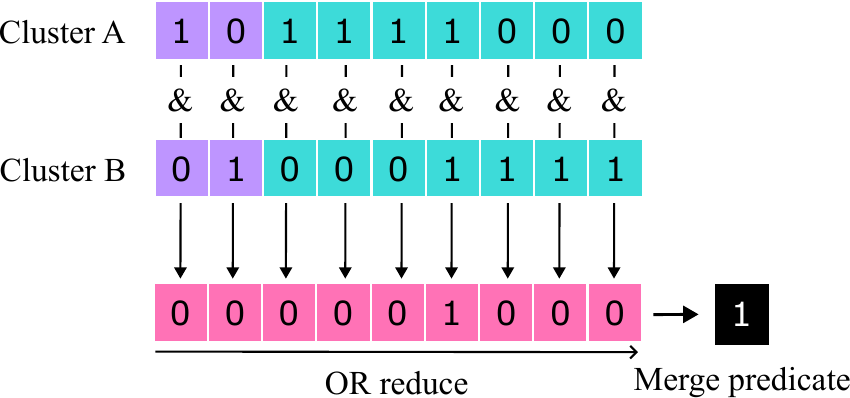}
    (c)\includegraphics[width=0.3\textwidth]{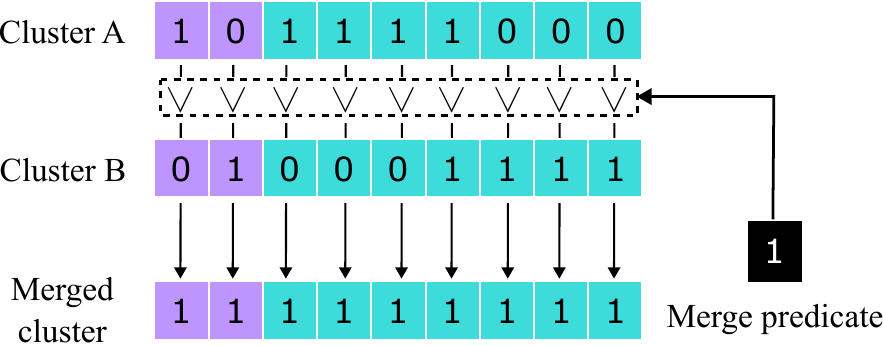}
    \hspace*{\fill}
    \caption{
        (a) Two clusters share a vertex and must be merged.  
        (b) Merge predicate: compute $m = \mathrm{OR\_reduce}(b \ \& \ b')$; if $m = 1$, the clusters overlap.  
        (c) Merge operation (if $m=1$): update $b \leftarrow b \ | \ b'$ to produce the merged bitmap.  
    }
    \label{fig:cluster_merge}
\end{figure}

\subsection{Determining interior fault nodes}
\label{sec:uf-interior}

\begin{figure}
    \centering
    (a)\includegraphics[width=0.45\linewidth]{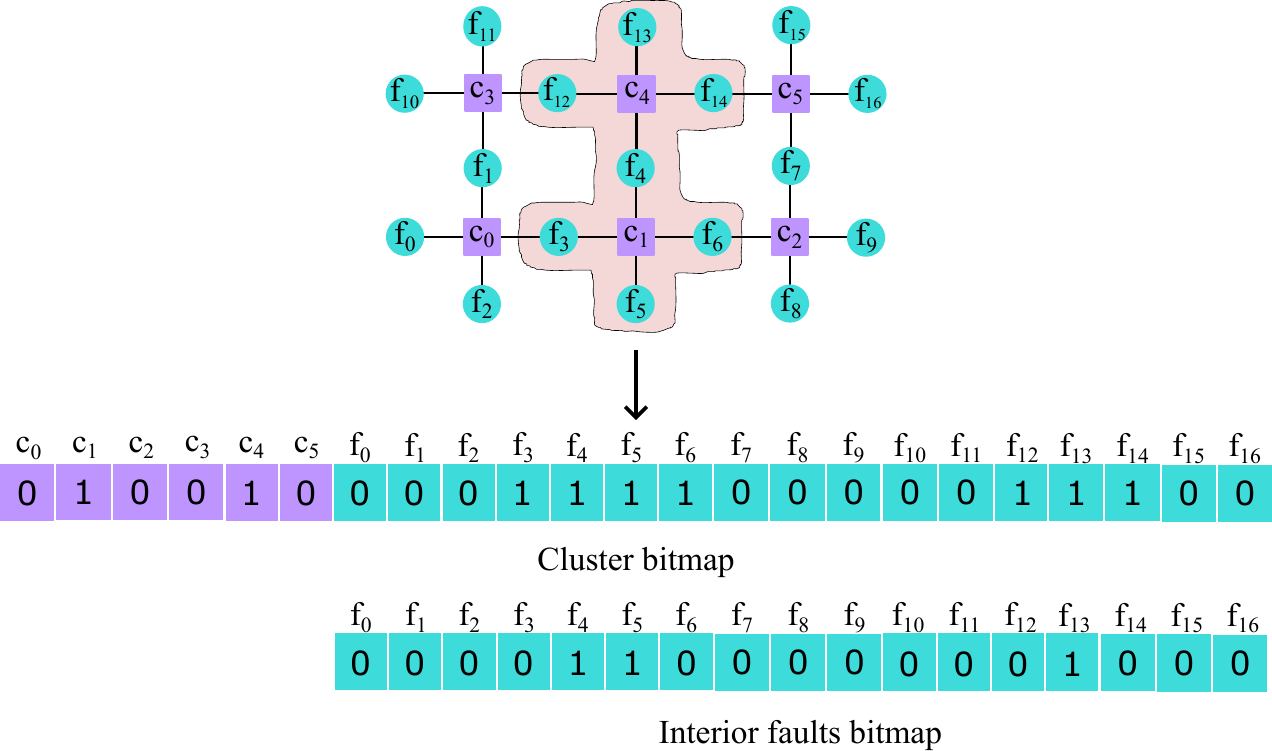}
    (b)\includegraphics[width=0.45\linewidth]{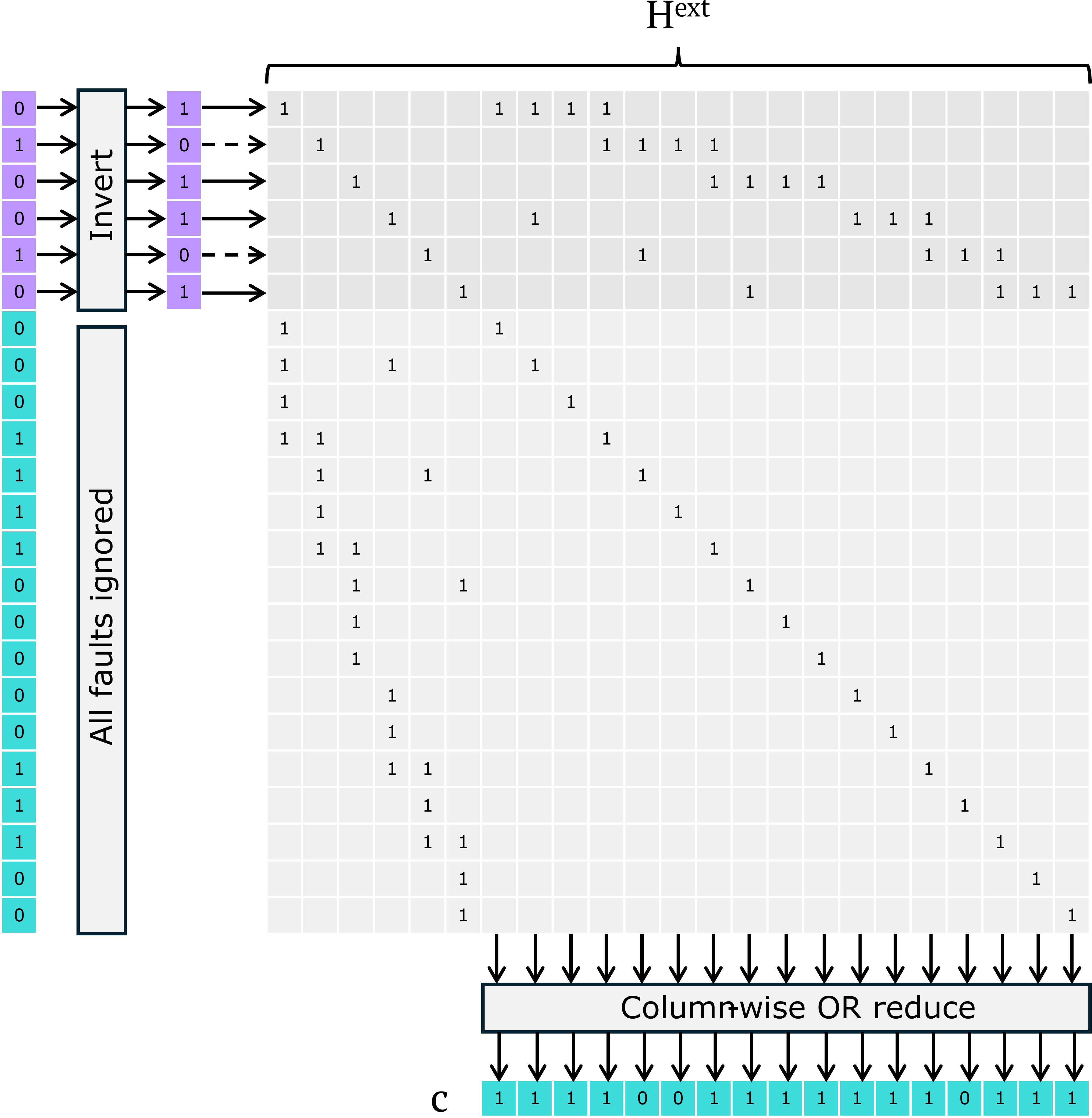}
    \caption{(a) An example cluster with an initial all-zero interior faults bitmap; 
    (b) Computing the intermediate non-interior faults bitmap $c$, which can then be used to update the interior faults bitmap by computing the bitwise AND of $\neg c$ and $t$ (zero entries in $H^\mathrm{ext}$ have been left blank). The resultant interior faults of the cluster shown in (a) will be $f_4,f_5,f_{13}$.}
    \label{fig:int_faults}
\end{figure}

As described in \sec{union-find-algo}, verifying the validity of a cluster requires identifying those fault nodes whose neighboring checks all lie within the cluster.
We capture this information in the \emph{interior faults bitmap}, which marks precisely these nodes.
In what follows, we describe how the interior faults bitmap can be computed from the cluster bitmap and the extended adjacency matrix $H^\mathrm{ext}$, as illustrated in \fig{int_faults}.

Given a cluster bitmap $b \in \mathbb{F}_2^{M+N}$, let $s \in \mathbb{F}_2^M$ denote its checks sub-bitmap and $t \in \mathbb{F}_2^N$ denote its faults sub-bitmap (the last $N$ entries of $b$).
The interior faults bitmap is determined as follows:
\begin{enumerate}
    \item Use the negated checks bitmap $\neg s$ to select rows of $H^\mathrm{ext}$ corresponding to checks not in the cluster.
    \item Apply a column-wise OR to these selected rows, yielding a non-interior faults bitmap $c \in \mathbb{F}_2^N$.
    \item Compute the interior faults bitmap as $(\neg c) \ \& \ t$.
\end{enumerate}

\clearpage
\subsection{Cluster validity (systolic solver)}
\label{sec:uf-validity}

Determining whether a cluster is \emph{valid} is the most computationally intensive step in cluster decoding for qLDPC codes.  
For the surface code, validity can be checked by counting enclosed syndrome vertices.  
For qLDPC codes, however, this requires solving a linear system whose variables correspond to the cluster’s \emph{interior fault nodes} and whose constraints correspond to its \emph{enclosed check nodes}.

Let $b \in \mathbb{F}_2^{M+N}$ be the bitmap of a cluster, as defined in \sec{data-structures}, with the first $M$ bits indexing check nodes and the remaining $N$ bits indexing fault nodes.  
Let $H^\mathrm{ext} \in \mathbb{F}_2^{(M+N) \times (M+N)}$ be the extended adjacency matrix of the Tanner graph $\mathcal{G}$ (\sec{uf-growth}).  
The \emph{interior fault nodes} are those fault nodes in the cluster whose entire neighborhood lies within the cluster.  
While these could be derived directly from $b$ using $H^\mathrm{ext}$, in our implementation a separate bitmap is maintained for interior fault nodes (and updated incrementally during growth and merge operations as we have seen).

\begin{figure}[ht!]
    \centering
    (a)\includegraphics[width=0.5\textwidth]{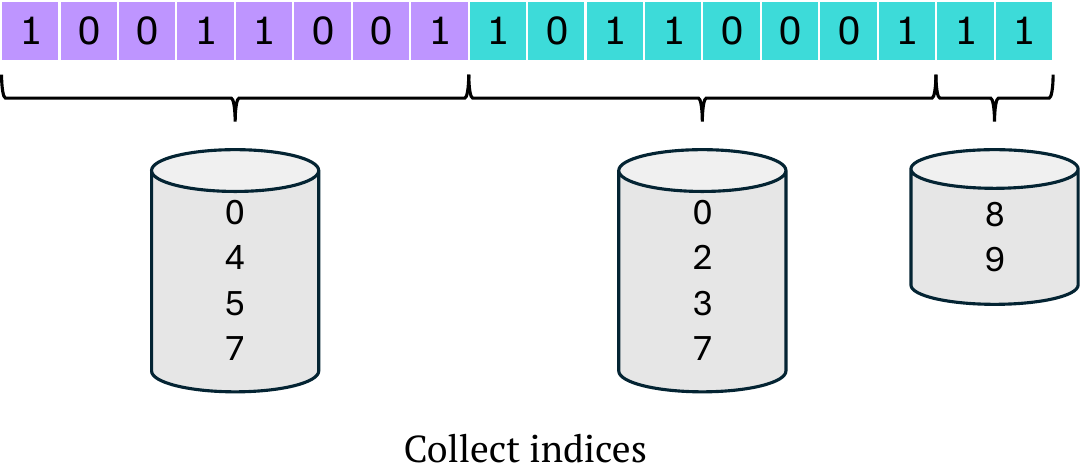}
    (b)\includegraphics[width=0.9\textwidth]{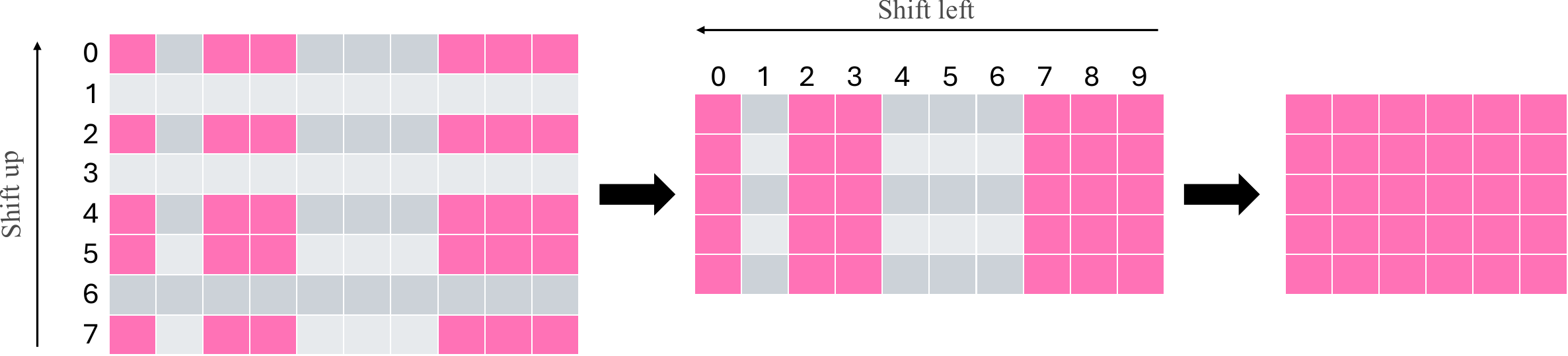}
    \caption{
    (a) Index extraction from a cluster bitmap: the bitmap is divided into buckets; $1$-bit positions are found sequentially within each bucket while buckets are processed in parallel.
    (b) Sub-matrix extraction from $H$: rows indexed by enclosed check nodes and columns indexed by interior fault nodes are selected using a shift-based method, reducing the worst-case cost to $M+N$ cycles. 
    During a shift up operation, the first row's index is compared with the first element in the buckets. If the row index matches the first index in the bucket, the row is kept and the index is popped from the buckets. 
    }
    \label{fig:cluster_validity}
\end{figure}

Validity checking then proceeds in three steps:
\begin{enumerate}
    \item \textbf{Index extraction:} Determine the row indices (enclosed check nodes) and column indices (interior fault nodes) from the first $M$ bits of $b$ and the column indices (interior fault nodes) from $b_\mathrm{int}$ (\fig{cluster_validity}b).
    \item \textbf{Submatrix extraction:} From the fixed decoding matrix $H \in \mathbb{F}_2^{M \times N}$, select the enclosed check node rows and interior fault node columns indexed above to form the submatrix $H_\text{sub} \in \mathbb{F}_2^{m \times n}$.
    \item \textbf{Solution existence test:} Use the systolic array algorithm from \sec{gateware_lse} to determine if $H_\text{sub} x = y$ has a solution. 
\end{enumerate}

\paragraph{Index extraction from the cluster bitmap.}
Determining the row and column indices for step~(2) reduces to finding the indices of all $1$-bits in a bitmap $b \in \{0,1\}^{M+N}$.
Separate lists for checks and faults are required, and as such one can first separate $b$ into $b_S\in\{0,1\}^M$ and $b_F\in\{0,1\}^N$ and handle them independently.
A naïve sequential scan over all bits would be possible, but would require about as many cycles as there are bits.
In what follows we briefly outline an alternative approach that reduces this slow-down (but we do not provide low-level alorithm details).
One can seperate the bits of $b$ into consecutive buckets each containing $\Delta$ bits, and process the buckets in parallel. 
Within the $i$th bucket, the $1$-bit indices $I_i \subset [i \Delta ,(i+1)\Delta-1]$ are found from among the $\Delta$ bits in the bucket and arranged into a sequential list.
The lists indices obtained for each bucket are concatenated together to form a list of global indices.
If buckets are emitted in increasing $i$, the concatenation $I_0\Vert I_1 \Vert\cdots$ is globally sorted. 
In a fully parallel implementation, the per-bucket scan would be expected to require around $\Delta$ cycles, and there would be some additional time overhead required to ensure the lists from buckets are emitted in order and merged, which we expect to be small for a moderate number of buckets.

\paragraph{Shift-based sub-matrix extraction.}
We have assumed $H$ is stored in memory.
Selecting arbitrary rows and columns via a large multiplexer would be fast but is impractical under FPGA timing constraints.  
Instead, we use a shift-based approach: rows are shifted upward until the top row matches a required row index, which is then copied to a temporary array. When a row index matches the first element, or head, of the list of indices in the buckets corresponding to the check nodes, the index is popped/removed from the buckets, making way for the next row to be found.
After collecting all rows, columns are selected by shifting left and discarding columns corresponding to non-matching indices in the faults buckets. 
While a naïve approach could require $mn$ cycles for an $m \times n$ submatrix, this shift-based method reduces the worst case to $M+N$ cycles and integrates straight-forwardly with systolic array linear system solving algorithm from \sec{gateware_lse}.

\paragraph{Solving the linear system.}
The final step in validity checking is to solve the system $H_\text{sub} x = y$, where $H_\text{sub}$ is the sub-matrix of $H$ indexed by the cluster’s enclosed check nodes and interior fault nodes, and $y$ is the corresponding measurement-outcome vector.  
A cluster is valid if and only if this system admits at least one solution.  
Since clusters can have arbitrary shapes during decoding, $H_\text{sub}$ may be highly under-determined ($n > m$) or over-determined ($m > n$), where $m$ and $n$ are the number of selected rows and columns, respectively.  
The generalized solver of \sec{gateware_lse} accommodates both cases.  
In our FPGA-tailored algorithm, for a given decoding problem (see \app{osd-resources}), we provision dedicated solvers for matrix sizes most frequently observed in the target operating regime; if a system exceeds these limits, the decoder declares failure.  
The choice of supported solver sizes thus entails a trade-off between resource usage and the resulting logical error rate.

\subsection{Algorithm output}
\label{sec:uf-workflow}

Before we describe the output of the algorithm, let us briefly discuss the overall workflow of the algorithm now that the constituent pieces have been described.
The workflow of the FPGA-tailored algorithm follows the high-level structure of the cluster decoder as described in \sec{union-find-algo}.  

The pre-decoder provides the input syndrome and erasure set, and the cluster initializer constructs the corresponding bitmaps in the cluster registers (\sec{uf-input}, \sec{uf-initialization}).  
The controller maintains a list of pointers to invalid clusters together with a set of pointers to all active clusters.  
At each iteration, the controller removes the first entry from the invalid-cluster list and passes the corresponding cluster through the functional units, updating its bitmap in place.
 
The cluster is first expanded by the growth unit (\sec{uf-growth}), then passed to the merge unit (\sec{uf-merge}), after which its interior faults bitmap is recomputed (\sec{uf-interior}).  
Finally, the controller assigns the cluster to an appropriately-sized systolic-array solver for validity checking (\sec{uf-validity}).  
If the cluster fails the validity test, it is appended to the end of the invalid-cluster list, producing a first-in-last-out behavior in which smaller clusters are typically processed before larger ones (observed to be beneficial in the Union Find algorithm, a cluster decoder for surface codes~\cite{delfosse2017almost}).

This update loop continues until no invalid clusters remain.  
At this point, every active cluster has a well-defined bitmap specifying its enclosed check and fault nodes.  
Each cluster’s interior faults are then used to compute a valid per-cluster correction bitmap $\tilde{F}^{(l)}$, and the final correction $\tilde{F}$ is obtained by aggregating these $L$ bitmaps via bitwise OR:
\[
\hat{F} \;=\; \tilde{F}^{(1)} | \tilde{F}^{(2)} | \dots | \tilde{F}^{(L)}.
\]
Since clusters have disjoint support by construction, this aggregation requires no further conflict resolution.

As we have seen, the controller orchestrates the entire workflow: it sequences clusters through the growth, merge, interior, and solver units, arbitrates access to the shared pool of systolic solvers, and monitors termination.

\clearpage 
\section{Results and outlook}
\label{sec:results}

This section presents FPGA resources in \sec{fpga-resources}, considerations for real-time decoding in \sec{real-time-decoding} followed by concluding remarks in \sec{conclusions}.
Our main performance and FPGA cycle count data are provided in \fig{intro} in \sec{intro}.

\subsection{FPGA resources and cutoff-time performance curves}
\label{sec:fpga-resources}

We benchmark all decoders with the QEC circuit decoding examples specified at the end of \sec{qec-background-general}, which have the following key parameters:
\begin{itemize}
    \item Gross code [[144, 12, 12]] and two-gross code [[288, 12, 18]].
    \item Circuit noise, $p=0.001$. 
    We decode the $Z$ component of $d$ noisy QEC rounds and a perfect round to project into the code space and test for logical failure.
    \item Decoding matrix dimensions: $936 \times 8784$ for gross and $2736 \times 26208$ for two-gross.
\end{itemize}

We estimate (without synthesis into an explicit FPGA implementation) the utilization of various FPGA resources, including flip-flops (FFs), two types of on-chip RAM (BRAM and URAM), and look-up tables (LUTs). 
Resource usage is reported as a percentage of the available capacity on a VU19P device (which contains 8,171,520 FFs, 2,160 BRAMs, 320 URAMs, and 4,085,760 LUTs), shown in \tab{fpga_resources}. 
For performance evaluation, we conduct Monte Carlo simulations to estimate the clock-cycle count and logical failure probability for each decoding problem; the corresponding results are summarized in \fig{intro} in \sec{intro}.

\begin{table}[h]
\centering
\begin{tabular}{@{}lccccc@{}}
\toprule
\textbf{Decoder} & \textbf{Code} & \textbf{Flip-flops} & \textbf{LUTs} & \textbf{BRAMs} & \textbf{URAMs} \\ 
\midrule
OSD~\cite{bascones2025exploring}        & gross     & 25\%  & 39\%  & 25\%  & 0\% \\
Filtered-OSD                           & gross     & 9\%  & 19\% & 14\% & 0\% \\
Clustering                             & gross     & 9\%  & 6\%  & 25\% & 0.0\% \\
Relay~\cite{muller2025improved}        & gross     & 7\%  & 52\% & 1\%  & 0\% \\
Filtered-OSD                           & two-gross & 20\% & 19\% & 10\% & 83\% \\
Clustering                             & two-gross & 69\% & 88\% & 53\% & 83\% \\ 
\bottomrule
\end{tabular}
\caption{FPGA resource utilization estimates for different decoders and code types.
}
\label{tab:fpga_resources}
\end{table}

We briefly review some basic features of the three kinds of FPGA memory resources that we count to provide context for some of the choices made in our implementations.
BRAM blocks are small and spread across the chip, while URAM blocks are larger but fewer. 
In practice: FFs are used for tiny per-node registers; BRAMs are used for many small, nearby buffers (e.g., per-edge BP messages); while URAMs are used for large tables and deep queues.
At rough scales, a typical BRAM block stores $\sim\!36\,\mathrm{Kb}$ whereas a typical URAM block stores $\sim\!288\,\mathrm{Kb}$.
BRAMs can be set to a chosen \emph{width} (bits per access) and \emph{depth} (number of entries) and can usually be both read and modified during a single FPGA cycle. 
URAMs have fixed \emph{width} (bits per access) and \emph{depth} (number of entries) and are thus less configurable.
Larger memories can be built by combining blocks: for example by placing them side-by-side to read/write more bits per cycle (widening) or chaining them end-to-end to store more entries (deepening). 
One can also use several blocks in parallel to store identical copies of data, referred to as \emph{banking} so different parts of the algorithm can access the same memory at the same time. 

Our predecoder, for both filtered-OSD and cluster decoder, is 80 iterations of MemBP (matching the first leg of Relay).
We observed that the output of MemBP is not a great source of LLRs for the next stage of decoding.
For this reason, when MemBP fails to converge, we generate the LLRs passed to filtered-OSD and Cluster decoding by applying 25 rounds of standard BP on the corresponding failing syndrome.
We include the cost of these 25 rounds in our analysis in \fig{intro}.
Details of how these FPGA resource estimates were obtained for the filtered-OSD and cluster decoders are provided in \app{osd-resources} and \app{cluster-resources}, and for the standard OSD and Relay decoders in \app{relay-resources}.
In \fig{intro}, we follow the independent-XZ decoding setting of Ref.~\cite{beverland2025fail}, reporting $X$-component logical failures over $d$ QEC rounds; by contrast, Ref.~\cite{muller2025improved} also considers correlated-XYZ decoding and reports logical error rates per QEC cycle.

\subsection{Real-time decoding despite latency tails}
\label{sec:real-time-decoding}

We now consider real-time decoding for Relay on our two example decoding problems.  
Assuming an FPGA cycle time of 12\,ns~\cite{maurer2025realtimedecoding}, the $d=12$ QEC cycles of the gross-code decoding problem correspond to $T_{\text{gen}} = 1000$ FPGA cycles when each QEC cycle lasts one microsecond.  
The FPGA cycle time for the two-gross code was not analyzed in Ref.~\cite{maurer2025realtimedecoding} and is likely longer due to its larger decoding matrix, while its decoding window spans $d=18$ QEC cycles rather than 12.  
For simplicity, we assume these two effects roughly offset one another and again take $T_{\text{gen}} = 1000$ FPGA cycles for the two-gross code.

In \fig{intro}, we presented the logical error rates achievable with different maximum budgets of FPGA cycles.  
Although Relay is substantially faster than the other decoders we evaluated, it exhibits tail latencies exceeding 1000 cycles, which do not satisfy the real-time constraint.
If we insisted on a strict decoding budget of 1000 cycles, the backlog problem would be directly avoided, but the logical error rates achievable would be very limited.

\paragraph{Decoding-latency tail condition.}
In \sec{real-time-background} we analyzed real-time decoding using a variable-latency model in which a decoder completes within a typical time $T_{\text{ref}} < T_{\text{gen}}$ with probability $1-\epsilon$, but with probability $\epsilon$ requires a longer time up to $T_{\text{max}} > T_{\text{gen}}$.
As we showed in \sec{real-time-background}, to ensure that the backlog problem is avoided in this setting for a computation with $C$ code blocks, it is sufficient to show that the latency-tail conditions \eq{latency-tail-requirement} hold.

We are free to choose $T_{\text{max}}$, the maximum decoding budget, which determines the logical error rate as shown in \fig{intro}. 
If the decoder has not identified a correction after $T_{\text{max}}$ FPGA cycles, we declare failure.
For the gross code we take $T_{\text{max}} = 6000$, yielding a logical error rate of $4\times 10^{-7}$.  
For the two-gross code we take $T_{\text{max}} = 36160$, the maximum runtime Relay can reach under the settings assumed here (see \app{relay-resources}).  
Although our Monte Carlo data did not extend far enough to estimate the corresponding logical error rate, the same decoding problem and Relay settings were analyzed in Ref.~\cite{beverland2025fail}, which predicts logical error rates below $10^{-9}$.

\begin{figure}[h]
    \centering
    (a)\includegraphics[width=0.46\linewidth]{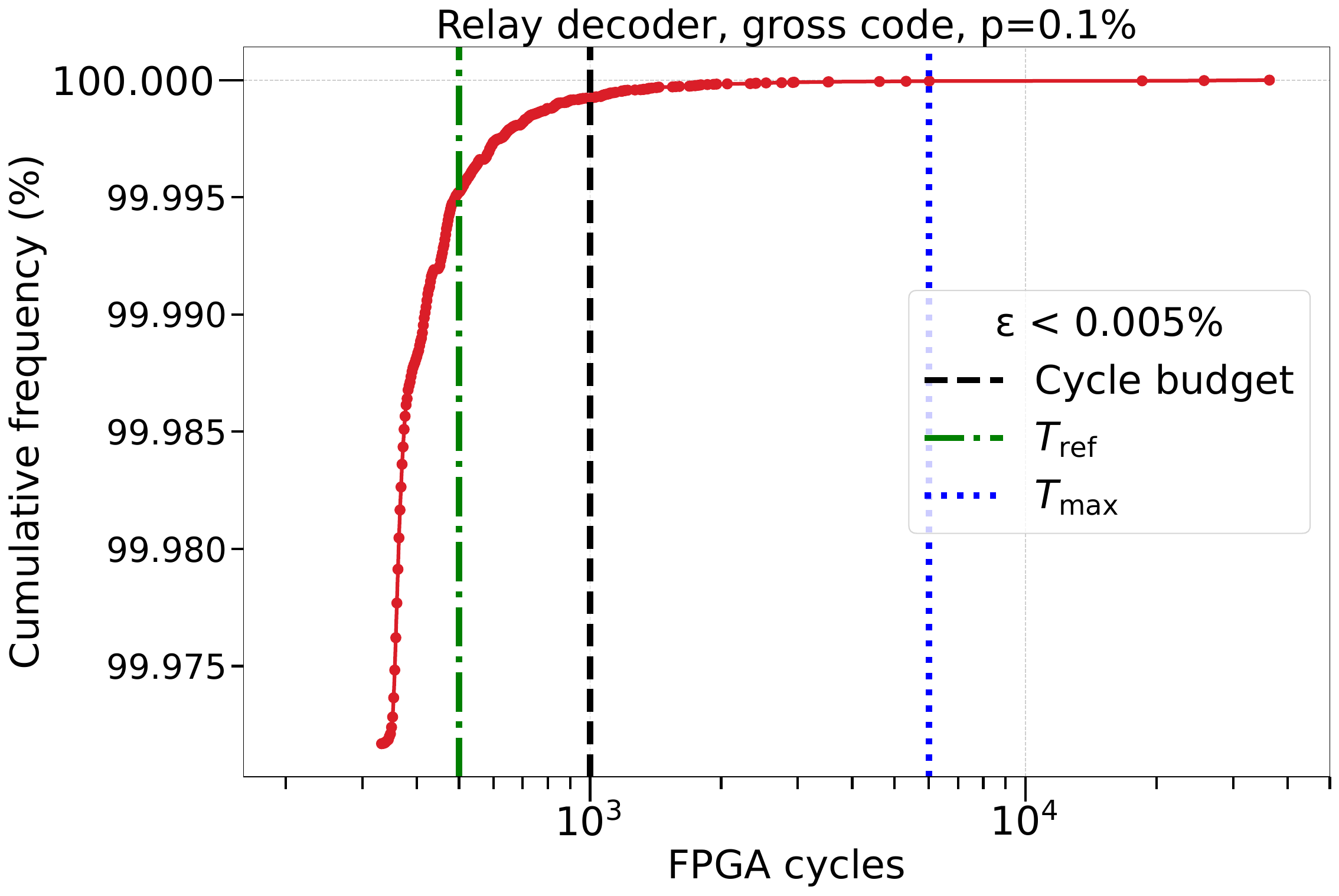}
    (b)\includegraphics[width=0.46\linewidth]{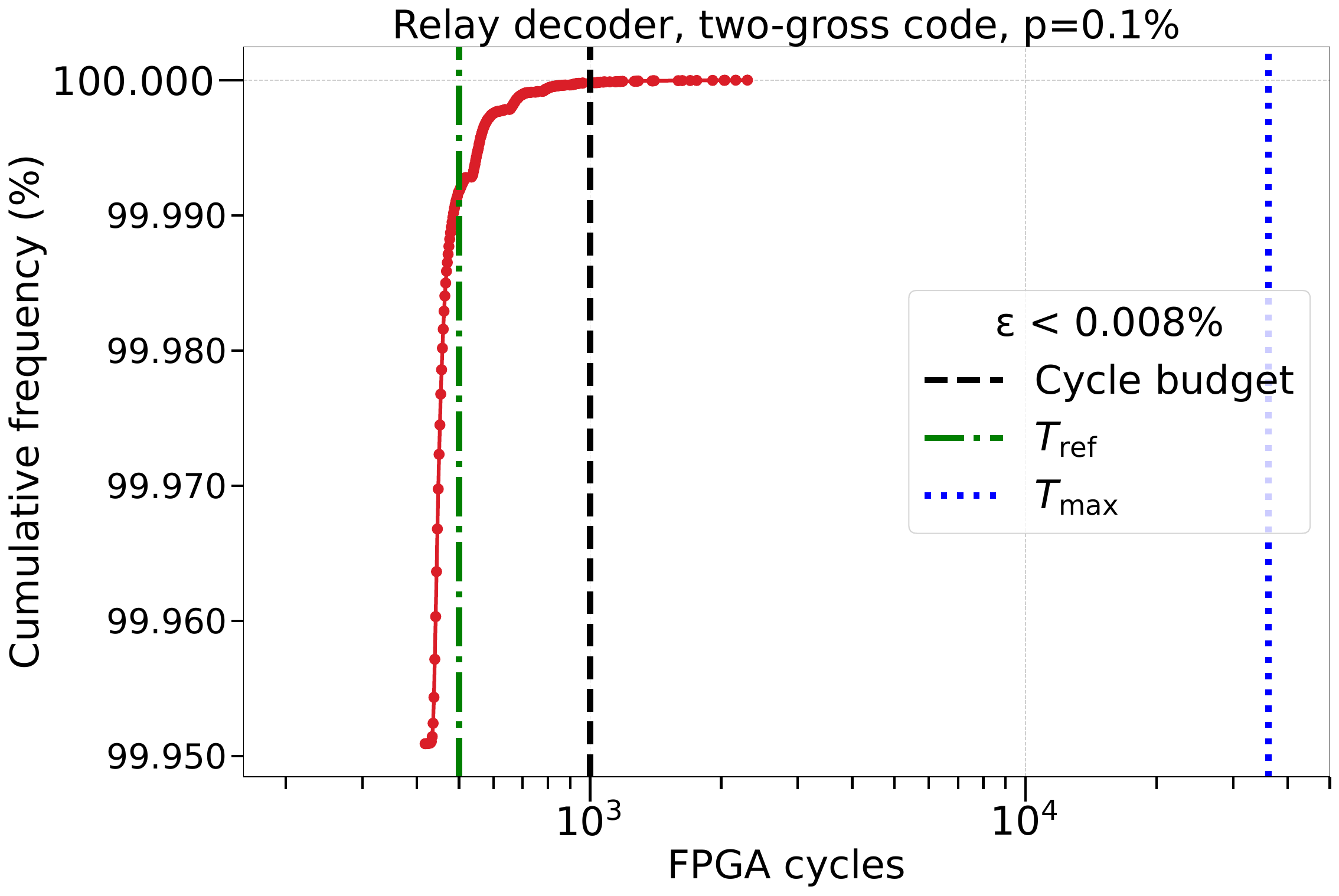}
    \caption{
    \textbf{Latency tail analysis for Relay.}
    We show the cumulative percentage of instances that Relay has finished decoding within a given budget of FPGA cycles for the gross and two-gross codes.
    We select a cutoff $T_{\text{max}}$ sufficient to achieve our desired logical error rates, and select a $T_{\text{ref}}$ value which allows us to show that the latency-tail condition is satisfied, ensuring real-time decoding with modest slowdown.
    }
    \label{fig:relay_dist}
\end{figure}

We are also free to choose any $T_{\text{ref}}<T_{\text{dec}}$ which yields a $\epsilon$ satisfying the latency-tail condition (although different choices may give more or less tight slowdown bounds) and can in principle choose whichever value gives the tightest slowdown bound and satisfies the latency-tail condition.
For simplicity, we take $T_{\text{ref}} = 500$ for both gross and two-gross, which from \fig{relay_dist} we see that $\epsilon < 5 \times 10^{-5}$ and $\epsilon < 8 \times 10^{-5}$ respectively. 
The latency-tail condition is satisfied provided that $C < 1333$ for gross and $C < 138$ for two-gross. 
For context, $C$ is less than 1000 for all algorithm examples considered in Ref.~\cite{yoder2025tour}, suggesting that a lower $T_{\text{ref}}$ may be beneficial for two-gross at the cost of a slight degradation of logical error rate.

Using \eq{latency-tail-slowdown}, we estimate the slowdown for a system with a modest number of code blocks ($C=10$, corresponding to roughly 120 logical qubits).
For Gross, the bound gives $\langle R \rangle < 1.62$, and for two-Gross, $\langle R \rangle < 1.74$.
For comparison, if $\epsilon = 0$ (so every decoder run finishes within $T_{\text{ref}} = 600$ FPGA cycles), the slowdown would be $R = 1.6$.

\subsection{Future directions}
\label{sec:conclusions}

We have shown that message passing decoders such as Relay~\cite{muller2025improved} remain the most practical option for real-time FPGA decoding of quantum LDPC codes at microsecond cycle times relevant for superconducting qubits. 
Several promising research directions remain.

\begin{enumerate}
    \item Decoders that we classified as too slow for real-time use in \fig{decoder-classes}, such as SAT-based search methods and neural network decoders, often achieve high accuracy and in some cases provably produce min-weight corrections. 
    Although unsuitable for microsecond-scale decoding, they may be viable for slower platforms such as trapped-ion or neutral-atom qubits, and remain valuable for validating fault-tolerant circuits or benchmarking other decoders. 
    Faster implementations of these approaches on specialized hardware such as FPGAs could broaden their applicability.

    \item Message passing decoders offer significant scope for refinement. 
    Improving update rules, graph representations, or scheduling, as well as exploring FPGA and ASIC specific optimizations in space and time, may yield substantial gains. 
    Both decoding examples used for benchmarks in this work focused on logical memory, and extending these ideas to the dynamically changing decoding graphs that arise during logical gates is an important next step that will result in much larger decoding matrices.

    \item There remains substantial opportunity to develop new FPGA-tailored decoders, whether through significant advances within the six existing classes we have identified in this work, or through the introduction of entirely new algorithmic paradigms not represented among current approaches.

    \item Lastly, we point out a caveat in our analysis here.
    Our decoding examples (specified in \sec{qec-background-general} and analyzed in \sec{real-time-decoding}) assume each decoding problem consists of a batch of $d$ consecutive QEC cycles of a code block and that a new batch is generated every $d$ cycles.
    This is an optimistic simplification: in practice: decoding typically requires overlapping windows of $T_{\text{commit}} + T_{\text{overlap}}$ cycles, committing only the first $T_{\text{commit}}$ before advancing by that amount to the next window.  
    Theoretical analyses that guarantee circuit-distance-$d$ fault tolerance often set $T_{\text{commit}} = T_{\text{overlap}} = d$, producing decoding problems of size $2d$ rather than $d$, while still generating a new problem every $d$ cycles.
    Furthermore, decoding logical operations that entangle code blocks requires decoding those blocks collectively.
    A more complete analysis would incorporate window decoding more fully.

\end{enumerate}

\paragraph{Acknowledgments.}
We thank Ted Yoder, Abdullah Khalid, Tristan M\"uller and Lev Bishop for comments on the manuscript and Scott Lekuch for encouragement in the early stages of this work.

\clearpage
\appendix
\section{Appendices}

\subsection{Analysis details for filtered ordered statistics decoding}
\label{app:osd-resources}

To estimate the FPGA footprint of the ordered statistics decoder described in \sec{fpga-osd}, we consider the following contributors to the overall resource utilization:

\begin{itemize}
    \item Storage of $H$.
    \item Truncated fault ranking unit.
    \item Permuted sub-matrix extraction unit.
    \item Systolic solver.
    \item Inverse-permutation.
\end{itemize}

\paragraph{Storage of $H$}
As outlined in \sec{fpga-osd}, we store $H$ in on-chip RAM in column-major form (one address per column), so a read returns the $M$ bits of that column. 
Using BRAM configured for $36$ bits per read and depth $1024$, one column of length $M$ bits is obtained by placing $\lceil M/36\rceil$ BRAM blocks side-by-side (widening), and all $N$ columns are covered by chaining $\lceil N/1024\rceil$ such groups end-to-end (deepening). 
The total BRAM count is $\lceil M/36\rceil\,\lceil N/1024\rceil$: about $234$ BRAMs for the gross code (fits on VU19P) and about $1976$ BRAMs for the two-gross code (exceeds VU19P). 
For the two-gross case we therefore instead store $H$ in URAM, configured for $72$ bits per read and depth $4096$, which requires $\lceil M/72\rceil\,\lceil N/4096\rceil\approx 266$ URAMs, which is within the VU19P capacity.

\paragraph{Truncated fault ranking unit}
We store each of the $N$ LLRs $\Lambda_j$ with 32-bit precision in an array of $36\times 1024$ BRAMs, requiring $\lceil 32N/(36\times 1024)\rceil$ BRAMs.
The truncated-ranking unit reads the BRAMs in dual-port mode (two values per cycle), filters entries with $\Lambda_j<\Lambda_{\text{confident}}$, and writes the surviving $\Lambda_j$ and their indices $j$ to two auxiliary BRAMs (one for values, one for indices). 
Doing this serially can be very slow (e.g. for the two-gross code, a dual-port configuration would require $\simeq13000$ cycles).
This large cycle count can be reduced by banking the BRAMs that store the input LLRs such that the list is divided into equally sized sub-lists and these sub-lists are stored in independent BRAMs. 
This allows multiple values to be read in the same cycle, boosting the throughput at the cost of higher BRAM requirements. 
We assume an 8-way banking, which increases the number of BRAMs by a factor of 8, but also reduces the number of cycles required for truncating the list of LLRs by a factor of 8. 
One $36\times 1024$ BRAM per list suffices for our decoding examples, so the total BRAM usage is $\lceil 8 \times \frac{32\times N}{36\times 1024}\rceil + 2$. 

We assume 32-bit LLR $\Lambda_j$ and 16-bit index $j$ entries are sorted into a register built out of FFs.
Let $|R|$ be the number of faults for which $\Lambda_j < \Lambda_\text{confident}$. 
The total number of cycles required to sort is $|R|$.
If on any given run, $|R| > |R|_\text{max}$, we say the decoder fails.
Storing these 32-bit $\Lambda_j$ and 16-bit index $j$ entries for the sorting network would then require $|R|_\text{max} \times (32 + 16)$ flip-flops.
Additionally, each entry of the sorting network also requires a 16-bit comparator for accepting/rejecting the broadcasted input.
We conservatively estimate that every 16-bit comparator requires 16 LUTs -- yielding a total cost of $|R|_\text{max} \times 16$ LUTs for the sorter.
We assume $|R| \leq |R|_\text{max}$, where $|R|_\text{max} = 500$ for both gross and two-gross.

\paragraph{Permuted sub-matrix extraction unit}
After ranking, we stream the sorted indices $R$ and read the corresponding columns of $H$ to form the permuted sub-matrix. 
Sorted indices $R$ stream from the sorter output to address columns of $H$ to construct the permuted sub-matrix. 
With dual-port BRAM/URAM (two columns per cycle), extraction completes in $\lceil |R|/2\rceil$ cycles. 
The permuted sub-matrix is stored in flip-flops as an $M\times|R|$ bit array, requiring $M\times|R|$ flip-flops.

The sub-matrix extracted can be very tall and thin -- more rows than columns. 
Many rows of this sub-matrix will be all zeros, making them redundant for determining the correction. 
At the cost of more RAMs and a few extra cycles, we can filter out these all zero rows from the sub-matrix to make the LSE solver take fewer cycles in the next step. 
For the gross code, we process the sub-matrix with 9 blocks checking for all zero rows in parallel, requiring 9 extra cascaded BRAMs. A cascaded BRAM is essentially multiple BRAMs chained together to get the required width of the matrix (number of columns).  
For the two-gross code, we process the sub-matrix with 27 blocks checking for all zero rows in parallel, requiring 27 extra cascaded BRAMs.
Each block of rows processed will also require a comparator for checking for all zero rows. 
The comparator is assumed to require $|R|_\text{max}$ LUTs, totaling $9|R|_\text{max}$ and $27|R|_\text{max}$ extra LUTs for the gross and two-gross codes. 
The BRAMs that store the rows of the sub-matrix after this filtering step can then be read by the systolic array in the next step. Since multiple BRAMs are used, some bookkeeping will be necessary for ensuring that the rows are provided to the systolic array in the right order. 

\paragraph{Systolic solver}
Since we assume $|R| \le |R|_{\text{max}}$, we use a $|R|_{\text{max}} \times (|R|_{\text{max}}+1)$ array of processing elements (PEs) for the systolic solver. 
Each PE stores one bit, so the flip-flop count is $|R|_{\text{max}}(|R|_{\text{max}}+1)$. 
The per-PE combinational logic (multiplexers and Boolean operations) is modeled as $\approx 3$ LUTs, giving a LUT estimate of $3\,|R|_{\text{max}}(|R|_{\text{max}}+1)$. 
We make an optimistic assumption that the systolic solver presented in \sec{gateware_lse} can be modified to deal with input matrices smaller than the maximum supported size, and it thus completes in $M + 3|R|-1$ cycles.

\paragraph{Inverse-permutation.}
To store the final output correction, we assume a single BRAM that contains $N$ entries.
The indices generated during the ranking step can be reused for accessing the BRAM sequentially to assign every bit of the $|R|$-bit correction to the final correction in the BRAM. 
Using the BRAM in dual-port mode produces the correction in $|R|/2$ cycles.

The filtered-OSD utilization in \tab{fpga_resources} is the sum of these per-component estimates.

\begin{figure}[h]
    \centering
    \includegraphics[width=0.5\linewidth]{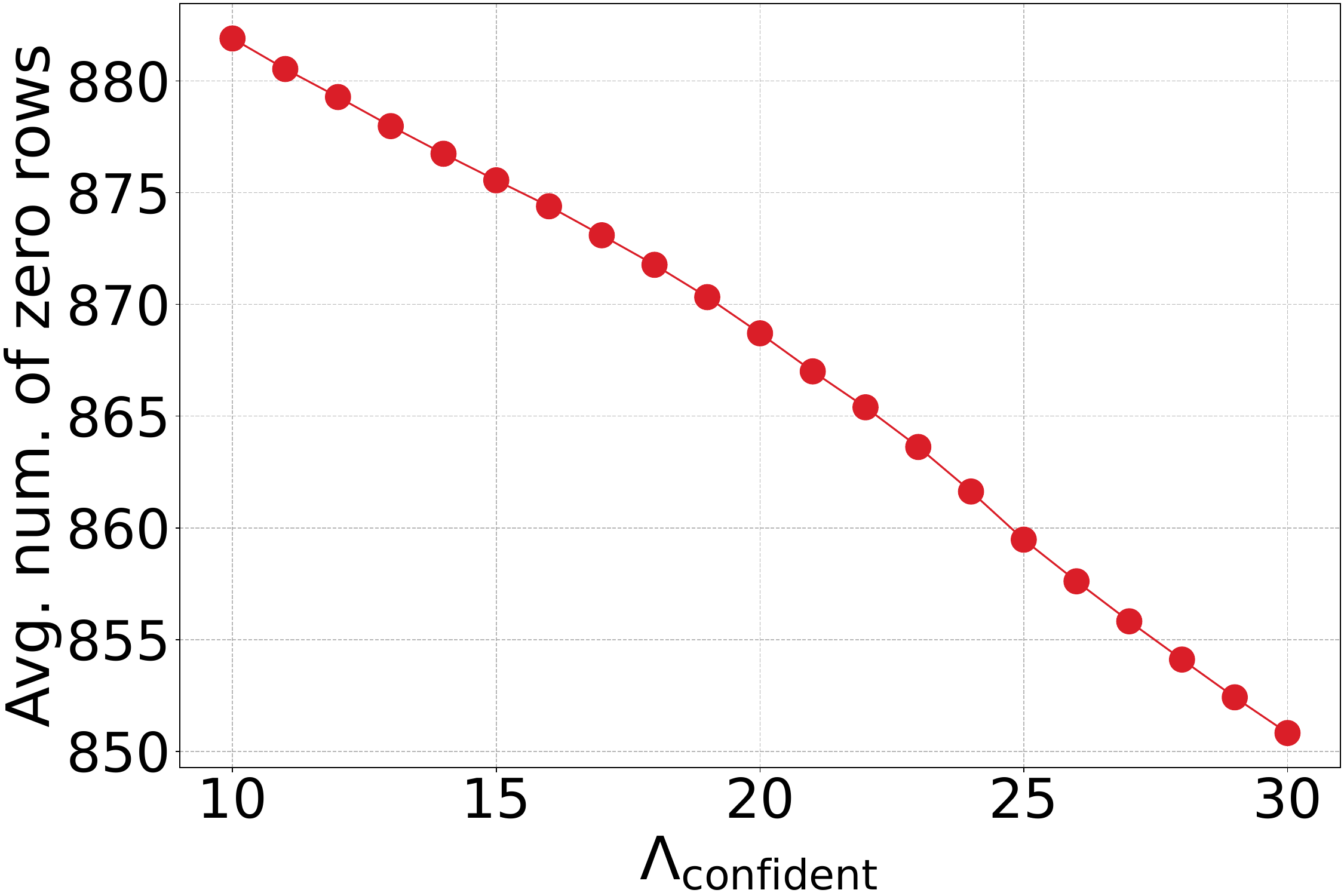}
    \caption{
    Number of all zero rows in $H_\text{R}$ for different values of $\Lambda_\text{confident}$ (gross code, $p=0.1\%$). The total number of rows in $H_\text{R}$ is 936.
    }
    \label{fig:osd_zero_rows}
\end{figure}

\paragraph{Exploiting the sparsity of the decoding matrix for filtered-OSD.}
A key optimization for the filtered-OSD decoder presented in \sec{filtered-osd-decoding} is filtering all zero rows to produce $H_\text{R}'$. 
This optimization saves a few hundred cycles during the operation of the systolic solver. \fig{osd_zero_rows} shows the number of all zero rows out of the total 936 rows of $H_\text{R}$ for different values of $\Lambda_\text{confident}$. Less than a 100 rows are useful for the final decoding result, highlighting the necessity of filtering out all zero rows from $H_\text{R}$.

\subsection{Analysis details for cluster decoding}
\label{app:cluster-resources}

To estimate the FPGA footprint of the cluster decoder, we consider the following contributors to the overall resource utilization: 
\begin{itemize}
    \item Cluster storage.
    \item Storage of decoding matrix and extended decoding matrix.
    \item Cluster growth unit.
    \item Cluster merge unit.
    \item Interior faults bitmap.
    \item Systolic solvers.
\end{itemize}

\paragraph{Cluster storage}
Every cluster bitmap needs $M+N$ bits of storage. 
A maximum of $N_\text{clus}$ clusters can be processed by the cluster decoder, such that the total number of flip-flops required to store the cluster bitmaps is $N_\text{clus} \times (M + N)$. 
We assume $N_\text{clus} = 50$ for gross and $N_\text{clus} = 100$ for two-gross. 

\paragraph{Storage for decoding matrix and extended decoding matrix}
The cluster validity stage requires the original decoding matrix $H$, while the growth and interior faults computation stages require the extended decoding matrix $H^\mathrm{ext}$. 
We assume the full decoding matrix $H$ is stored as a read-only memory in RAM as for the ordered statistics decoder: 
we assume $\lceil M/36\rceil\,\lceil N/1024\rceil \approx 234$ BRAMs are used for the gross code, and that $\lceil M/72\rceil\,\lceil N/4096\rceil\approx 266$ URAMs are used for the two-gross code.

$H^\mathrm{ext}$ is used for growing clusters and finding interior fault nodes. Since it is constant and experiences no modifications, it can be synthesized as a block of constants on an FPGA, requiring no storage overhead~\cite{amd_synthesis}.
This assumes that the decoding matrix is fixed and does not change during the decoding process. 
Any changes to the decoding graph will require at least a partial re-synthesis of the logic on the FPGA.

\paragraph{Cluster growth unit}
The growth unit requires an OR-reduction over $k=M+N$ bits. 
The VU19P device has 6-input LUTs which can OR up to 6 bits in one cycle, so we reduce in stages: first OR groups of up to 6 to get $\lceil k/6\rceil$ partial results, then repeat on these partials until one bit remains. 
Reusing the same LUTs each stage, this uses about $\lceil k/6 \rceil$ LUTs and takes $ \lceil \log_{6} k \rceil$ cycles. 
A fully parallel single cycle version would be possible but consumes more LUTs.

\paragraph{Cluster merge unit}
To merge clusters, bitwise \textbf{AND} and \textbf{OR} operations are used to generate the merge predicate and the resulting merged cluster, respectively. 
We make a conservative estimate that each LUT can perform one bitwise AND or OR operation, so computing the bitwise AND/OR between two cluster bitmaps of size $M + N$ requires $M + N$ LUTs. 
We allocate separate sets of $M + N$ LUTs for the AND and OR operations to enable parallel computation. 

The merge unit also requires a OR-reduction. 
The OR-reduction used for the growth stage can be reused since cluster growth and merge operations will not overlap. 

\paragraph{Interior faults bitmap}
Determining the interior faults bitmap requires an OR-reduction as well as bitwise operations between the intermediate bitmap $c$ and the faults bitmap. 
The network for computing the bitwise AND operation in the cluster merge unit can be reused, along with the reduce network used by the growth stage.

\paragraph{Systolic solvers}
For a systolic solver configured to use a matrix of size $M' \times N'$, the total number of nodes in the systolic array can be computed as: $N'\times (N'+1)$. 
Since each node requires a single bit of storage, this number also represents the number of flip-flops required for the systolic solver. 
Within each node, multiplexers and boolean logic are assumed to require three look-up tables (LUTs). 
For the cluster decoder, we assume the availability of systolic solvers of different sizes: (18, 18), (18, 36), (36, 72), (72, 144), and (144, 288) for the gross code and (18, 18), (18, 36), (36, 72), (72, 144), (144, 288), (288, 576), (576, 1152) for the two-gross code. The matrix sizes supported can be changed/tuned for an increase/decrease in the overall decoder accuracy, but the larger size requirements of the two-gross code increase its FPGA footprint.
We also provision for $M' \times N'$ flip-flops to hold the extracted sub-matrix before it is processed by the systolic solver.
For the solvers used by the cluster decoders, we again make the optimistic assumption made in \app{osd-resources} that the solvers can handle input matrices smaller than the maximum size supported by the solver. 

We also allocate temporary storage for each solver for (i) the indices of 1s in the cluster bitmap and (ii) a scratch matrix used when extracting the cluster sub-matrix. 
First, for generating and storing the indices from the cluster bitmap, we assume buckets that each span non-overlapping 32-bits of the cluster bitmap, thus requiring 32-cycles to generate all indices (since every bucket can be filled in parallel). 
We assume the storage required for every bucket is in FPGA BRAM, with every bucket requiring 32 16-bit entries for the index.
Since not every bit of the cluster bitmap will be one, we do not need to provision for all $N$ possible indices. 
We provision for $N/4$ indices, thus requiring $(N/4)\times16$ bits of total BRAM storage and $\lceil \frac{(N/4)\times16~bits}{36 \times 1024~bits}\rceil$ BRAMs.

The rows of the decoding matrix $H$ will be extracted first (\fig{cluster_validity}), the intermediate matrix requires $M' \times N$ bits of storage. 
As the rows of the matrix are extracted sequentially, this intermediate matrix can be stored in Block-RAMs (BRAMs) too. To allow the matrix columns to be accessible in the second stage of the sub-matrix extraction process, each row of the intermediate matrix can have its own BRAM. For a 36 KBit BRAM configured with a port width of 1 bit, the BRAM has a depth of 36K elements, which is enough to store all the column bits for a row of the intermediate matrix. The total number of BRAMs required for this intermediate matrix storage is the sum of the number of rows supported by all provisioned systolic solver sizes: $(18 + 18+36+72+144) = 288$ BRAMs. 
To access a column of this intermediate matrix, the same access address for every row BRAM can be used. 

The cluster decoder utilization in \tab{fpga_resources} is the sum of these per-component estimates.

\begin{figure}[h]
    \centering
    \includegraphics[width=0.7\linewidth]{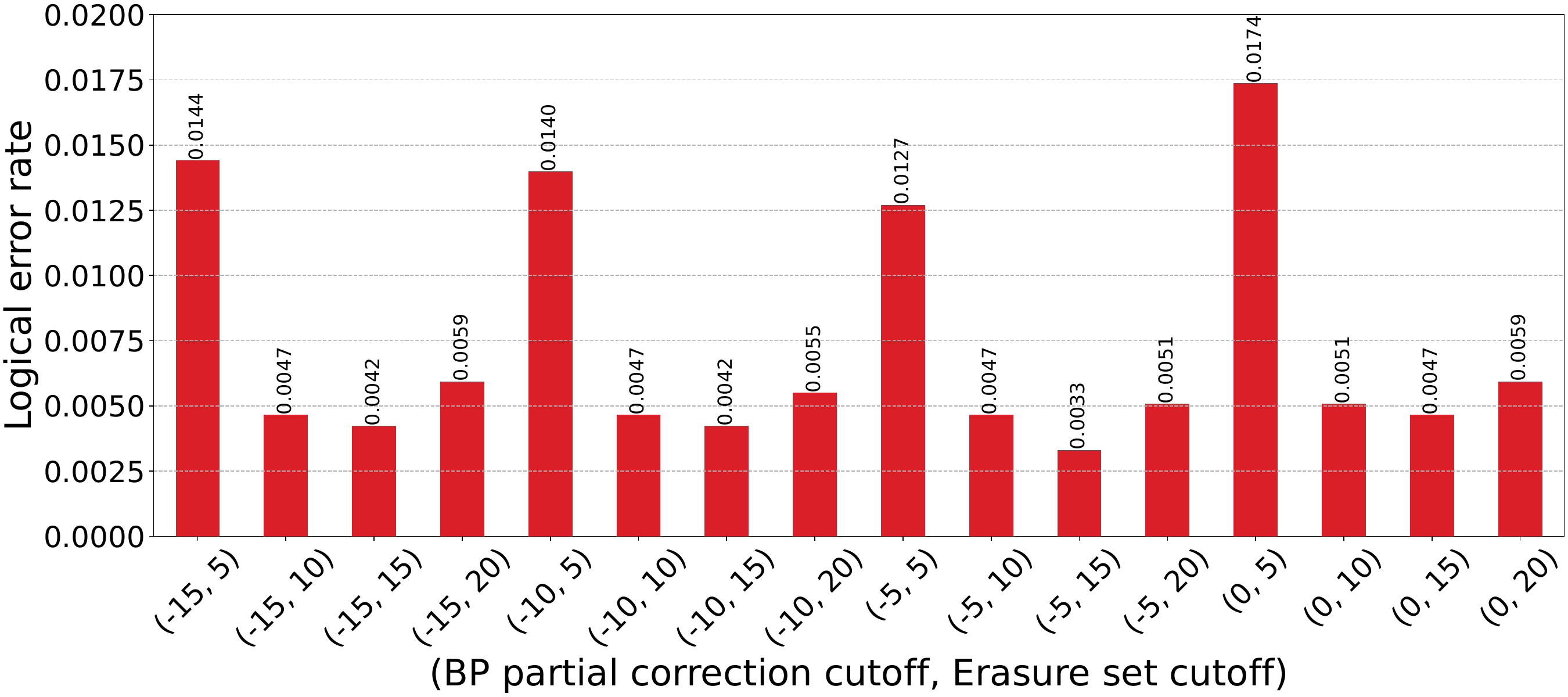}
    \caption{Different choices of the partial correction and erasure set cutoffs yield differences in the accuracy of the decoder ($p=0.1\%$). }
    \label{fig:bp_corr_cutoff}
\end{figure}

\paragraph{Hyperparameter optimization}
The partial-correction cutoff should be chosen to maximize the reduction in syndrome weight, thereby reducing the initial number of clusters presented to the cluster decoder.  
Similarly, the erasure-set cutoff should be chosen to minimize the number of additional clusters introduced due to erasures.  
\fig{bp_corr_cutoff} illustrates the effect of varying these cutoffs: selecting an erasure-set cutoff that is too low (e.g., $5$ in this example) significantly degrades performance, whereas the decoder is less sensitive to the partial-correction cutoff.

\begin{figure}[h]
    \centering
    (a)\includegraphics[width=0.8\linewidth]{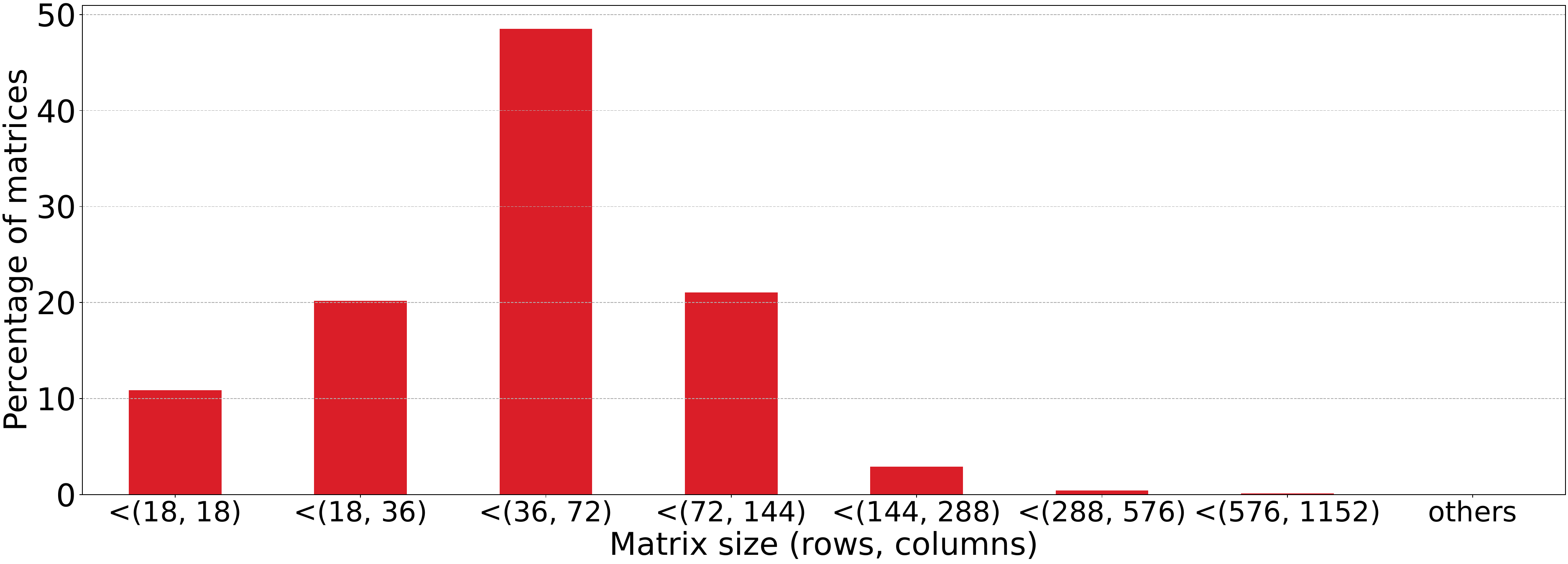}
    \\
    (b)\includegraphics[width=0.8\linewidth]{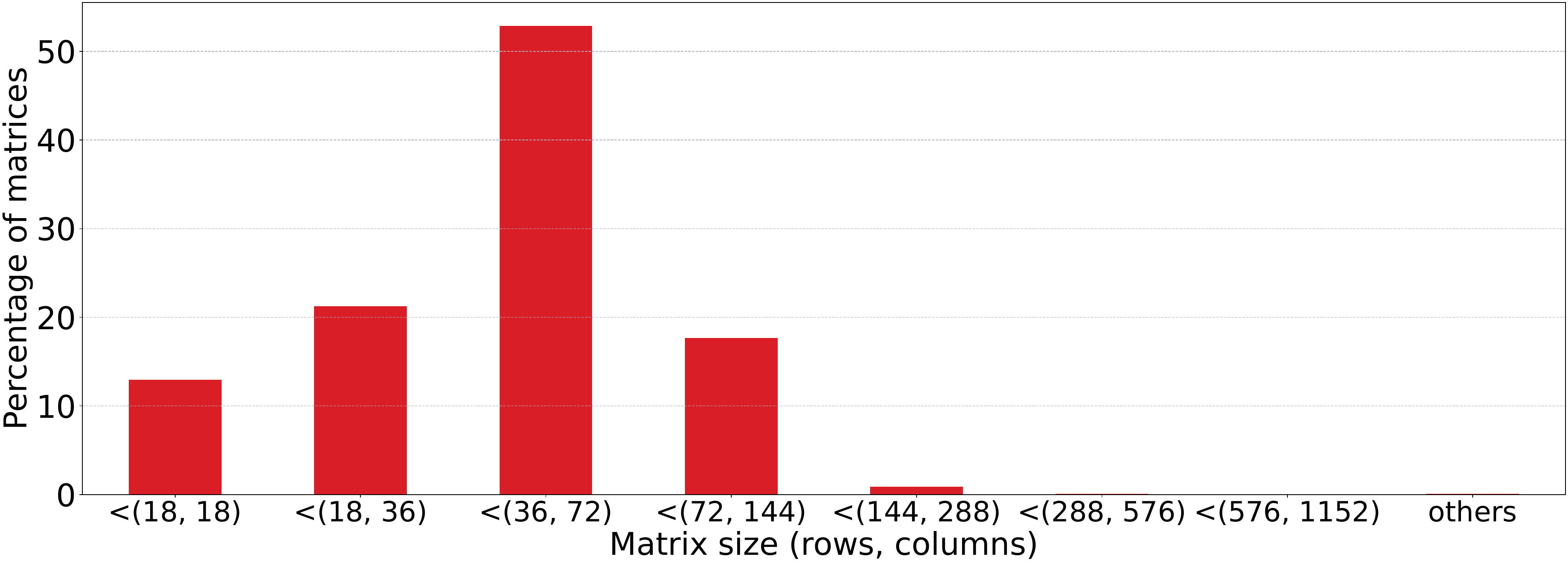}
    \caption{Distribution of matrix sizes encountered by the cluster decoder for (a) the gross code, (b) the two-gross code ($p=0.1\%$).}
    \label{fig:problem-sizes}
\end{figure}

\paragraph{Problem sizes encountered with the cluster decoder}
An advantage offered by the cluster decoder is that it can reduce the decoding problem to more manageable sizes. \fig{problem-sizes} shows the distribution of matrix sizes (as percentage of all matrices seen) encountered by the cluster decoder. More than 90\% of the matrices encountered by the decoder can fit within a $72\times 144$ matrix. 

\subsection{Analysis details for Relay decoding and standard ordered statistics decoding}
\label{app:relay-resources}

Our cycle estimates for standard OSD are derived from Ref.~\cite{bascones2025exploring}, which defines a deterministic cycle count of $3M + 2N$ cycles. 
Ref.~\cite{bascones2025exploring} used a different FPGA for their analysis, we thus make a rough translation of the utilization reported in~\cite{bascones2025exploring} to our chosen FPGA, the AMD VU19P.
This translation was performed by taking the number of LUTs, flip-flops, BRAMs reported in~\cite{bascones2025exploring} and dividing those numbers by the total available resources on the VU19P.

Our cycle count estimates for Relay assume $\alpha=2$ FPGA cycles per Relay iteration, consistent with the analysis in Ref.~\cite{maurer2025realtimedecoding}. 
For this analysis, Relay parameters are $S=1$, $T=80$ for the first leg and $T=60$ thereafter, $R=300$, and $\gamma_0=0.125$ for the first leg, with constants as defined in \sec{bp-decoding}.
For subsequent legs, $\gamma$ is drawn uniformly from $[-0.24,\,0.66]$ (gross) or $[-0.161,\,0.815]$ (two-gross). 
We retrieve the resource utilization reported in Ref.~\cite{maurer2025realtimedecoding} in \tab{fpga_resources}, since the same FPGA was used for that work.

Both standard OSD and Relay did not report resource utilization for the two-gross code, which is why they are not reported in \tab{fpga_resources}.

\bibliographystyle{alpha}
\bibliography{references}

@article{toshio2025decoder,
  title={Decoder Switching: Breaking the Speed-Accuracy Tradeoff in Real-Time Quantum Error Correction},
  author={Toshio, Riki and Kishi, Kaito and Fujisaki, Jun and Oshima, Hirotaka and Sato, Shintaro and Fujii, Keisuke},
  journal={arXiv preprint arXiv:2510.25222},
  year={2025}
}

@article{kurman2025benchmarking,
  title={Benchmarking the ability of a controller to execute quantum error corrected non-Clifford circuits},
  author={Kurman, Yaniv and Ella, Lior and Szmuk, Ramon and Wertheim, Oded and Dorschner, Benedikt and Stanwyck, Sam and Cohen, Yonatan},
  journal={IEEE Transactions on Quantum Engineering},
  year={2025},
  publisher={IEEE}
}

@article{skoric2023parallel,
  title={Parallel window decoding enables scalable fault tolerant quantum computation},
  author={Skoric, Luka and Browne, Dan E and Barnes, Kenton M and Gillespie, Neil I and Campbell, Earl T},
  journal={Nature Communications},
  volume={14},
  number={1},
  pages={7040},
  year={2023},
  publisher={Nature Publishing Group UK London}
}

@article{tan2023scalable,
  title={Scalable surface-code decoders with parallelization in time},
  author={Tan, Xinyu and Zhang, Fang and Chao, Rui and Shi, Yaoyun and Chen, Jianxin},
  journal={PRX Quantum},
  volume={4},
  number={4},
  pages={040344},
  year={2023},
  publisher={APS}
}

@article{beverland2025fail,
  author       = {Michael E. Beverland and Malcolm Carroll and Andrew W. Cross and Theodore J. Yoder},
  title        = {Fail fast: techniques to probe rare events in quantum error correction},
  journal      = {arXiv preprint arXiv:2511.15177},
  year         = {2025},
  url          = {https://arxiv.org/abs/2511.15177},
  doi          = {},
  note         = {Submitted 20 November 2025}
}

@article{caune2023belief,
  title={Belief propagation as a partial decoder},
  author={Caune, Laura and Reid, Brendan and Camps, Joan and Campbell, Earl},
  journal={arXiv preprint arXiv:2306.17142},
  year={2023}
}

@article{liu2019neural,
  title={Neural belief-propagation decoders for quantum error-correcting codes},
  author={Liu, Ye-Hua and Poulin, David},
  journal={Physical review letters},
  volume={122},
  number={20},
  pages={200501},
  year={2019},
  publisher={APS}
}

@article{yoder2025tour,
  title={Tour de gross: A modular quantum computer based on bivariate bicycle codes},
  author={Yoder, Theodore J and Schoute, Eddie and Rall, Patrick and Pritchett, Emily and Gambetta, Jay M and Cross, Andrew W and Carroll, Malcolm and Beverland, Michael E},
  journal={arXiv preprint arXiv:2506.03094},
  year={2025}
}

@article{blue2025machine,
  title={Machine Learning Decoding of Circuit-Level Noise for Bivariate Bicycle Codes},
  author={Blue, John and Avlani, Harshil and He, Zhiyang and Ziyin, Liu and Chuang, Isaac L},
  journal={arXiv preprint arXiv:2504.13043},
  year={2025}
}

@article{hu2025efficient,
  title={Efficient and universal neural-network decoder for stabilizer-based quantum error correction},
  author={Hu, Gengyuan and Ouyang, Wanli and Lu, Chao-Yang and Lin, Chen and Zhong, Han-Sen},
  journal={arXiv preprint arXiv:2502.19971},
  year={2025}
}

@article{wu2025minimum,
  title={Minimum-weight parity factor decoder for quantum error correction},
  author={Wu, Yue and Li, Binghong and Chang, Kathleen and Puri, Shruti and Zhong, Lin},
  journal={arXiv preprint arXiv:2508.04969},
  year={2025}
}

@article{murphy2013loopy,
  title={Loopy belief propagation for approximate inference: An empirical study},
  author={Murphy, Kevin and Weiss, Yair and Jordan, Michael I},
  journal={arXiv preprint arXiv:1301.6725},
  year={2013}
}

@article{chytas2025enhanced,
  title={Enhanced min-sum decoding of quantum codes using previous iteration dynamics},
  author={Chytas, Dimitris and Raveendran, Nithin and Vasic, Bane},
  journal={arXiv preprint arXiv:2501.05021},
  year={2025}
}

@article{koutsioumpas2025automorphism,
  title={Automorphism Ensemble Decoding of Quantum LDPC Codes},
  author={Koutsioumpas, Stergios and Sayginel, Hasan and Webster, Mark and Browne, Dan E},
  journal={arXiv preprint arXiv:2503.01738},
  year={2025}
}

@article{yin2024symbreak,
  title={SymBreak: Mitigating Quantum Degeneracy Issues in QLDPC Code Decoders by Breaking Symmetry},
  author={Yin, Keyi and Fang, Xiang and Ruan, Jixuan and Zhang, Hezi and Tullsen, Dean and Sornborger, Andrew and Liu, Chenxu and Li, Ang and Humble, Travis and Ding, Yufei},
  journal={arXiv preprint arXiv:2412.02885},
  year={2024}
}

@article{demarti2024almost,
  title={An almost-linear time decoding algorithm for quantum LDPC codes under circuit-level noise},
  author={deMarti iOlius, Antonio and Etxezarreta Martinez, Imanol and Roffe, Joschka and Etxezarreta Martinez, Josu},
  journal={arXiv e-prints},
  pages={arXiv--2409},
  year={2024}
}

@article{wolanski2024ambiguity,
  title={Ambiguity Clustering: an accurate and efficient decoder for qLDPC codes},
  author={Wolanski, Stasiu and Barber, Ben},
  journal={arXiv preprint arXiv:2406.14527},
  year={2024}
}

@article{hillmann2024localized,
  title={Localized statistics decoding: A parallel decoding algorithm for quantum low-density parity-check codes},
  author={Hillmann, Timo and Berent, Lucas and Quintavalle, Armanda O and Eisert, Jens and Wille, Robert and Roffe, Joschka},
  journal={arXiv preprint arXiv:2406.18655},
  volume={10},
  year={2024}
}

@article{demarti2024closed,
  title={The closed-branch decoder for quantum LDPC codes},
  author={deMarti iOlius, Antonio and Etxezarreta Martinez, Josu},
  journal={arXiv e-prints},
  pages={arXiv--2402},
  year={2024}
}

@inproceedings{yao2024belief,
  title={Belief propagation decoding of quantum LDPC codes with guided decimation},
  author={Yao, Hanwen and Laban, Waleed Abu and H{\"a}ger, Christian and i Amat, Alexandre Graell and Pfister, Henry D},
  booktitle={2024 IEEE International Symposium on Information Theory (ISIT)},
  pages={2478--2483},
  year={2024},
  organization={IEEE}
}

@article{du2024check,
  title={Check-agnosia based post-processor for message-passing decoding of quantum LDPC codes},
  author={Du Crest, Julien and Garcia-Herrero, Francisco and Mhalla, Mehdi and Savin, Valentin and Valls, Javier},
  journal={Quantum},
  volume={8},
  pages={1334},
  year={2024},
  publisher={Verein zur F{\"o}rderung des Open Access Publizierens in den Quantenwissenschaften}
}

@inproceedings{du2022stabilizer,
  title={Stabilizer inactivation for message-passing decoding of quantum LDPC codes},
  author={Du Crest, Julien and Mhalla, Mehdi and Savin, Valentin},
  booktitle={2022 IEEE Information Theory Workshop (ITW)},
  pages={488--493},
  year={2022},
  organization={IEEE}
}

@article{gong2024toward,
  title={Toward low-latency iterative decoding of QLDPC codes under circuit-level noise},
  author={Gong, Anqi and Cammerer, Sebastian and Renes, Joseph M},
  journal={arXiv preprint arXiv:2403.18901},
  year={2024}
}

@article{tsubouchi2025degeneracy,
  title={Degeneracy Cutting: A Local and Efficient Post-Processing for Belief Propagation Decoding of Quantum Low-Density Parity-Check Codes},
  author={Tsubouchi, Kento and Yamasaki, Hayata and Tamiya, Shiro},
  journal={arXiv preprint arXiv:2510.08695},
  year={2025}
}

@article{Berent_2024,
   title={Decoding quantum color codes with MaxSAT},
   volume={8},
   ISSN={2521-327X},
   url={http://dx.doi.org/10.22331/q-2024-10-23-1506},
   DOI={10.22331/q-2024-10-23-1506},
   journal={Quantum},
   publisher={Verein zur Forderung des Open Access Publizierens in den Quantenwissenschaften},
   author={Berent, Lucas and Burgholzer, Lukas and Derks, Peter-Jan H.S. and Eisert, Jens and Wille, Robert},
   year={2024},
   month=oct, pages={1506} }

@misc{noormandipour2024maxsatdecodersarbitrarycss,
      title={MaxSAT decoders for arbitrary CSS codes}, 
      author={Mohammadreza Noormandipour and Tobias Haug},
      year={2024},
      eprint={2410.01673},
      archivePrefix={arXiv},
      primaryClass={quant-ph},
      url={https://arxiv.org/abs/2410.01673}, 
}

@article{shutty2022decoding,
  title={Decoding merged color-surface codes and finding fault-tolerant Clifford circuits using solvers for satisfiability modulo theories},
  author={Shutty, Noah and Chamberland, Christopher},
  journal={Physical Review Applied},
  volume={18},
  number={1},
  pages={014072},
  year={2022},
  publisher={APS}
}

@article{higgott2023improved,
  title={Improved decoding of circuit noise and fragile boundaries of tailored surface codes},
  author={Higgott, Oscar and Bohdanowicz, Thomas C and Kubica, Aleksander and Flammia, Steven T and Campbell, Earl T},
  journal={Physical Review X},
  volume={13},
  number={3},
  pages={031007},
  year={2023},
  publisher={APS}
}

@article{beni2025tesseract,
  title={Tesseract: A search-based decoder for quantum error correction},
  author={Beni, Laleh Aghababaie and Higgott, Oscar and Shutty, Noah},
  journal={arXiv preprint arXiv:2503.10988},
  year={2025}
}

@article{vassilevska2023omega,
  title     = {New Bounds for Matrix Multiplication: from Alpha to Omega},
  author    = {Vassilevska Williams, Virginia and Xu, Yinzhan and Xu, Zixuan and Zhou, Renfei},
  journal   = {SIAM Journal on Computing},
  year      = {2023},
  note      = {arXiv:2307.07970},
  url       = {https://arxiv.org/abs/2307.07970},
  eprint    = {2307.07970},
  archivePrefix = {arXiv},
  primaryClass  = {cs.DS}
}

@article{maan2025decoding,
  title         = {Decoding Correlated Errors in Quantum {LDPC} Codes},
  author        = {Maan, Arshpreet Singh and Garcia Herrero, Francisco and Paler, Alexandru and Savin, Valentin},
  year          = {2025},
  eprint        = {2510.14060},
  archivePrefix = {arXiv},
  primaryClass  = {quant-ph},
  doi           = {10.48550/arXiv.2510.14060},
  url           = {https://arxiv.org/abs/2510.14060},
  note          = {arXiv:2510.14060}
}

@article{bascones2025exploring,
  title = {Exploring the {FPGA} and {ASIC} design space of belief propagation and ordered statistics decoders for Quantum Error Correction Codes},
  url = {http://dx.doi.org/10.21203/rs.3.rs-6420548/v1},
  DOI = {10.21203/rs.3.rs-6420548/v1},
  publisher = {Springer Science and Business Media LLC},
  author = {Bascones,  Daniel and Garcia-Herrero,  Francisco and Valls,  Javier},
  year = {2025},
  month = jul 
}

@inproceedings{liyanage2023scalable,
  title={Scalable quantum error correction for surface codes using {FPGA}},
  author={Liyanage, Namitha and Wu, Yue and Deters, Alexander and Zhong, Lin},
  booktitle={2023 IEEE International Conference on Quantum Computing and Engineering (QCE)},
  volume={1},
  pages={916--927},
  year={2023},
  organization={IEEE}
}

@article{ziad2024local,
  title={Local Clustering Decoder: a fast and adaptive hardware decoder for the surface code},
  author={Ziad, Abbas B and Zalawadiya, Ankit and Topal, Canberk and Camps, Joan and Geh{\'e}r, Gy{\"o}rgy P and Stafford, Matthew P and Turner, Mark L},
  journal={arXiv preprint arXiv:2411.10343},
  year={2024}
}

@inproceedings{wu2025micro,
  title={Micro blossom: Accelerated minimum-weight perfect matching decoding for quantum error correction},
  author={Wu, Yue and Liyanage, Namitha and Zhong, Lin},
  booktitle={Proceedings of the 30th ACM International Conference on Architectural Support for Programming Languages and Operating Systems, Volume 2},
  pages={639--654},
  year={2025}
}

@inproceedings{vittal2023astrea,
  title={Astrea: Accurate quantum error-decoding via practical minimum-weight perfect-matching},
  author={Vittal, Suhas and Das, Poulami and Qureshi, Moinuddin},
  booktitle={Proceedings of the 50th Annual International Symposium on Computer Architecture},
  pages={1--16},
  year={2023}
}

@article{muller2025improved,
  title={Improved belief propagation is sufficient for real-time decoding of quantum memory},
  author={M{\"u}ller, Tristan and Alexander, Thomas and Beverland, Michael E and B{\"u}hler, Markus and Johnson, Blake R and Maurer, Thilo and Vandeth, Drew},
  journal={arXiv preprint arXiv:2506.01779},
  year={2025}
}

@article{fossorier2002iterative,
  title={Iterative reliability-based decoding of low-density parity check codes},
  author={Fossorier, Marc PC},
  journal={IEEE Journal on selected Areas in Communications},
  volume={19},
  number={5},
  pages={908--917},
  year={2002},
  publisher={IEEE}
}

@article{fossorier1995soft,
  title={Soft-decision decoding of linear block codes based on ordered statistics},
  author={Fossorier, Marc PC and Lin, Shu},
  journal={IEEE Transactions on information Theory},
  volume={41},
  number={5},
  pages={1379--1396},
  year={1995},
  publisher={IEEE}
}

@inproceedings{papaphilippou2020adaptable,
  title={An adaptable high-throughput {FPGA} merge sorter for accelerating database analytics},
  author={Papaphilippou, Philippos and Brooks, Chris and Luk, Wayne},
  booktitle={2020 30th International Conference on Field-Programmable Logic and Applications (FPL)},
  pages={65--72},
  year={2020},
  organization={IEEE}
}

@article{zuluaga2016streaming,
  title={Streaming sorting networks},
  author={Zuluaga, Marcela and Milder, Peter and P{\"u}schel, Markus},
  journal={ACM Transactions on Design Automation of Electronic Systems (TODAES)},
  volume={21},
  number={4},
  pages={1--30},
  year={2016},
  publisher={ACM New York, NY, USA}
}

@article{li2020extended,
  title={An extended nonstrict partially ordered set-based configurable linear sorter on {FPGA}s},
  author={Li, Dalin and Huang, Lan and Gao, Teng and Feng, Yang and Tavares, Adriano and Wang, Kangping},
  journal={IEEE Transactions on Computer-Aided Design of Integrated Circuits and Systems},
  volume={39},
  number={5},
  pages={1031--1044},
  year={2020},
  publisher={IEEE}
}

@article{ortiz2011streaming,
  title={A Streaming High-Throughput Linear Sorter System with Contention Buffering},
  author={Ortiz, Jorge and Andrews, David},
  journal={International Journal of Reconfigurable Computing},
  volume={2011},
  number={1},
  pages={963539},
  year={2011},
  publisher={Wiley Online Library}
}

@article{lee1995shift,
  title={A shift register architecture for high-speed data sorting},
  author={Lee, Chen-Yi and Tsai, Jer-Min},
  journal={Journal of VLSI signal processing systems for signal, image and video technology},
  volume={11},
  number={3},
  pages={273--280},
  year={1995},
  publisher={Springer}
}

@inproceedings{Zhou2025,
  series = {MICRO 2025},
  title = {Vegapunk: Accurate and Fast Decoding for Quantum LDPC Codes with Online Hierarchical Algorithm and Sparse Accelerator},
  url = {http://dx.doi.org/10.1145/3725843.3756084},
  DOI = {10.1145/3725843.3756084},
  booktitle = {Proceedings of the 2025 58th IEEE/ACM International Symposium on Microarchitecture},
  publisher = {ACM},
  author = {Zhou,  Kaiwen and Lu,  Liqiang and Xiang,  Debin and Tao,  Chenning and Wu,  Anbang and Leng,  Jingwen and Liu,  Fangxin and Chen,  Mingshuai and Yin,  Jianwei},
  year = {2025},
  month = oct,
  pages = {719–732},
  collection = {MICRO 2025}
}

@article{panteleev2021degenerate,
  title={Degenerate quantum LDPC codes with good finite length performance},
  author={Panteleev, Pavel and Kalachev, Gleb},
  journal={Quantum},
  volume={5},
  pages={585},
  year={2021},
  publisher={Verein zur F{\"o}rderung des Open Access Publizierens in den Quantenwissenschaften}
}

@inproceedings{Ninkovic2024,
  title = {Decoding Quantum LDPC Codes Using Graph Neural Networks},
  url = {http://dx.doi.org/10.1109/GLOBECOM52923.2024.10901425},
  DOI = {10.1109/globecom52923.2024.10901425},
  booktitle = {GLOBECOM 2024 - 2024 IEEE Global Communications Conference},
  publisher = {IEEE},
  author = {Ninkovic,  Vukan and Kundacina,  Ognjen and Vukobratovic,  Dejan and H\"{a}ger,  Christian and Amat,  Alexandre Graell i},
  year = {2024},
  month = dec,
  pages = {3479–3484}
}

@article{khalid2025impacts,
  title={Impacts of Decoder Latency on Utility-Scale Quantum Computer Architectures},
  author={Khalid, Abdullah and Silva, Allyson and Dagnew, Gebremedhin A and Dvir, Tom and Wertheim, Oded and Gruda, Motty and Kong, Xiangzhou and Kramer, Mia and Webb, Zak and Scherer, Artur and others},
  journal={arXiv preprint arXiv:2511.10633},
  year={2025}
}

@article{Maurya2024,
  doi = {10.48550/ARXIV.2406.17995},
  url = {https://arxiv.org/abs/2406.17995},
  journal={arXiv preprint arXiv:2406.17995},
  author = {Maurya,  Satvik and Molavi,  Abtin and Albarghouthi,  Aws and Tannu,  Swamit},
  title = {Managing Classical Processing Requirements for Quantum Error Correction},
  publisher = {arXiv},
  year = {2024},
  copyright = {Creative Commons Attribution Non Commercial No Derivatives 4.0 International}
}

@inproceedings{Gong2024,
  title = {Graph Neural Networks for Enhanced Decoding of Quantum LDPC Codes},
  url = {http://dx.doi.org/10.1109/ISIT57864.2024.10619589},
  DOI = {10.1109/isit57864.2024.10619589},
  booktitle = {2024 IEEE International Symposium on Information Theory (ISIT)},
  publisher = {IEEE},
  author = {Gong,  Anqi and Cammerer,  Sebastian and Renes,  Joseph M.},
  year = {2024},
  month = jul,
  pages = {2700–2705}
}

@misc{bravyi2024github,
  author       = {Sergey Bravyi},
  title        = {BivariateBicycleCodes},
  year         = {2024},
  howpublished = {\url{https://github.com/sbravyi/BivariateBicycleCodes}},
  note         = {GitHub repository; commit fa77e33; accessed 19~Sep~2025}
}

@misc{amd_synthesis,
  author       = {AMD},
  title        = {Vivado Design Suite User Guide},
  year         = {2022},
  howpublished = {\url{https://www.xilinx.com/support/documents/sw_manuals/xilinx2022_2/ug901-vivado-synthesis.pdf}},
  note         = {Accessed 19~Sep~2025}
}

@article{bravyi2024high,
  title={High-threshold and low-overhead fault-tolerant quantum memory},
  author={Bravyi, Sergey and Cross, Andrew W and Gambetta, Jay M and Maslov, Dmitri and Rall, Patrick and Yoder, Theodore J},
  journal={Nature},
  volume={627},
  number={8005},
  pages={778--782},
  year={2024},
  publisher={Nature Publishing Group UK London}
}

@article{Fossorier1995,
  author={Fossorier, M.P.C. and Shu Lin},
  journal={IEEE Transactions on Information Theory}, 
  title={Soft-decision decoding of linear block codes based on ordered statistics}, 
  year={1995},
  volume={41},
  number={5},
  pages={1379-1396},
  doi={10.1109/18.412683}}

@article{terhal2015quantum,
  title={Quantum error correction for quantum memories},
  author={Terhal, Barbara M},
  journal={Reviews of Modern Physics},
  volume={87},
  number={2},
  pages={307--346},
  year={2015},
  publisher={APS}
}

@article{delfosse2022toward,
  title={Toward a union-find decoder for quantum LDPC codes},
  author={Delfosse, Nicolas and Londe, Vivien and Beverland, Michael E},
  journal={IEEE Transactions on Information Theory},
  volume={68},
  number={5},
  pages={3187--3199},
  year={2022},
  publisher={IEEE}
}

@inproceedings{gentleman1982matrix,
  title={Matrix triangularization by systolic arrays},
  author={Gentleman, W Morven and Kung, HT},
  booktitle={Real-time signal processing IV},
  volume={298},
  pages={19--26},
  year={1982},
  organization={SPIE}
}

@article{wang1993systolic,
  author    = {C.-L. Wang and J.-L. Lin},
  title     = {A Systolic Architecture for Computing Inverses in GF(2m)},
  journal   = {IEEE Transactions on Computers},
  year      = {1993}
}

@article{rupp2011hardware,
  title={Hardware SLE solvers: Efficient building blocks for cryptographic and cryptanalyticapplications},
  author={Rupp, Andy and Eisenbarth, Thomas and Bogdanov, Andrey and Grieb, Oliver},
  journal={Integration},
  volume={44},
  number={4},
  pages={290--304},
  year={2011},
  publisher={Elsevier}
}

@inproceedings{jasinski2010gf2,
  title={An improved GF (2) matrix inverter with linear time complexity},
  author={Jasinski, Ricardo P and Pedroni, Volnei A and Gortan, Antonio and Godoy Jr, Walter},
  booktitle={2010 International Conference on Reconfigurable Computing and {FPGA}s},
  pages={322--327},
  year={2010},
  organization={IEEE}
}

@article{hu2024universal,
  title={Universal Gaussian elimination hardware for cryptographic purposes},
  author={Hu, Jingwei and Wang, Wen and Gaj, Kris and Chen, Donglong and Wang, Huaxiong},
  journal={Journal of Cryptographic Engineering},
  volume={14},
  number={2},
  pages={383--397},
  year={2024},
  publisher={Springer}
}

@inproceedings{geiselmann2003hardware,
  author    = {T. Geiselmann and R. Steinwandt},
  title     = {Hardware to Solve Sparse Systems of Linear Equations over GF(2)},
  booktitle = {CHES 2003, LNCS 2779},
  pages     = {51--61},
  year      = {2003}
}

@inproceedings{scholl2013hardware,
  title={Hardware implementations of Gaussian elimination over GF (2) for channel decoding algorithms},
  author={Scholl, Stefan and Stumm, Christopher and Wehn, Norbert},
  booktitle={2013 Africon},
  pages={1--5},
  year={2013},
  organization={IEEE}
}

@article{valls2021syndrome,
  title={Syndrome-based min-sum vs OSD-0 decoders: {FPGA} implementation and analysis for quantum LDPC codes},
  author={Valls, Javier and Garcia-Herrero, Francisco and Raveendran, Nithin and Vasi{\'c}, Bane},
  journal={IEEE Access},
  volume={9},
  pages={138734--138743},
  year={2021},
  publisher={IEEE}
}

@inproceedings{scholl2014hardware,
  title={Hardware implementation of a Reed-Solomon soft decoder based on information set decoding},
  author={Scholl, Stefan and Wehn, Norbert},
  booktitle={2014 Design, Automation \& Test in Europe Conference \& Exhibition (DATE)},
  pages={1--6},
  year={2014},
  organization={IEEE}
}

@article{parkinson1984compact,
  author    = {R. Parkinson and B. Wunderlich},
  title     = {A Compact Algorithm for Gaussian Elimination over GF(2) on Highly Parallel Computers},
  journal   = {Parallel Computing},
  volume    = {1},
  number    = {1},
  year      = {1984}
}

@article{Ott2025,
  title={Decision-tree decoders for general quantum LDPC codes},
  author={Ott, Kai R and Het{\'e}nyi, Bence and Beverland, Michael E},
  journal={arXiv preprint arXiv:2502.16408},
  year={2025}
}

@article{Hochet1989,
  title = {Systolic Gaussian elimination over GF(p) with partial pivoting},
  volume = {38},
  ISSN = {0018-9340},
  url = {http://dx.doi.org/10.1109/12.29471},
  DOI = {10.1109/12.29471},
  number = {9},
  journal = {IEEE Transactions on Computers},
  publisher = {Institute of Electrical and Electronics Engineers (IEEE)},
  author = {Hochet,  B. and Quinton,  P. and Robert,  Y.},
  year = {1989},
  pages = {1321–1324}
}

@inproceedings{Rupp2006,
  title = {A Parallel Hardware Architecture for fast Gaussian Elimination over GF(2)},
  volume = {3659},
  url = {http://dx.doi.org/10.1109/FCCM.2006.12},
  DOI = {10.1109/fccm.2006.12},
  booktitle = {2006 14th Annual IEEE Symposium on Field-Programmable Custom Computing Machines},
  publisher = {IEEE},
  author = {Rupp,  A. and Pelzl,  J. and Paar,  C. and Mertens,  M.C. and Bogdanov,  A.},
  year = {2006},
  month = apr,
  pages = {237–248}
}

@inproceedings{wang2016solving,
  title={Solving large systems of linear equations over GF (2) on {FPGA}s},
  author={Wang, Wen and Szefer, Jakub and Niederhagen, Ruben},
  booktitle={2016 International Conference on ReConFigurable Computing and {FPGA}s (ReConFig)},
  pages={1--7},
  year={2016},
  organization={IEEE}
}

@inproceedings{pamuk2011fpga,
  title={An {FPGA} implementation architecture for decoding of polar codes},
  author={Pamuk, Alptekin},
  booktitle={2011 8th International symposium on wireless communication systems},
  pages={437--441},
  year={2011},
  organization={IEEE}
}

@inproceedings{wang2017fpga,
  title={{FPGA}-based key generator for the Niederreiter cryptosystem using binary Goppa codes},
  author={Wang, Wen and Szefer, Jakub and Niederhagen, Ruben},
  booktitle={International Conference on Cryptographic Hardware and Embedded Systems},
  pages={253--274},
  year={2017},
  organization={Springer}
}

@article{Shoufan2010,
  title = {A Novel Cryptoprocessor Architecture for the McEliece Public-Key Cryptosystem},
  volume = {59},
  ISSN = {0018-9340},
  url = {http://dx.doi.org/10.1109/TC.2010.115},
  DOI = {10.1109/tc.2010.115},
  number = {11},
  journal = {IEEE Transactions on Computers},
  publisher = {Institute of Electrical and Electronics Engineers (IEEE)},
  author = {Shoufan,  Abdulhadi and Wink,  Thorsten and Molter,  H. Gregor and Huss,  Sorin A. and Kohnert,  Eike},
  year = {2010},
  month = nov,
  pages = {1533–1546}
}

@article{delfosse2021almost,
  title={Almost-linear time decoding algorithm for topological codes},
  author={Delfosse, Nicolas and Nickerson, Naomi H},
  journal={Quantum},
  volume={5},
  pages={595},
  year={2021},
  publisher={Verein zur F{\"o}rderung des Open Access Publizierens in den Quantenwissenschaften}
}

@article{calderbank1996good,
	Author = {Calderbank, A Robert and Shor, Peter W},
	Journal = {Physical Review A},
	Number = {2},
	Pages = {1098},
	Publisher = {APS},
	Title = {Good quantum error-correcting codes exist},
	Volume = {54},
	Year = {1996}}

@article{delfosse2017almost,
  title={Almost-linear time decoding algorithm for topological codes},
  author={Delfosse, Nicolas and Nickerson, Naomi H},
  journal={arXiv preprint arXiv:1709.06218},
  year={2017}
}

@article{fowler2012surface,
	Author = {Fowler, Austin G and Mariantoni, Matteo and Martinis, John M and Cleland, Andrew N},
	Journal = {Physical Review A},
	Number = {3},
	Pages = {032324},
	Publisher = {APS},
	Title = {Surface codes: Towards practical large-scale quantum computation},
	Volume = {86},
	Year = {2012}}

@article{kovalev2013fault,
	Author = {Kovalev, Alexey A and Pryadko, Leonid P},
	Journal = {Physical Review A},
	Number = {2},
	Pages = {020304},
	Publisher = {APS},
	Title = {Fault tolerance of quantum low-density parity check codes with sublinear distance scaling},
	Volume = {87},
	Year = {2013},
	doi={10.1103/PhysRevA.87.020304}}

@article{roffe2020,
  title={Decoding across the quantum low-density parity-check code landscape},
  author={Roffe, Joschka and White, David R and Burton, Simon and Campbell, Earl},
  journal={Physical Review Research},
  volume={2},
  number={4},
  pages={043423},
  year={2020},
  publisher={APS}
}

@article{poulin2008on,
  title={On the iterative decoding of sparse quantum codes},
  author={Poulin, David and Chung, Yeojin},
  journal={Quantum Information \& Computation},
  volume={8},
  number={10},
  pages={987--1000},
  year={2008},
  publisher={Rinton Press, Incorporated}
}

@article{steane1996multiple,
	Author = {Steane, Andrew},
	Journal = {Proceedings of the Royal Society A},
	Number = {1954},
	Pages = {2551--2577},
	Publisher = {The Royal Society},
	Title = {Multiple-particle interference and quantum error correction},
	Volume = {452},
	Year = {1996}}

@incollection{HOLDSWORTH2002163,
	address = {Oxford},
	author = {B. HOLDSWORTH and R.C. WOODS},
	booktitle = {Digital Logic Design (Fourth Edition)},
	doi = {https://doi.org/10.1016/B978-075064582-9/50008-1},
	edition = {Fourth Edition},
	editor = {B. HOLDSWORTH and R.C. WOODS},
	isbn = {978-0-7506-4582-9},
	pages = {163-206},
	publisher = {Newnes},
	title = {7 - Counters and registers},
	url = {https://www.sciencedirect.com/science/article/pii/B9780750645829500081},
	year = {2002}}

@misc{FPGARouting,
    title = {A Tutorial on {FPGA} Routing},
    author = {Daniel Gomez-Prado  and Maciej Ciesielski},
    url = {http://www.gstitt.ece.ufl.edu/courses/fall19/eel4720_5721/reading/Routing.pdf},
    year = {2019},
}

@misc{IBMFPGA,
    title = {What is a field programmable gate array ({FPGA})?},
    author = {Josh Schneider and Ian Smalley},
    url = {https://www.ibm.com/think/topics/field-programmable-gate-arrays}}

@book{national1961modern,
	author = {National Physical Laboratory},
	publisher = {H.M. Stationery Office},
	series = {Notes on applied science},
	title = {Modern Computing Methods},
	url = {https://books.google.com.au/books?id=zxgb0QEACAAJ},
	year = {1961}
}

@inproceedings{cosnard1986matching,
  author       = {M. Cosnard and Y. Robert and M. Tchuente},
  title        = {Matching Parallel Algorithms with Architectures: A Case Study},
  booktitle    = {Proceedings of the IFIP Working Conference on Highly Parallel Computers},
  address      = {Nice, France},
  organization = {IFIP},
  month        = {March},
  year         = {1986},
  note         = {March 24--26, 1986}
}

@article{robert1985resolution,
  author  = {Y. Robert and M. Tchuente},
  title   = {R{\'e}solution systolique de syst\`emes lin{\'e}aires denses},
  journal = {RAIRO -- Mod{\'e}lisation et Analyse Num{\'e}rique},
  volume  = {19},
  number  = {2},
  pages   = {315--326},
  year    = {1985},
  language = {French}
}

@article{Battistel_2023,
   title={Real-time decoding for fault-tolerant quantum computing: progress, challenges and outlook},
   volume={7},
   ISSN={2399-1984},
   url={http://dx.doi.org/10.1088/2399-1984/aceba6},
   DOI={10.1088/2399-1984/aceba6},
   number={3},
   journal={Nano Futures},
   publisher={IOP Publishing},
   author={Battistel, F and Chamberland, C and Johar, K and Overwater, R W J and Sebastiano, F and Skoric, L and Ueno, Y and Usman, M},
   year={2023},
   month=aug, pages={032003} }

@online{intel_avoid_expensive_functions_2023,
  title        = {Avoid Expensive Functions},
  author       = {{Intel Corporation}},
  year         = {2023},
  month        = mar,
  date         = {2023-03-31},
  url          = {https://www.intel.com/content/www/us/en/docs/oneapi-fpga-add-on/optimization-guide/2023-1/avoid-expensive-functions.html},
  urldate      = {2025-10-19},
  organization = {Intel},
  note         = {Section of \emph{FPGA Optimization Guide for Intel\textregistered{} oneAPI Toolkits} (Doc ID 767853, version 2023.1)}
}

@misc{besta2019graphprocessingfpgastaxonomy,
      title={Graph Processing on {FPGA}s: Taxonomy, Survey, Challenges}, 
      author={Maciej Besta and Dimitri Stanojevic and Johannes De Fine Licht and Tal Ben-Nun and Torsten Hoefler},
      year={2019},
      eprint={1903.06697},
      archivePrefix={arXiv},
      primaryClass={cs.DC},
      url={https://arxiv.org/abs/1903.06697}, 
}

@inproceedings{Halstead2011,
author = {Halstead, Robert J. and Villarreal, Jason and Najjar, Walid},
title = {Exploring irregular memory accesses on {FPGA}s},
year = {2011},
isbn = {9781450311212},
publisher = {Association for Computing Machinery},
address = {New York, NY, USA},
url = {https://doi.org/10.1145/2089142.2089151},
doi = {10.1145/2089142.2089151},
booktitle = {Proceedings of the 1st Workshop on Irregular Applications: Architectures and Algorithms},
pages = {31–34},
numpages = {4},
location = {Seattle, Washington, USA},
series = {IA3 '11}
}

@INPROCEEDINGS{Luo2017,
  author={Luo, Yingyi and Wen, Xianshan and Yoshii, Kazutomo and Ogrenci-Memik, Seda and Memik, Gokhan and Finkel, Hal and Cappello, Franck},
  booktitle={2017 27th International Conference on Field Programmable Logic and Applications (FPL)}, 
  title={Evaluating irregular memory access on OpenCL {FPGA} platforms: A case study with XSBench}, 
  year={2017},
  volume={},
  number={},
  pages={1-4},
  doi={10.23919/FPL.2017.8056827}}

@misc{maurer2025realtimedecoding,
      title={Real-time decoding of the gross code memory with {FPGA}s}, 
      author={Thilo Maurer and Markus Bühler and Michael Kröner and Frank Haverkamp and Tristan Müller and Drew Vandeth and Blake R. Johnson},
      year={2025},
      eprint={2510.21600},
      archivePrefix={arXiv},
      primaryClass={quant-ph},
      url={https://arxiv.org/abs/2510.21600}, 
}

\end{document}